\documentclass[longauth]{aa} 

\usepackage{graphicx,lscape}
\usepackage{txfonts}
\usepackage{multirow}
\usepackage{booktabs}
\usepackage{siunitx}
\usepackage{soul}
\usepackage{lscape}

\def\ms{\hbox{\,m\,s$^{-1}$}}         

\begin{document}

   \title{The GAPS Programme at TNG}
   \subtitle{XXIII. HD\,164922\,d: a close-in super-Earth discovered with HARPS-N in a system with a long-period Saturn mass companion 
   \thanks{Based on observations made with the Italian Telescopio Nazionale Galileo (TNG) operated on the island of La Palma by the Fundaci\'on Galileo Galilei of the INAF (Istituto Nazionale di Astrofisica) at the Spanish Observatorio del Roque de los Muchachos (ORM) of the IAC.}}

\author{S. Benatti,\inst{1}
        M. Damasso,\inst{2}
        S. Desidera,\inst{3}
        F. Marzari,\inst{4}
        K. Biazzo,\inst{5}
        R. Claudi,\inst{3}
        M.P. Di Mauro,\inst{6} 
        A.F. Lanza,\inst{7}
        M. Pinamonti,\inst{2}
        D. Barbato,\inst{2,8}
        L. Malavolta,\inst{4}
        E. Poretti,\inst{9}
        A. Sozzetti,\inst{2}
        L. Affer,\inst{1}
        A. Bignamini,\inst{10}
        A. S. Bonomo,\inst{2} 
        F. Borsa,\inst{11}
        M. Brogi,\inst{12,2,13}
        G. Bruno,\inst{7}
        I. Carleo,\inst{14}
        R. Cosentino,\inst{9}
        E. Covino,\inst{15}
        G. Frustagli,\inst{11,16}
        P. Giacobbe,\inst{2} 
        M. Gonzalez,\inst{9}
        R. Gratton,\inst{3}
        A. Harutyunyan,\inst{9} 
        C. Knapic,\inst{10}
        G. Leto,\inst{7}
        M. Lodi,\inst{9}
        A. Maggio,\inst{1} 
        J. Maldonado,\inst{1} 
        L. Mancini,\inst{17,18,2}  
        A. Martinez Fiorenzano,\inst{9}
        G. Micela,\inst{1} 
        E. Molinari,\inst{19}
        M. Molinaro,\inst{10}
        D. Nardiello,\inst{20,3}
        V. Nascimbeni,\inst{3} 
        I. Pagano,\inst{7}
        M. Pedani,\inst{9}
        G. Piotto,\inst{4} 
        M. Rainer,\inst{21}
        G. Scandariato,\inst{7}  
        }
   \institute{INAF -- Osservatorio Astronomico di Palermo, Piazza del Parlamento 1, I-90134, Palermo, Italy 
      \email{serena.benatti@inaf.it}
   \and INAF -- Osservatorio Astrofisico di Torino, Via Osservatorio 20, I-10025, Pino Torinese (TO), Italy 
   \and INAF -- Osservatorio Astronomico di Padova, Vicolo dell'Osservatorio 5, I-35122, Padova, Italy 
   \and Dipartimento di Fisica e Astronomia "G. Galilei"-- Universt\`a degli Studi di Padova, Vicolo dell'Osservatorio 3, I-35122 Padova 
   \and INAF -- Osservatorio Astronomico di Roma, Via Frascati 33, I-00040, Monte Porzio Catone (RM), Italy
   \and INAF -- Istituto di Astrofisica e Planetologia Spaziali, Via del Fosso del Cavaliere 100, I-00133, Roma, Italy
   \and INAF -- Osservatorio Astrofisico di Catania, Via S. Sofia 78, I-95123, Catania, Italy 
   \and Observatoire de Gen\`eve, Universit\'e de Gen\`eve, 51 Chemin des Maillettes, CH-1290 Sauverny, Switzerland
   \and Fundaci\'on Galileo Galilei - INAF, Rambla Jos\'e Ana Fernandez P\'erez 7, E-38712, Bre\~na Baja, TF - Spain  
   \and INAF -- Osservatorio Astronomico di Trieste, via Tiepolo 11, I-34143 Trieste, Italy  
   \and INAF -- Osservatorio Astronomico di Brera, Via E. Bianchi 46, I-23807 Merate (LC), Italy  
   \and Department of Physics, University of Warwick, Gibbet Hill Road, Coventry, CV4 7AL, UK
   \and Centre for Exoplanets and Habitability, University of Warwick, Gibbet Hill Road, Coventry, CV4 7AL, UK
   \and Astronomy Department and Van Vleck Observatory, Wesleyan University, Middletown, CT 06459, USA
   \and INAF -- Osservatorio Astronomico di Capodimonte, Salita Moiariello 16, I-80131 Napoli, Italy
   \and Dipartimento di Fisica G. Occhialini, Università degli Studi di Milano-Bicocca, Piazza della Scienza 3, I-20126 Milano, Italy
   \and Dipartimento di Fisica, Universit\`a di Roma ``Tor Vergata'', Via della Ricerca Scientifica 1, I-00133, Roma, Italy
   \and \,Max Planck Institute for Astronomy, K\"{o}nigstuhl 17, D-69117, Heidelberg, Germany 
   \and INAF -- Osservatorio Astronomico di Cagliari, Via della Scienza 5, I-09047, Selargius (CA), Italy
   \and Aix Marseille Univ, CNRS, CNES, LAM, Marseille, France 
   \and INAF -- Osservatorio Astrofisico di Arcetri, Largo Enrico Fermi, 5, I-50125 Firenze, Italy 
}

   \date{Received ; accepted }

\abstract{Observations of exoplanetary systems show that a wide variety of architectures are possible. 
Determining the rate of occurrence of Solar System analogs -- with inner terrestrial planets and outer gas giants -- is still an open question.} 
{In the framework of the Global Architecture of Planetary Systems (GAPS) project we collected more than 300 spectra with HARPS-N at the Telescopio Nazionale Galileo for the bright G9V star HD\,164922. This target is known to host one gas giant planet in a wide orbit ($P_{\rm b}\sim$1200 days, semi-major axis $\sim 2$ au) and a Neptune-mass planet with a period $P_{\rm c}\sim$76 days. Our aim was to investigate the presence of additional low-mass companions in the inner region of the system. }
{We compared the radial velocities (RV) and the activity indices derived from the HARPS-N time series to measure the rotation period of the star and used a Gaussian process regression to describe the behaviour of the stellar activity. We exploited this information in a combined model of planetary and stellar activity signals in an RV time-series composed of almost 700 high-precision RVs, both from HARPS-N and literature data. We performed a dynamical analysis to evaluate the stability of the system and the allowed regions for additional potential companions. We performed experiments of injection and recovery of additional planetary signals to gauge the sensitivity thresholds in minimum mass and orbital separation imposed by our data.}
{Thanks to the high sensitivity of the HARPS-N dataset, we detect an additional inner super-Earth with an RV semi-amplitude of $1.3\pm0.2$ m s$^{-1}$ and a minimum mass of m$_{\rm d}\sin i =4\pm1$ M$_{\oplus}$. It orbits HD\,164922 with a period of 12.458$\pm$0.003 days. We disentangle the planetary signal from activity and measure a stellar rotation period of $\sim 42$ days. The dynamical analysis shows the long term stability of the orbits of the three-planet system and allows us to identify the permitted regions for additional planets in the semi-major axis ranges 0.18 -- 0.21 au and 0.6 -- 1.4 au. The latter partially includes the habitable zone of the system. We did not detect any planet in these regions, down to minimum detectable masses of 5 and 18 M$_{\oplus}$, respectively. 
A larger region of allowed planets is expected beyond the orbit of planet b, where our sampling rules-out bodies with minimum mass > 50 M$_{\oplus}$. The planetary orbital parameters and the location of the snow line suggest that this system has been shaped by a gas disk migration process that halted after its dissipation.}
{}


\titlerunning{The GAPS programme: a super-Earth around HD\,164922}
\authorrunning{S. Benatti et al.}
   \maketitle
%

\section{Introduction}
\label{sec:intro}
Giant planets are known to play a dominant role in the dynamical evolution of the planetary systems, since they produce gravitational interactions able to compromise the presence of low mass planets, even preventing their formation \citep{2003ApJ...582L..47A,2013ApJ...767..129M,2015ApJ...800L..22I,2018MNRAS.481.1281S}. This is supported by the observational evidence, in particular from transit surveys, that terrestrial and super-Earth planets are generally found in multiple, often packed systems with other low-mass companions (e.g., \citealt{2013ApJS..204...24B,2018AJ....155...48W}), while it seems more challenging to find systems showing both low-mass and giant planets (see e.g. the summary in \citealt{2017MNRAS.471.4494A}). Up to a few years ago, these results strengthened the idea that the Solar System architecture is not common, even if this could be the result of observing bias. Transit surveys are more sensitive toward packed coplanar configurations, disfavouring the detection of multiple systems containing giant planets.
Theoretical simulations on dynamics and orbital stability show that, under specific conditions, inner rocky planet formation and survival in systems with an outer gas giant is actually allowed, even within the habitable zone  \citep{2002A&A...384..594T,2007ApJ...660..823M,2017MNRAS.471.4494A,2018MNRAS.481.4680A}. For exoplanetary systems, only recently these simulations have been corroborated by observations (e.g.,  \citealt{2018AJ....156...92Z,2019AJ....157...52B}), favoured by a long observation baseline and cutting-edge instrumentation.
During the first decade of the exoplanets search, a large fraction of giant planets in intermediate orbit (larger than 1 au) were detected without the support of high-precision radial velocities (RV), since their RV signal allowed their detection even with medium performance instrumentation (e.g. \citealt{2000ApJ...544L.145H}: coud\'e
spectrograph at CFHT, Hamilton at Lick Observatory, coud\'e spectrograph at ESO-La Silla, coud\'e spectrograph McDonald Observatory, RV precisions between 11 and 22 m s$^{-1}$; \citealt{2004A&A...415..391M}: CORALIE at ESO-La Silla, RV precision of 3 m s$^{-1}$). On the other hand, due to the low sensitivity of those data and their sparse sampling, the possible presence of additional low-mass companions could not be revealed.
Indeed, with the state-of-the-art instrumentation currently operational (e.g. the spectrographs HIRES, HARPS, HARPS-N, ESPRESSO) it is possible to search for inner low-mass planets in systems with gas giant companions (see e.g., \citealt{2018A&A...615A.175B}). The increasing sensitivity also implies reaching the current limit in low-mass planet detection, represented by the achievable RV precision and the stellar activity ``noise''. In recent years, a lot of computational efforts have been made to develop methods to efficiently remove the stellar contribution, that can be significant even in quiet stars. One of the most employed techniques involves Gaussian process (GP) regression (e.g. \citealt{2014MNRAS.443.2517H} and references therein). 
An optimised observing strategy is also crucial both to reduce the impact of the stellar noise (e.g. \citealt{2011A&A...527A..82D,2011A&A...525A.140D}) and to improve the efficiency of the modelling methods. A well sampled and large dataset enhances our ability to model and thus mitigate the contribution of the stellar activity on the RV time series.

Since 2012 the Global Architecture of Planetary System (GAPS, \citealt{2013A&A...554A..28C,2016frap.confE..69B}) project is carrying out a radial velocity survey with the High Accuracy Radial velocity Planet Searcher (HARPS-N) high-resolution spectrograph \citep{2014SPIE.9147E..8CC} mounted at the Italian Telescopio Nazionale Galileo (TNG, La Palma, Canary Islands).
One of the first goals of the GAPS programme was the search for close-in, low-mass planets in systems hosting giant planets at wide separation. 
In this framework, we observed the bright G9V star HD\,164922 (m$_{\rm V}$ = 6.99), known to host a long period 
gas giant (planet b, $P_{\rm b} = 1201$ days, $a_b\sim$2.2 au, $m_{\rm b}\sin i_b = 107.6$ M$_{\oplus}$ $ \sim 0.3$ M$_{\rm Jup}$, detected by \cite{2006ApJ...646..505B} and a temperate Neptune-mass planet (planet c, $P_{\rm c} = 75.76$ days, $a_c \sim$0.35 au, and $m_{\rm c}\sin i_c = 12.9$ M$_{\oplus}$, announced by \citealt{2016ApJ...830...46F}, hereinafter F2016) after we started to follow-up the star. The mean value of the chromospheric activity index $\log$R$^{\rm \prime}_{\rm HK}$ is $\sim -5.05$, as reported by several authors. This indicates that the target is rather quiet from the viewpoint of stellar activity, also confirmed by a comparison with the same index for the quiet Sun ($\log$R$^{\rm \prime}_{\rm HK, min \odot} = -4.98 \pm 0.01$, see e.g. \citealt{2017ApJ...835...25E}).
Nevertheless, the robust detection of the planet c required a considerable amount of RV measurements, for a total of almost 400, mainly provided by the HIRES spectrograph
at the Keck Observatory, and with a dedicated monitoring with the Levy spectrograph mounted at the Automated Planet Finder (APF) of the Lick Observatory. 
In this work we present the detection of a super-Earth orbiting at a distance of 0.1 au with a dataset of more than 300 high-precision RVs measured from HARPS-N spectra, supported by $\sim 400$ additional RVs from literature.

This paper is organised as follows: in Section \ref{sec:obs} we describe the characteristics of our HARPS-N observations, from which we obtained the stellar parameters presented in Section \ref{sec:star}. The first evaluation of the frequency content in our data and a comparison with those available in the literature, are presented in Section \ref{sec:tsa}.
In Section \ref{sec:gpanalisi} we show the best-fit results of our RV modelling through Gaussian process regression.
In Section \ref{sec:architecture} we provide an evaluation on the system stability with some clues on the planet migration history from the present architecture, and draw our conclusions in Section \ref{sec:conc}. 


\section{HARPS-N Dataset}
\label{sec:obs}
\begin{figure}
\includegraphics[width=9cm,angle=0]{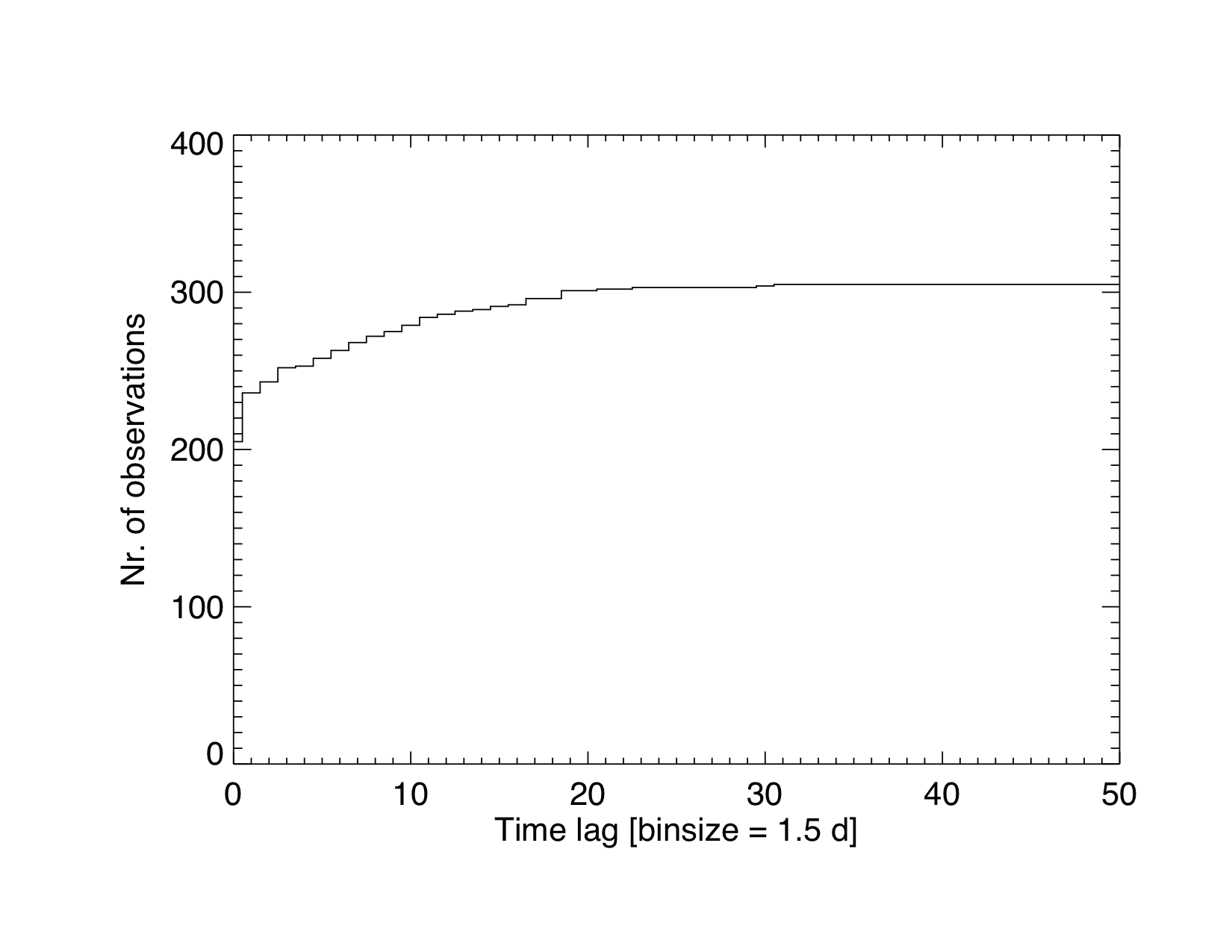}
\caption{\label{fig:hist} Cumulative histogram of the time lag between consecutive HARPS-N spectra (only data points in the same season are considered). }
\end{figure}

We observed at high cadence HD\,164922 with HARPS-N for five seasons, from March 2013 to October 2017, with an additional set of 10 spectra collected between May and June 2018. HARPS-N is a cross-dispersed high-resolution (R=115000) echelle spectrograph, in the range 3830 -- 6930 \AA. A constant control of its pressure and temperature prevents spectral drifts due to environmental parameters variations. HARPS-N is fed by two fibers: the target feeds the first fiber while the second one can be illuminated either by the sky or a simultaneous calibration source. To obtain high precision in our RV measurements, we exploited the latter technique.
We collected a total of 314 spectra with a mean signal-to-noise ratio (S/N) of 200 at 5500 \AA\, and a mean integration time of 600 s.
Figure \ref{fig:hist} shows the cumulative distribution of the time lag between consecutive data points in our sampling. To account for spectra collected with a time difference slightly larger than one day, as it can happen the night following one observation, we binned the distribution at 1.5 days. Within the same season, $\sim$80\% of our spectra were collected in the same week, while 65\% were taken in contiguous nights or in the same night. The largest difference is 46 days while in the second, third and fourth seasons we collected two spectra in the same night, with a time lag of about three hours.

The spectra were reduced through the HARPS-N Data Reduction Software (DRS), including the radial velocity extraction based on the Cross-Correlation Function (CCF) method (see \citealt{2002A&A...388..632P} and the references therein), where the observed spectrum is correlated with a K5 template, called ``mask'', depicting the features of a star with spectral type K5V. The offline version of the HARPS-N DRS, which is implemented at the INAF Trieste Observatory\footnote{http://ia2.inaf.it}, can be run by the users to re-process the data through the YABI workflow interface \citep{YABI}, with the possibility of modifying the default parameters obtaining thus a custom reduction.
The resulting CCF is the weighted and scaled average of the lines contained in the correlation mask, and can be used as a proxy for common line profile changes. For this reason it can be analysed through several diagnostics in order to investigate effects related to stellar activity.  
For instance, \cite{2018A&A...616A.155L} describe a procedure for deriving a set of CCF asymmetry indicators such as the Bisector Inverse Slope (BIS, e.g., \citealt{2001A&A...379..279Q}), the V$_{\rm asy(mod)}$ (re-adapted from the original definition by \citealt{2013A&A...557A..93F}), and the $\Delta V$ \citep{2006A&A...453..309N}. 
With YABI it is also possible to calculate the chromospheric activity index from the H\&K lines of Ca II  (S-index and $\log$R'$_{\rm HK}$) as introduced by \cite{1984ApJ...279..763N}, through a procedure  
described in \cite{2011arXiv1107.5325L}. Finally, we obtained the activity index based on the H$\alpha$ line following the recipe in \cite{2011A&A...534A..30G} and references therein.
The HARPS-N RVs, $\log R'_{\rm HK}$ and BIS time series are listed in Table \ref{RVharps} with the corresponding uncertainties.
We excluded from the time series one outlier (BJD$_{\rm TDB}$ = 2456410.6424, with an RV deviating of about 20 m s$^{-1}$ from the expected distribution) and one spectrum due to poor S/N (BJD$_{\rm TDB}$ = 2457154.5097, S/N = 43, RV$_{\rm Err}$ = 1.9 m s$^{-1}$).
 
\begin{figure}
\centering
\includegraphics[width=9cm,angle=0]{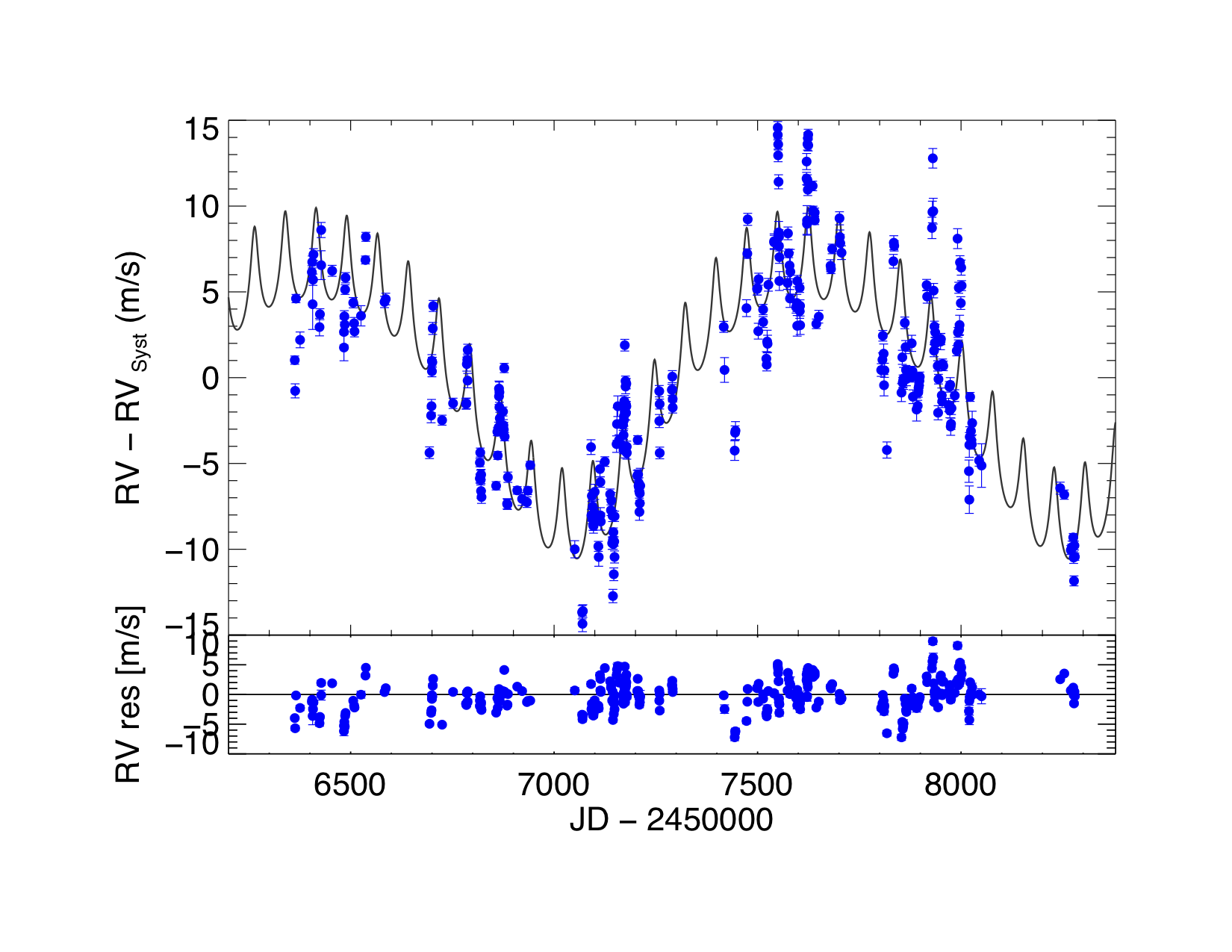}
\caption{\label{fig:rvharps} HARPS-N RV time series overimposed on the two planets model (gray line) obtained from the orbital solution presented by F2016 and the corresponding residuals (lower panel). The correction for the perspective acceleration is applied.}
\end{figure}

By using the definition in \cite{1999A&A...348.1040D} we were able to estimate the perspective acceleration of HD\,164922, adopting the parallax and the proper motions listed in Table \ref{param}. We therefore noticed that this target shows the non-negligible value of 0.26 m s$^{-1}$ yr$^{-1}$, so we decided to apply this correction to our RVs (this is not taken into account by the HARPS-N DRS). The following analysis is therefore performed on the corrected RVs, shown in Figure \ref{fig:rvharps}. The mean uncertainty of the RV measurements is 0.4 m s$^{-1}$.
\longtab{
\begin{landscape}
\begin{longtable}{ccccccccccccc}
\caption{\label{RVharps} Time series of the HARPS-N spectra for HD\,164922: radial velocity (RV), $\log R'_{\rm HK}$ and bisector span (BIS) indices with the corresponding uncertainties.}\\
\hline\hline
BJD$_{\rm TDB}$--2\,450\,000 & RV & RV$_{\rm Err}$ & $\log R'_{\rm HK}$& $\log R'_{\rm HK\,Err}$ & BIS & BIS$_{\rm Err}$ & $\Delta$V & $\Delta$V$_{\rm Err}$ & V$_{\rm asy(mod)}$ & V$_{\rm asy (mod)\,Err}$ & H$\alpha$ & H$\alpha_{\rm Err}$ \\
 & [m s$^{-1}$] & [m s$^{-1}$] & & &[m s$^{-1}$]  & [m s$^{-1}$] & & & & & & \\
\hline
\endfirsthead
\noalign{\smallskip}
\caption{continued.}\\
\hline\hline
BJD$_{\rm TDB}$--2\,450\,000 & RV & RV$_{\rm Err}$ & $\log R'_{\rm HK}$& $\log R'_{\rm HK\,Err}$ & BIS & BIS$_{\rm Err}$ & $\Delta$V & $\Delta$V$_{\rm Err}$ & V$_{\rm asy(mod)}$ & V$_{\rm asy (mod)\,Err}$ & H$\alpha$ & H$\alpha_{\rm Err}$ \\
 & [m s$^{-1}$] & [m s$^{-1}$] & & &[m s$^{-1}$]  & [m s$^{-1}$] & & & & & & \\
\hline
\endhead
\hline
\noalign{\smallskip}
\endfoot
6362.761620 & 20.3668 & 0.2410 & -5.0539 & 0.0016 &   -3.9170 &  1.4220 & -0.0073 &  0.0075 &  0.0120 &  0.0020 &  0.4333 &  0.0023 \\
6363.696888 & 20.3650 & 0.4140 & -5.0649 & 0.0036 &   -3.9440 &  1.7310 & -0.0078 &  0.0076 &  0.0133 &  0.0023 &  0.4331 &  0.0041 \\
6365.760607 & 20.3704 & 0.2360 & -5.0556 & 0.0016 &   -3.0150 &  1.3040 & -0.0071 &  0.0075 &  0.0113 &  0.0021 &  0.4336 &  0.0024 \\
6375.730565 & 20.3680 & 0.4590 & -5.0693 & 0.0040 &   -6.5920 &  1.4980 & -0.0093 &  0.0076 &  0.0152 &  0.0023 &  0.4340 &  0.0045 \\
6404.703976 & 20.3719 & 0.3710 & -5.0687 & 0.0029 &   -4.8830 &  1.6760 & -0.0083 &  0.0076 &  0.0146 &  0.0022 &  0.4328 &  0.0035 \\
6405.637835 & 20.3725 & 0.3490 & -5.0585 & 0.0026 &   -3.6740 &  1.4560 & -0.0075 &  0.0074 &  0.0128 &  0.0022 &  0.4328 &  0.0031 \\
6406.626674 & 20.3701 & 1.4720 & -5.1498 & 0.0271 &   -3.5670 &  3.6800 & -0.0104 &  0.0076 &  0.0149 &  0.0039 &  0.4330 &  0.0105 \\
6407.564872 & 20.3715 & 0.3310 & -5.0566 & 0.0025 &   -4.0530 &  1.1550 & -0.0075 &  0.0076 &  0.0118 &  0.0021 &  0.4315 &  0.0028 \\
6408.732249 & 20.3730 & 0.3410 & -5.0505 & 0.0025 &   -1.9270 &  1.8780 & -0.0063 &  0.0075 &  0.0107 &  0.0022 &  0.4327 &  0.0028 \\
6410.642418 & 20.3541 & 0.5640 & -5.0667 & 0.0053 &   -0.4020 &  1.4110 & -0.0043 &  0.0077 &  0.0035 &  0.0023 &  0.4328 &  0.0044 \\
6423.539517 & 20.3687 & 0.5110 & -5.0850 & 0.0047 &   -2.1470 &  1.7630 & -0.0058 &  0.0077 &  0.0094 &  0.0024 &  0.4325 &  0.0052 \\
6424.603068 & 20.3695 & 0.3240 & -5.0777 & 0.0024 &   -2.3670 &  1.3740 & -0.0063 &  0.0075 &  0.0102 &  0.0021 &  0.4322 &  0.0034 \\
6427.690042 & 20.3723 & 0.8390 & -5.1085 & 0.0102 &   -0.0480 &  2.0970 & -0.0033 &  0.0081 &  0.0030 &  0.0028 &  0.4345 &  0.0091 \\
6427.700564 & 20.3744 & 0.4390 & -5.0918 & 0.0037 &   -2.9240 &  1.0980 & -0.0064 &  0.0076 &  0.0098 &  0.0022 &  0.4346 &  0.0053 \\
6454.609161 & 20.3720 & 0.3180 & -5.0861 & 0.0024 &   -1.7140 &  2.0120 & -0.0059 &  0.0075 &  0.0115 &  0.0021 &  0.4340 &  0.0036 \\
6483.623256 & 20.3676 & 0.7690 & -5.1285 & 0.0100 &   -6.5890 &  2.1530 & -0.0112 &  0.0077 &  0.0179 &  0.0027 &  0.4342 &  0.0073 \\
6483.629865 & 20.3685 & 0.7570 & -5.1116 & 0.0095 &   -1.8240 &  1.8920 & -0.0051 &  0.0079 &  0.0098 &  0.0027 &  0.4339 &  0.0071 \\
6484.600379 & 20.3694 & 0.3350 & -5.0836 & 0.0025 &   -2.4370 &  1.5910 & -0.0062 &  0.0076 &  0.0104 &  0.0021 &  0.4333 &  0.0032 \\
6485.609861 & 20.3689 & 0.3980 & -5.0819 & 0.0032 &   -3.2780 &  1.3180 & -0.0073 &  0.0076 &  0.0115 &  0.0022 &  0.4328 &  0.0037 \\
6486.538857 & 20.3710 & 0.2820 & -5.0818 & 0.0020 &   -3.6930 &  1.6160 & -0.0074 &  0.0076 &  0.0118 &  0.0021 &  0.4331 &  0.0032 \\
6487.567712 & 20.3717 & 0.3020 & -5.0817 & 0.0022 &   -3.0600 &  1.6900 & -0.0063 &  0.0074 &  0.0104 &  0.0021 &  0.4334 &  0.0030 \\
6506.411771 & 20.3702 & 0.3030 & -5.0762 & 0.0021 &   -2.8970 &  1.7100 & -0.0066 &  0.0074 &  0.0117 &  0.0021 &  0.4333 &  0.0034 \\
6508.548814 & 20.3690 & 0.3360 & -5.0770 & 0.0025 &   -4.7110 &  1.0750 & -0.0080 &  0.0076 &  0.0117 &  0.0021 &  0.4321 &  0.0032 \\
6509.541862 & 20.3686 & 0.3210 & -5.0773 & 0.0024 &   -3.0040 &  1.6880 & -0.0072 &  0.0074 &  0.0111 &  0.0021 &  0.4326 &  0.0031 \\
6525.466046 & 20.3695 & 0.5900 & -5.0831 & 0.0058 &   -3.8200 &  1.6070 & -0.0061 &  0.0073 &  0.0106 &  0.0024 &  0.4315 &  0.0049 \\
6536.438353 & 20.3727 & 0.2300 & -5.0732 & 0.0015 &   -2.0220 &  1.3520 & -0.0050 &  0.0073 &  0.0088 &  0.0020 &  0.4324 &  0.0024 \\
6537.428167 & 20.3741 & 0.2620 & -5.0718 & 0.0018 &   -1.8300 &  1.4990 & -0.0055 &  0.0072 &  0.0089 &  0.0021 &  0.4323 &  0.0025 \\
6543.352067 & 20.3724 & 0.2870 & -5.0783 & 0.0020 &   -3.0260 &  1.4280 & -0.0072 &  0.0074 &  0.0109 &  0.0021 &  0.4341 &  0.0031 \\
6583.309196 & 20.3703 & 0.3350 & -5.0837 & 0.0025 &   -1.6710 &  0.8370 & -0.0058 &  0.0073 &  0.0077 &  0.0021 &  0.4334 &  0.0041 \\
6586.306001 & 20.3705 & 0.3420 & -5.0784 & 0.0026 &   -2.4270 &  1.6620 & -0.0062 &  0.0074 &  0.0105 &  0.0021 &  0.4339 &  0.0031 \\
6693.786703 & 20.3616 & 0.3430 & -5.0717 & 0.0027 &   -1.3680 &  1.5040 & -0.0050 &  0.0077 &  0.0086 &  0.0021 &  0.4337 &  0.0030 \\
6697.781048 & 20.3638 & 0.4200 & -5.0703 & 0.0034 &   -3.8020 &  1.2430 & -0.0075 &  0.0076 &  0.0115 &  0.0022 &  0.4336 &  0.0041 \\
6698.795195 & 20.3643 & 0.3890 & -5.0711 & 0.0031 &   -2.7030 &  1.4880 & -0.0054 &  0.0078 &  0.0102 &  0.0021 &  0.4333 &  0.0037 \\
6699.717729 & 20.3670 & 0.3550 & -5.0690 & 0.0029 &   -2.0800 &  1.6940 & -0.0067 &  0.0076 &  0.0116 &  0.0021 &  0.4330 &  0.0031 \\
6699.788508 & 20.3664 & 0.3180 & -5.0750 & 0.0024 &   -2.0400 &  1.7300 & -0.0052 &  0.0076 &  0.0093 &  0.0021 &  0.4338 &  0.0029 \\
6700.772227 & 20.3669 & 0.2920 & -5.0706 & 0.0021 &   -1.3380 &  1.4770 & -0.0059 &  0.0078 &  0.0092 &  0.0020 &  0.4339 &  0.0030 \\
6701.751757 & 20.3689 & 0.3500 & -5.0722 & 0.0027 &   -3.2730 &  1.5610 & -0.0068 &  0.0077 &  0.0116 &  0.0021 &  0.4339 &  0.0033 \\
6702.793327 & 20.3702 & 0.3160 & -5.0693 & 0.0022 &   -1.7250 &  1.6000 & -0.0065 &  0.0077 &  0.0104 &  0.0021 &  0.4336 &  0.0027 \\
6724.774273 & 20.3635 & 0.2910 & -5.0782 & 0.0021 &   -2.1210 &  1.1120 & -0.0059 &  0.0076 &  0.0086 &  0.0020 &  0.4338 &  0.0031 \\
6751.721120 & 20.3645 & 0.3010 & -5.0803 & 0.0024 &   -2.4770 &  1.4910 & -0.0058 &  0.0077 &  0.0105 &  0.0022 &  0.4334 &  0.0035 \\
6783.531829 & 20.3645 & 0.3060 & -5.0791 & 0.0026 &   -1.8490 &  1.7730 & -0.0061 &  0.0076 &  0.0109 &  0.0022 &  0.4330 &  0.0030 \\
6784.518425 & 20.3645 & 0.3120 & -5.0751 & 0.0026 &   -1.7730 &  1.4240 & -0.0057 &  0.0078 &  0.0098 &  0.0022 &  0.4320 &  0.0032 \\
6785.527949 & 20.3668 & 0.4120 & -5.0839 & 0.0039 &   -1.8470 &  1.1290 & -0.0056 &  0.0079 &  0.0095 &  0.0023 &  0.4321 &  0.0043 \\
6786.622326 & 20.3671 & 0.3420 & -5.0744 & 0.0029 &   -0.7350 &  1.9380 & -0.0054 &  0.0077 &  0.0095 &  0.0022 &  0.4336 &  0.0038 \\
6787.534102 & 20.3659 & 0.4190 & -5.0906 & 0.0041 &   -1.1950 &  1.7440 & -0.0058 &  0.0079 &  0.0106 &  0.0023 &  0.4339 &  0.0042 \\
6787.723369 & 20.3677 & 0.3300 & -5.0785 & 0.0028 &   -2.0570 &  1.8240 & -0.0059 &  0.0076 &  0.0112 &  0.0022 &  0.4337 &  0.0035 \\
6817.527124 & 20.3602 & 0.2630 & -5.0697 & 0.0019 &   -3.2130 &  1.6970 & -0.0066 &  0.0074 &  0.0119 &  0.0022 &  0.4336 &  0.0027 \\
6817.633167 & 20.3611 & 0.2450 & -5.0761 & 0.0018 &   -1.7220 &  1.8430 & -0.0053 &  0.0074 &  0.0102 &  0.0021 &  0.4336 &  0.0025 \\
6818.530390 & 20.3617 & 0.2570 & -5.0724 & 0.0019 &   -2.3640 &  1.8360 & -0.0057 &  0.0074 &  0.0106 &  0.0021 &  0.4336 &  0.0025 \\
6818.627232 & 20.3602 & 0.2660 & -5.0753 & 0.0020 &   -2.0510 &  2.0180 & -0.0055 &  0.0075 &  0.0116 &  0.0022 &  0.4338 &  0.0029 \\
6819.573957 & 20.3604 & 0.2920 & -5.0804 & 0.0023 &   -2.3810 &  1.5780 & -0.0053 &  0.0075 &  0.0103 &  0.0022 &  0.4334 &  0.0034 \\
6819.646886 & 20.3601 & 0.3010 & -5.0730 & 0.0023 &   -2.6810 &  1.7050 & -0.0061 &  0.0075 &  0.0110 &  0.0022 &  0.4329 &  0.0032 \\
6820.476260 & 20.3595 & 0.4300 & -5.0806 & 0.0041 &   -2.0550 &  1.9590 & -0.0052 &  0.0075 &  0.0104 &  0.0023 &  0.4319 &  0.0042 \\
6820.681320 & 20.3604 & 0.3030 & -5.0749 & 0.0024 &   -1.2960 &  1.5480 & -0.0054 &  0.0074 &  0.0098 &  0.0022 &  0.4319 &  0.0029 \\
6821.614860 & 20.3591 & 0.3730 & -5.0825 & 0.0033 &   -1.7090 &  1.6590 & -0.0052 &  0.0076 &  0.0099 &  0.0023 &  0.4329 &  0.0039 \\
6857.561399 & 20.3598 & 0.2410 & -5.0785 & 0.0018 &   -1.8550 &  1.4430 & -0.0057 &  0.0076 &  0.0096 &  0.0021 &  0.4337 &  0.0031 \\
6860.501442 & 20.3630 & 0.2310 & -5.0819 & 0.0017 &   -2.1920 &  1.6510 & -0.0057 &  0.0076 &  0.0102 &  0.0021 &  0.4338 &  0.0028 \\
6861.533968 & 20.3616 & 0.2300 & -5.0779 & 0.0017 &   -3.2770 &  1.6130 & -0.0070 &  0.0075 &  0.0118 &  0.0021 &  0.4341 &  0.0026 \\
6863.589643 & 20.3632 & 0.3000 & -5.0807 & 0.0024 &   -3.2680 &  1.7020 & -0.0068 &  0.0075 &  0.0127 &  0.0022 &  0.4337 &  0.0035 \\
6864.593615 & 20.3655 & 0.4400 & -5.0726 & 0.0040 &   -2.2780 &  1.5140 & -0.0059 &  0.0075 &  0.0101 &  0.0023 &  0.4340 &  0.0042 \\
6864.597492 & 20.3650 & 0.4760 & -5.0704 & 0.0045 &   -3.1030 &  1.7740 & -0.0069 &  0.0076 &  0.0134 &  0.0024 &  0.4335 &  0.0044 \\
6865.586694 & 20.3644 & 0.6520 & -5.0675 & 0.0070 &   -1.9140 &  1.6290 & -0.0048 &  0.0077 &  0.0109 &  0.0027 &  0.4334 &  0.0051 \\
6865.590432 & 20.3653 & 0.5910 & -5.0730 & 0.0061 &   -2.6840 &  1.7360 & -0.0072 &  0.0076 &  0.0127 &  0.0026 &  0.4336 &  0.0048 \\
6866.557713 & 20.3637 & 0.3490 & -5.0755 & 0.0029 &   -4.1420 &  1.8860 & -0.0081 &  0.0076 &  0.0149 &  0.0023 &  0.4343 &  0.0034 \\
6874.485969 & 20.3641 & 0.2620 & -5.0756 & 0.0020 &   -4.5720 &  1.7610 & -0.0081 &  0.0077 &  0.0144 &  0.0022 &  0.4342 &  0.0031 \\
6875.509363 & 20.3633 & 0.3270 & -5.0752 & 0.0027 &   -3.3470 &  1.3070 & -0.0059 &  0.0076 &  0.0122 &  0.0022 &  0.4328 &  0.0037 \\
6876.449090 & 20.3631 & 0.4160 & -5.0751 & 0.0038 &   -3.5940 &  1.8220 & -0.0073 &  0.0077 &  0.0140 &  0.0023 &  0.4315 &  0.0038 \\
6877.495861 & 20.3667 & 0.2720 & -5.0714 & 0.0021 &   -2.4520 &  1.7030 & -0.0061 &  0.0075 &  0.0111 &  0.0022 &  0.4323 &  0.0028 \\
6878.501938 & 20.3627 & 0.2410 & -5.0754 & 0.0018 &   -3.3320 &  1.7590 & -0.0067 &  0.0076 &  0.0124 &  0.0021 &  0.4338 &  0.0029 \\
6884.464710 & 20.3588 & 0.3010 & -5.0756 & 0.0024 &   -3.6920 &  1.9490 & -0.0069 &  0.0076 &  0.0139 &  0.0022 &  0.4340 &  0.0032 \\
6885.464695 & 20.3587 & 0.2880 & -5.0786 & 0.0023 &   -2.4540 &  1.5740 & -0.0058 &  0.0076 &  0.0112 &  0.0022 &  0.4340 &  0.0033 \\
6886.463522 & 20.3603 & 0.3050 & -5.0801 & 0.0024 &   -1.4490 &  1.6040 & -0.0053 &  0.0075 &  0.0102 &  0.0022 &  0.4340 &  0.0038 \\
6909.369759 & 20.3596 & 0.2200 & -5.0770 & 0.0016 &   -1.5400 &  1.5580 & -0.0053 &  0.0075 &  0.0096 &  0.0021 &  0.4340 &  0.0029 \\
6921.383481 & 20.3591 & 0.3820 & -5.0744 & 0.0033 &   -3.3330 &  1.8620 & -0.0065 &  0.0077 &  0.0122 &  0.0023 &  0.4339 &  0.0040 \\
6933.329026 & 20.3589 & 0.3150 & -5.0712 & 0.0025 &   -3.5880 &  1.3790 & -0.0073 &  0.0075 &  0.0120 &  0.0022 &  0.4340 &  0.0033 \\
6935.322257 & 20.3596 & 0.2320 & -5.0658 & 0.0016 &   -2.6460 &  1.6680 & -0.0066 &  0.0075 &  0.0115 &  0.0021 &  0.4336 &  0.0023 \\
6941.317821 & 20.3611 & 0.2230 & -5.0779 & 0.0016 &   -2.6040 &  1.7810 & -0.0062 &  0.0075 &  0.0117 &  0.0021 &  0.4334 &  0.0027 \\
7050.778430 & 20.3562 & 0.4980 & -5.0761 & 0.0052 &   -1.1180 &  2.0010 & -0.0043 &  0.0077 &  0.0097 &  0.0025 &  0.4337 &  0.0045 \\
7068.788105 & 20.3526 & 0.3890 & -5.0826 & 0.0035 &   -3.0750 &  0.9730 & -0.0055 &  0.0077 &  0.0088 &  0.0022 &  0.4338 &  0.0039 \\
7069.780635 & 20.3519 & 0.4560 & -5.0721 & 0.0044 &   -1.0630 &  1.3280 & -0.0062 &  0.0076 &  0.0086 &  0.0023 &  0.4331 &  0.0046 \\
7070.747122 & 20.3526 & 0.3740 & -5.0708 & 0.0032 &   -3.9970 &  1.2630 & -0.0063 &  0.0078 &  0.0113 &  0.0023 &  0.4336 &  0.0036 \\
7090.755284 & 20.3622 & 0.4260 & -5.0746 & 0.0039 &   -1.7820 &  1.8170 & -0.0041 &  0.0077 &  0.0091 &  0.0023 &  0.4333 &  0.0038 \\
7091.737418 & 20.3582 & 0.3350 & -5.0754 & 0.0028 &   -2.1290 &  1.7080 & -0.0072 &  0.0078 &  0.0117 &  0.0023 &  0.4335 &  0.0030 \\
7092.748430 & 20.3594 & 0.3250 & -5.0723 & 0.0027 &   -2.3610 &  1.0470 & -0.0056 &  0.0077 &  0.0088 &  0.0022 &  0.4337 &  0.0029 \\
7093.738146 & 20.3581 & 0.2850 & -5.0727 & 0.0022 &   -1.8950 &  1.7660 & -0.0057 &  0.0076 &  0.0110 &  0.0022 &  0.4333 &  0.0030 \\
7094.743036 & 20.3588 & 0.3260 & -5.0728 & 0.0026 &   -2.1530 &  1.5850 & -0.0053 &  0.0078 &  0.0100 &  0.0022 &  0.4332 &  0.0029 \\
7095.773831 & 20.3577 & 0.3230 & -5.0721 & 0.0026 &   -2.6620 &  1.6630 & -0.0063 &  0.0076 &  0.0115 &  0.0022 &  0.4334 &  0.0031 \\
7096.764658 & 20.3576 & 0.4140 & -5.0675 & 0.0038 &   -0.8450 &  1.7060 & -0.0048 &  0.0077 &  0.0091 &  0.0023 &  0.4333 &  0.0041 \\
7097.737707 & 20.3580 & 0.3090 & -5.0723 & 0.0024 &   -2.3120 &  1.9670 & -0.0056 &  0.0076 &  0.0107 &  0.0022 &  0.4336 &  0.0031 \\
7098.736845 & 20.3584 & 0.2690 & -5.0694 & 0.0020 &   -2.9710 &  1.6430 & -0.0065 &  0.0078 &  0.0116 &  0.0022 &  0.4337 &  0.0026 \\
7099.732082 & 20.3596 & 0.4420 & -5.0788 & 0.0042 &   -2.2630 &  2.0510 & -0.0067 &  0.0077 &  0.0127 &  0.0024 &  0.4330 &  0.0042 \\
7108.754064 & 20.3565 & 0.2880 & -5.0737 & 0.0023 &   -3.4690 &  1.4540 & -0.0064 &  0.0076 &  0.0113 &  0.0021 &  0.4333 &  0.0029 \\
7109.675072 & 20.3558 & 0.5350 & -5.0797 & 0.0058 &   -0.2100 &  2.0460 & -0.0046 &  0.0076 &  0.0093 &  0.0025 &  0.4331 &  0.0048 \\
7112.689327 & 20.3610 & 0.4490 & -5.0853 & 0.0044 &   -2.4630 &  1.6560 & -0.0062 &  0.0077 &  0.0116 &  0.0024 &  0.4351 &  0.0057 \\
7113.681218 & 20.3583 & 0.4500 & -5.0847 & 0.0045 &   -0.0380 &  1.8760 & -0.0039 &  0.0077 &  0.0088 &  0.0024 &  0.4337 &  0.0044 \\
7113.747982 & 20.3602 & 0.3030 & -5.0846 & 0.0025 &   -3.1400 &  1.4500 & -0.0065 &  0.0075 &  0.0117 &  0.0022 &  0.4343 &  0.0039 \\
7114.714756 & 20.3579 & 0.4990 & -5.0767 & 0.0050 &   -1.9310 &  1.7060 & -0.0058 &  0.0077 &  0.0103 &  0.0024 &  0.4330 &  0.0051 \\
7124.715849 & 20.3614 & 0.2800 & -5.0818 & 0.0022 &   -2.9610 &  2.0230 & -0.0065 &  0.0077 &  0.0132 &  0.0022 &  0.4336 &  0.0032 \\
7137.737450 & 20.3595 & 0.2950 & -5.0687 & 0.0023 &   -1.9360 &  1.7930 & -0.0052 &  0.0076 &  0.0104 &  0.0022 &  0.4334 &  0.0026 \\
7139.732011 & 20.3586 & 0.3090 & -5.0768 & 0.0025 &   -1.4430 &  1.8890 & -0.0059 &  0.0076 &  0.0104 &  0.0022 &  0.4337 &  0.0034 \\
7140.573109 & 20.3592 & 0.4650 & -5.0608 & 0.0043 &   -3.7790 &  1.8020 & -0.0068 &  0.0077 &  0.0126 &  0.0024 &  0.4336 &  0.0037 \\
7142.665322 & 20.3567 & 0.3550 & -5.0807 & 0.0031 &   -1.6570 &  1.5370 & -0.0052 &  0.0077 &  0.0101 &  0.0022 &  0.4339 &  0.0045 \\
7143.681230 & 20.3582 & 0.3270 & -5.0901 & 0.0028 &   -1.3400 &  1.5710 & -0.0047 &  0.0076 &  0.0090 &  0.0022 &  0.4336 &  0.0041 \\
7144.581494 & 20.3536 & 0.3890 & -5.0814 & 0.0037 &   -1.0710 &  1.4070 & -0.0049 &  0.0075 &  0.0090 &  0.0023 &  0.4338 &  0.0042 \\
7144.674681 & 20.3569 & 0.3280 & -5.0841 & 0.0028 &   -2.2650 &  1.8600 & -0.0058 &  0.0077 &  0.0116 &  0.0022 &  0.4338 &  0.0041 \\
7145.635365 & 20.3566 & 0.3000 & -5.0806 & 0.0024 &   -2.6120 &  2.0110 & -0.0061 &  0.0077 &  0.0116 &  0.0022 &  0.4324 &  0.0036 \\
7145.728251 & 20.3573 & 0.2480 & -5.0722 & 0.0019 &   -2.2380 &  1.8750 & -0.0055 &  0.0077 &  0.0106 &  0.0021 &  0.4326 &  0.0029 \\
7146.736969 & 20.3549 & 0.3670 & -5.0713 & 0.0032 &   -4.8630 &  1.2210 & -0.0085 &  0.0076 &  0.0130 &  0.0022 &  0.4314 &  0.0034 \\
7147.718624 & 20.3568 & 0.2720 & -5.0731 & 0.0021 &   -1.9820 &  2.2700 & -0.0062 &  0.0077 &  0.0125 &  0.0022 &  0.4329 &  0.0028 \\
7148.532470 & 20.3559 & 0.3490 & -5.0803 & 0.0032 &   -2.2590 &  1.4100 & -0.0058 &  0.0076 &  0.0098 &  0.0022 &  0.4328 &  0.0037 \\
7148.674004 & 20.3582 & 0.2970 & -5.0853 & 0.0024 &   -2.4700 &  1.6660 & -0.0061 &  0.0077 &  0.0110 &  0.0022 &  0.4337 &  0.0037 \\
7153.543791 & 20.3624 & 0.4800 & -5.0720 & 0.0049 &   -3.9160 &  2.6130 & -0.0067 &  0.0077 &  0.0147 &  0.0024 &  0.4334 &  0.0045 \\
7154.509654 & 20.3591 & 1.9270 & -5.1984 & 0.0572 &  -12.9690 &  4.8180 & -0.0174 &  0.0081 &  0.0278 &  0.0047 &  0.4341 &  0.0132 \\
7154.541426 & 20.3636 & 0.9480 & -5.0800 & 0.0139 &   -0.9270 &  2.3700 & -0.0037 &  0.0077 &  0.0086 &  0.0030 &  0.4321 &  0.0068 \\
7156.543614 & 20.3646 & 0.6030 & -5.0636 & 0.0065 &   -3.3650 &  1.5070 & -0.0076 &  0.0075 &  0.0126 &  0.0026 &  0.4316 &  0.0046 \\
7159.538915 & 20.3627 & 0.3120 & -5.0789 & 0.0026 &   -1.7310 &  2.0640 & -0.0057 &  0.0076 &  0.0117 &  0.0022 &  0.4323 &  0.0036 \\
7169.474275 & 20.3623 & 0.2880 & -5.0738 & 0.0023 &   -2.5110 &  2.0300 & -0.0056 &  0.0077 &  0.0120 &  0.0022 &  0.4337 &  0.0031 \\
7169.575655 & 20.3636 & 0.3230 & -5.0735 & 0.0027 &   -2.9260 &  1.7780 & -0.0055 &  0.0077 &  0.0124 &  0.0022 &  0.4339 &  0.0033 \\
7170.543366 & 20.3621 & 0.2490 & -5.0704 & 0.0018 &   -2.4520 &  1.6470 & -0.0061 &  0.0076 &  0.0114 &  0.0021 &  0.4333 &  0.0026 \\
7170.672050 & 20.3630 & 0.2640 & -5.0677 & 0.0020 &   -1.6880 &  1.7060 & -0.0056 &  0.0077 &  0.0109 &  0.0022 &  0.4331 &  0.0027 \\
7171.457719 & 20.3641 & 0.7310 & -5.0709 & 0.0098 &   -2.0880 &  1.8280 & -0.0048 &  0.0078 &  0.0085 &  0.0027 &  0.4334 &  0.0068 \\
7171.581808 & 20.3638 & 0.3470 & -5.0749 & 0.0030 &   -3.5660 &  1.5810 & -0.0074 &  0.0076 &  0.0132 &  0.0022 &  0.4338 &  0.0037 \\
7172.588396 & 20.3650 & 0.3230 & -5.0686 & 0.0026 &   -3.4510 &  1.5980 & -0.0065 &  0.0078 &  0.0120 &  0.0022 &  0.4337 &  0.0035 \\
7173.515919 & 20.3644 & 0.4550 & -5.0673 & 0.0043 &   -1.5630 &  1.7770 & -0.0045 &  0.0078 &  0.0097 &  0.0023 &  0.4338 &  0.0048 \\
7173.644059 & 20.3682 & 0.3370 & -5.0783 & 0.0028 &   -0.2700 &  1.8800 & -0.0039 &  0.0077 &  0.0085 &  0.0022 &  0.4341 &  0.0042 \\
7174.452307 & 20.3621 & 0.4110 & -5.0709 & 0.0040 &   -4.5680 &  2.0880 & -0.0079 &  0.0076 &  0.0158 &  0.0023 &  0.4339 &  0.0045 \\
7174.566442 & 20.3643 & 0.4340 & -5.0693 & 0.0041 &   -2.6160 &  1.9590 & -0.0063 &  0.0078 &  0.0123 &  0.0023 &  0.4338 &  0.0043 \\
7175.500884 & 20.3661 & 0.2950 & -5.0726 & 0.0023 &   -0.8940 &  1.9890 & -0.0046 &  0.0076 &  0.0106 &  0.0022 &  0.4335 &  0.0033 \\
7175.646396 & 20.3658 & 0.4160 & -5.0697 & 0.0038 &   -5.0100 &  1.6730 & -0.0065 &  0.0077 &  0.0133 &  0.0023 &  0.4335 &  0.0046 \\
7176.538244 & 20.3648 & 0.3970 & -5.0637 & 0.0035 &   -1.3440 &  1.8360 & -0.0044 &  0.0077 &  0.0095 &  0.0023 &  0.4332 &  0.0041 \\
7176.673884 & 20.3643 & 0.4070 & -5.0698 & 0.0037 &   -2.3630 &  1.4130 & -0.0057 &  0.0077 &  0.0105 &  0.0023 &  0.4332 &  0.0044 \\
7177.480845 & 20.3660 & 0.3040 & -5.0717 & 0.0025 &   -2.2080 &  1.8480 & -0.0052 &  0.0078 &  0.0102 &  0.0022 &  0.4335 &  0.0036 \\
7177.563798 & 20.3647 & 0.2890 & -5.0763 & 0.0023 &   -1.6160 &  1.5420 & -0.0049 &  0.0077 &  0.0096 &  0.0021 &  0.4332 &  0.0032 \\
7178.533608 & 20.3623 & 0.3230 & -5.0734 & 0.0027 &   -1.0540 &  1.6340 & -0.0047 &  0.0077 &  0.0096 &  0.0022 &  0.4332 &  0.0035 \\
7178.678379 & 20.3620 & 0.3710 & -5.0710 & 0.0032 &   -1.5700 &  1.6130 & -0.0056 &  0.0077 &  0.0107 &  0.0022 &  0.4328 &  0.0038 \\
7204.480621 & 20.3606 & 0.2960 & -5.0691 & 0.0023 &   -3.3950 &  2.0300 & -0.0066 &  0.0077 &  0.0131 &  0.0022 &  0.4335 &  0.0028 \\
7205.484594 & 20.3627 & 0.2540 & -5.0831 & 0.0020 &   -1.7310 &  1.7420 & -0.0050 &  0.0077 &  0.0107 &  0.0022 &  0.4340 &  0.0031 \\
7206.485278 & 20.3607 & 0.2670 & -5.0893 & 0.0021 &   -1.3410 &  1.8290 & -0.0056 &  0.0077 &  0.0109 &  0.0022 &  0.4343 &  0.0035 \\
7207.498566 & 20.3600 & 0.3020 & -5.0865 & 0.0025 &   -1.8680 &  1.6550 & -0.0048 &  0.0077 &  0.0098 &  0.0022 &  0.4338 &  0.0035 \\
7208.560671 & 20.3602 & 0.3390 & -5.0818 & 0.0029 &   -2.7070 &  1.6740 & -0.0056 &  0.0077 &  0.0112 &  0.0022 &  0.4337 &  0.0036 \\
7208.592221 & 20.3600 & 0.2880 & -5.0747 & 0.0022 &   -3.9470 &  1.5870 & -0.0071 &  0.0077 &  0.0130 &  0.0022 &  0.4337 &  0.0030 \\
7209.560310 & 20.3597 & 0.3890 & -5.0766 & 0.0035 &   -3.1290 &  1.6330 & -0.0065 &  0.0077 &  0.0121 &  0.0023 &  0.4338 &  0.0040 \\
7209.680795 & 20.3585 & 0.4910 & -5.0844 & 0.0052 &   -2.2160 &  1.7240 & -0.0070 &  0.0077 &  0.0125 &  0.0024 &  0.4335 &  0.0048 \\
7210.555643 & 20.3596 & 0.4060 & -5.0789 & 0.0037 &   -3.1200 &  1.3460 & -0.0062 &  0.0077 &  0.0110 &  0.0023 &  0.4328 &  0.0040 \\
7210.613779 & 20.3590 & 0.4790 & -5.0809 & 0.0048 &   -1.7210 &  1.6370 & -0.0058 &  0.0078 &  0.0107 &  0.0024 &  0.4325 &  0.0046 \\
7211.664677 & 20.3601 & 1.0140 & -5.0931 & 0.0168 &   -7.3350 &  2.5340 & -0.0099 &  0.0080 &  0.0184 &  0.0033 &  0.4315 &  0.0086 \\
7258.366043 & 20.3656 & 0.3300 & -5.0790 & 0.0028 &   -0.7480 &  2.1760 & -0.0052 &  0.0075 &  0.0104 &  0.0022 &  0.4336 &  0.0036 \\
7258.553209 & 20.3639 & 0.3830 & -5.0791 & 0.0036 &   -0.8590 &  1.7080 & -0.0044 &  0.0076 &  0.0098 &  0.0023 &  0.4334 &  0.0037 \\
7259.364159 & 20.3648 & 0.2920 & -5.0797 & 0.0023 &   -1.7860 &  1.7360 & -0.0048 &  0.0074 &  0.0095 &  0.0022 &  0.4340 &  0.0032 \\
7259.542413 & 20.3620 & 0.3360 & -5.0759 & 0.0029 &   -2.8990 &  1.4090 & -0.0065 &  0.0073 &  0.0113 &  0.0022 &  0.4337 &  0.0033 \\
7289.344087 & 20.3657 & 0.3210 & -5.0806 & 0.0026 &   -2.0040 &  1.4500 & -0.0050 &  0.0073 &  0.0090 &  0.0022 &  0.4329 &  0.0035 \\
7290.381509 & 20.3657 & 0.6670 & -5.0839 & 0.0076 &   -2.0780 &  1.6660 & -0.0055 &  0.0075 &  0.0100 &  0.0027 &  0.4332 &  0.0068 \\
7290.466723 & 20.3665 & 0.3530 & -5.0658 & 0.0031 &   -2.0250 &  1.7010 & -0.0057 &  0.0074 &  0.0109 &  0.0022 &  0.4324 &  0.0032 \\
7291.337432 & 20.3652 & 0.5300 & -5.0841 & 0.0052 &   -0.4160 &  1.5420 & -0.0040 &  0.0073 &  0.0080 &  0.0024 &  0.4335 &  0.0062 \\
7291.464565 & 20.3647 & 0.3180 & -5.0667 & 0.0026 &   -0.8280 &  1.7300 & -0.0045 &  0.0074 &  0.0090 &  0.0022 &  0.4324 &  0.0030 \\
7416.791183 & 20.3695 & 0.3180 & -5.0799 & 0.0027 &   -2.1300 &  1.2380 & -0.0051 &  0.0075 &  0.0090 &  0.0022 &  0.4328 &  0.0035 \\
7418.791692 & 20.3670 & 0.7200 & -5.0841 & 0.0095 &   -0.6460 &  2.1160 & -0.0054 &  0.0078 &  0.0104 &  0.0028 &  0.4341 &  0.0069 \\
7443.762794 & 20.3623 & 0.5670 & -5.0737 & 0.0062 &   -2.7220 &  1.5810 & -0.0063 &  0.0077 &  0.0111 &  0.0025 &  0.4336 &  0.0054 \\
7444.746683 & 20.3633 & 0.3970 & -5.0667 & 0.0034 &   -2.4080 &  1.7660 & -0.0054 &  0.0076 &  0.0104 &  0.0023 &  0.4333 &  0.0035 \\
7445.743744 & 20.3634 & 0.5290 & -5.0702 & 0.0055 &   -4.7560 &  1.9270 & -0.0085 &  0.0075 &  0.0144 &  0.0025 &  0.4333 &  0.0051 \\
7472.739054 & 20.3706 & 0.4880 & -5.0868 & 0.0049 &   -1.9010 &  1.8870 & -0.0063 &  0.0076 &  0.0123 &  0.0024 &  0.4338 &  0.0054 \\
7474.753014 & 20.3738 & 0.2810 & -5.0752 & 0.0022 &   -0.2640 &  1.7010 & -0.0045 &  0.0076 &  0.0087 &  0.0022 &  0.4326 &  0.0030 \\
7475.755740 & 20.3758 & 0.3500 & -5.0806 & 0.0029 &   -2.3410 &  1.7090 & -0.0055 &  0.0076 &  0.0108 &  0.0023 &  0.4338 &  0.0040 \\
7498.746583 & 20.3717 & 0.3480 & -5.0784 & 0.0029 &   -2.0660 &  1.5670 & -0.0055 &  0.0076 &  0.0101 &  0.0022 &  0.4331 &  0.0037 \\
7499.654862 & 20.3718 & 0.4000 & -5.0844 & 0.0036 &   -2.9580 &  1.6100 & -0.0072 &  0.0078 &  0.0118 &  0.0023 &  0.4305 &  0.0046 \\
7501.669117 & 20.3693 & 0.4550 & -5.0864 & 0.0045 &   -1.2430 &  1.8990 & -0.0059 &  0.0077 &  0.0105 &  0.0024 &  0.4335 &  0.0048 \\
7502.677053 & 20.3723 & 0.3530 & -5.0884 & 0.0031 &   -2.2270 &  1.6470 & -0.0062 &  0.0079 &  0.0111 &  0.0022 &  0.4330 &  0.0042 \\
7513.611168 & 20.3705 & 0.2980 & -5.0863 & 0.0024 &   -2.8450 &  1.5800 & -0.0058 &  0.0076 &  0.0114 &  0.0022 &  0.4327 &  0.0036 \\
7513.660522 & 20.3698 & 0.3160 & -5.0847 & 0.0026 &   -1.9820 &  2.0390 & -0.0054 &  0.0077 &  0.0114 &  0.0022 &  0.4326 &  0.0035 \\
7521.536743 & 20.3677 & 0.4050 & -5.0823 & 0.0038 &   -1.4580 &  1.9280 & -0.0055 &  0.0078 &  0.0110 &  0.0023 &  0.4326 &  0.0042 \\
7522.547443 & 20.3673 & 0.3590 & -5.0842 & 0.0031 &   -1.4140 &  1.5970 & -0.0048 &  0.0075 &  0.0087 &  0.0022 &  0.4325 &  0.0038 \\
7523.551752 & 20.3687 & 0.6640 & -5.0934 & 0.0082 &   -0.7890 &  1.9850 & -0.0041 &  0.0076 &  0.0099 &  0.0026 &  0.4334 &  0.0068 \\
7524.533340 & 20.3685 & 0.5000 & -5.0809 & 0.0052 &   -4.3430 &  1.7560 & -0.0082 &  0.0079 &  0.0147 &  0.0024 &  0.4320 &  0.0050 \\
7526.531539 & 20.3720 & 0.3720 & -5.0849 & 0.0033 &   -3.6530 &  2.0070 & -0.0067 &  0.0076 &  0.0138 &  0.0023 &  0.4329 &  0.0042 \\
7540.648303 & 20.3745 & 0.4350 & -5.0755 & 0.0040 &   -2.9790 &  1.6090 & -0.0063 &  0.0077 &  0.0119 &  0.0024 &  0.4326 &  0.0040 \\
7540.712448 & 20.3745 & 0.4080 & -5.0637 & 0.0035 &   -3.4080 &  1.5110 & -0.0063 &  0.0076 &  0.0118 &  0.0023 &  0.4318 &  0.0032 \\
7549.520925 & 20.3812 & 0.3520 & -5.0656 & 0.0028 &   -4.3100 &  1.6890 & -0.0067 &  0.0075 &  0.0131 &  0.0023 &  0.4310 &  0.0027 \\
7549.628750 & 20.3807 & 0.2950 & -5.0638 & 0.0022 &   -3.1360 &  1.6430 & -0.0067 &  0.0076 &  0.0116 &  0.0022 &  0.4309 &  0.0023 \\
7550.507442 & 20.3796 & 0.3720 & -5.0680 & 0.0030 &   -4.7260 &  1.6030 & -0.0071 &  0.0075 &  0.0139 &  0.0023 &  0.4306 &  0.0029 \\
7550.655742 & 20.3802 & 0.3290 & -5.0668 & 0.0026 &   -1.4760 &  2.0330 & -0.0049 &  0.0075 &  0.0107 &  0.0022 &  0.4311 &  0.0025 \\
7551.510428 & 20.3780 & 0.4070 & -5.0776 & 0.0036 &   -2.7540 &  1.3870 & -0.0067 &  0.0077 &  0.0116 &  0.0023 &  0.4313 &  0.0035 \\
7551.668312 & 20.3751 & 0.6360 & -5.0906 & 0.0075 &   -1.7060 &  2.7040 & -0.0049 &  0.0077 &  0.0121 &  0.0027 &  0.4322 &  0.0056 \\
7552.480519 & 20.3743 & 0.5780 & -5.0837 & 0.0065 &    0.3070 &  2.0350 & -0.0040 &  0.0075 &  0.0088 &  0.0026 &  0.4327 &  0.0055 \\
7552.648530 & 20.3748 & 0.4190 & -5.0745 & 0.0037 &   -2.9020 &  2.0300 & -0.0071 &  0.0075 &  0.0129 &  0.0024 &  0.4326 &  0.0038 \\
7553.481616 & 20.3722 & 0.5480 & -5.0927 & 0.0060 &   -0.6810 &  1.6100 & -0.0041 &  0.0076 &  0.0077 &  0.0024 &  0.4327 &  0.0052 \\
7553.644060 & 20.3736 & 0.3610 & -5.0849 & 0.0031 &   -2.3910 &  1.7630 & -0.0056 &  0.0077 &  0.0109 &  0.0023 &  0.4327 &  0.0038 \\
7573.663488 & 20.3721 & 0.3740 & -5.0734 & 0.0033 &   -2.4280 &  1.8300 & -0.0062 &  0.0075 &  0.0115 &  0.0022 &  0.4328 &  0.0040 \\
7574.654725 & 20.3750 & 0.3700 & -5.0703 & 0.0032 &   -1.9470 &  1.2110 & -0.0050 &  0.0075 &  0.0087 &  0.0022 &  0.4320 &  0.0033 \\
7576.625113 & 20.3739 & 0.2460 & -5.0693 & 0.0018 &   -1.8620 &  1.8980 & -0.0056 &  0.0077 &  0.0117 &  0.0022 &  0.4324 &  0.0024 \\
7578.580670 & 20.3732 & 0.7730 & -5.0813 & 0.0099 &   -1.0860 &  1.9310 & -0.0054 &  0.0078 &  0.0109 &  0.0028 &  0.4324 &  0.0071 \\
7579.613452 & 20.3712 & 0.5230 & -5.0764 & 0.0053 &   -2.2660 &  1.3670 & -0.0050 &  0.0075 &  0.0091 &  0.0024 &  0.4329 &  0.0049 \\
7580.529962 & 20.3728 & 1.2990 & -5.1144 & 0.0241 &   -5.0490 &  3.2490 & -0.0080 &  0.0080 &  0.0111 &  0.0037 &  0.4341 &  0.0114 \\
7594.530504 & 20.3707 & 0.3460 & -5.0693 & 0.0028 &   -1.7790 &  1.7370 & -0.0051 &  0.0075 &  0.0102 &  0.0022 &  0.4313 &  0.0032 \\
7596.563224 & 20.3710 & 0.4990 & -5.0718 & 0.0048 &   -4.9790 &  1.8030 & -0.0079 &  0.0077 &  0.0142 &  0.0024 &  0.4312 &  0.0045 \\
7597.395046 & 20.3697 & 0.5970 & -5.0884 & 0.0066 &   -2.4790 &  2.2770 & -0.0063 &  0.0076 &  0.0131 &  0.0026 &  0.4328 &  0.0055 \\
7597.572065 & 20.3723 & 0.3310 & -5.0635 & 0.0026 &   -2.5580 &  1.8100 & -0.0062 &  0.0075 &  0.0108 &  0.0023 &  0.4319 &  0.0029 \\
7603.406702 & 20.3719 & 0.2680 & -5.0623 & 0.0019 &   -2.2200 &  1.7210 & -0.0056 &  0.0075 &  0.0104 &  0.0021 &  0.4329 &  0.0026 \\
7603.584576 & 20.3705 & 0.4170 & -5.0617 & 0.0036 &   -2.8850 &  1.8320 & -0.0057 &  0.0076 &  0.0108 &  0.0023 &  0.4323 &  0.0033 \\
7604.410572 & 20.3708 & 0.3470 & -5.0748 & 0.0029 &   -2.9560 &  2.0920 & -0.0069 &  0.0076 &  0.0133 &  0.0023 &  0.4321 &  0.0038 \\
7604.589338 & 20.3697 & 0.5420 & -5.0625 & 0.0057 &   -2.9360 &  1.8010 & -0.0058 &  0.0076 &  0.0125 &  0.0025 &  0.4318 &  0.0053 \\
7620.364903 & 20.3782 & 0.3640 & -5.0814 & 0.0031 &    0.4150 &  2.0470 & -0.0041 &  0.0075 &  0.0087 &  0.0023 &  0.4328 &  0.0040 \\
7620.525855 & 20.3792 & 0.4660 & -5.0780 & 0.0046 &   -1.4540 &  2.1670 & -0.0052 &  0.0074 &  0.0118 &  0.0024 &  0.4328 &  0.0046 \\
7621.372546 & 20.3756 & 0.6080 & -5.0791 & 0.0067 &   -3.6430 &  1.5200 & -0.0077 &  0.0073 &  0.0117 &  0.0026 &  0.4329 &  0.0058 \\
7621.549955 & 20.3758 & 0.8550 & -5.0863 & 0.0126 &   -1.2550 &  3.3510 & -0.0062 &  0.0081 &  0.0145 &  0.0030 &  0.4335 &  0.0076 \\
7622.415336 & 20.3782 & 0.3130 & -5.0733 & 0.0024 &   -2.2270 &  1.8050 & -0.0055 &  0.0074 &  0.0109 &  0.0022 &  0.4320 &  0.0032 \\
7622.548292 & 20.3803 & 0.3540 & -5.0659 & 0.0029 &   -2.3410 &  1.5500 & -0.0054 &  0.0073 &  0.0103 &  0.0022 &  0.4312 &  0.0031 \\
7623.423696 & 20.3776 & 0.3390 & -5.0733 & 0.0027 &   -0.9030 &  1.7180 & -0.0045 &  0.0074 &  0.0084 &  0.0022 &  0.4325 &  0.0036 \\
7623.538562 & 20.3806 & 0.3350 & -5.0842 & 0.0029 &   -0.8410 &  1.8080 & -0.0039 &  0.0075 &  0.0093 &  0.0022 &  0.4326 &  0.0040 \\
7624.403734 & 20.3808 & 0.3090 & -5.0810 & 0.0024 &   -0.8410 &  1.8810 & -0.0043 &  0.0074 &  0.0097 &  0.0022 &  0.4330 &  0.0040 \\
7624.515325 & 20.3802 & 0.3310 & -5.0683 & 0.0026 &   -2.7480 &  1.6580 & -0.0056 &  0.0074 &  0.0111 &  0.0022 &  0.4314 &  0.0031 \\
7635.376766 & 20.3778 & 0.2750 & -5.0809 & 0.0021 &   -2.7420 &  2.0860 & -0.0064 &  0.0074 &  0.0128 &  0.0022 &  0.4332 &  0.0036 \\
7636.380096 & 20.3764 & 0.2490 & -5.0751 & 0.0018 &   -2.4120 &  1.5710 & -0.0061 &  0.0075 &  0.0110 &  0.0022 &  0.4333 &  0.0030 \\
7640.387558 & 20.3763 & 0.2560 & -5.0755 & 0.0019 &   -1.8620 &  1.5340 & -0.0052 &  0.0074 &  0.0098 &  0.0021 &  0.4322 &  0.0034 \\
7640.481314 & 20.3758 & 0.2460 & -5.0706 & 0.0018 &   -1.8750 &  1.5690 & -0.0051 &  0.0073 &  0.0100 &  0.0021 &  0.4319 &  0.0030 \\
7644.344638 & 20.3698 & 0.2820 & -5.0741 & 0.0021 &   -1.8270 &  1.4020 & -0.0046 &  0.0073 &  0.0095 &  0.0022 &  0.4327 &  0.0029 \\
7650.369881 & 20.3702 & 0.3660 & -5.0717 & 0.0031 &   -3.5060 &  1.8830 & -0.0068 &  0.0074 &  0.0130 &  0.0023 &  0.4330 &  0.0039 \\
7679.330548 & 20.3732 & 0.3530 & -5.0677 & 0.0029 &   -2.7760 &  1.8230 & -0.0057 &  0.0074 &  0.0115 &  0.0023 &  0.4316 &  0.0032 \\
7680.374550 & 20.3730 & 0.2670 & -5.0734 & 0.0020 &   -0.6890 &  1.6320 & -0.0048 &  0.0073 &  0.0093 &  0.0021 &  0.4327 &  0.0034 \\
7683.334125 & 20.3742 & 0.2800 & -5.0745 & 0.0021 &   -0.5230 &  2.1500 & -0.0042 &  0.0073 &  0.0095 &  0.0022 &  0.4326 &  0.0033 \\
7701.303844 & 20.3760 & 0.3780 & -5.0795 & 0.0032 &   -2.3990 &  1.7790 & -0.0046 &  0.0076 &  0.0093 &  0.0022 &  0.4326 &  0.0039 \\
7702.303348 & 20.3749 & 0.4120 & -5.0820 & 0.0037 &   -2.9010 &  1.1430 & -0.0067 &  0.0074 &  0.0103 &  0.0023 &  0.4335 &  0.0042 \\
7703.299960 & 20.3745 & 0.4950 & -5.0749 & 0.0049 &   -3.1220 &  1.4430 & -0.0060 &  0.0076 &  0.0110 &  0.0024 &  0.4330 &  0.0048 \\
7706.311386 & 20.3740 & 0.3980 & -5.0752 & 0.0035 &   -1.1890 &  2.1080 & -0.0057 &  0.0077 &  0.0111 &  0.0023 &  0.4324 &  0.0039 \\
7803.755945 & 20.3672 & 0.3710 & -5.0709 & 0.0031 &   -1.6410 &  1.9180 & -0.0052 &  0.0076 &  0.0102 &  0.0023 &  0.4325 &  0.0040 \\
7806.750744 & 20.3678 & 0.3990 & -5.0666 & 0.0034 &   -2.3800 &  1.5960 & -0.0058 &  0.0075 &  0.0105 &  0.0023 &  0.4322 &  0.0040 \\
7807.757167 & 20.3692 & 0.3100 & -5.0664 & 0.0024 &   -2.2780 &  2.0850 & -0.0055 &  0.0075 &  0.0115 &  0.0022 &  0.4330 &  0.0038 \\
7809.765814 & 20.3682 & 0.5610 & -5.0469 & 0.0054 &   -1.0040 &  1.7690 & -0.0048 &  0.0075 &  0.0092 &  0.0025 &  0.4326 &  0.0051 \\
7810.739969 & 20.3663 & 0.6220 & -5.0539 & 0.0067 &   -3.6370 &  1.7530 & -0.0064 &  0.0076 &  0.0107 &  0.0026 &  0.4325 &  0.0055 \\
7811.782971 & 20.3672 & 0.4080 & -5.0519 & 0.0034 &   -2.8710 &  1.5550 & -0.0052 &  0.0076 &  0.0104 &  0.0023 &  0.4321 &  0.0039 \\
7817.777919 & 20.3626 & 0.4660 & -5.0569 & 0.0042 &   -2.1250 &  1.1960 & -0.0047 &  0.0075 &  0.0084 &  0.0023 &  0.4323 &  0.0044 \\
7833.740669 & 20.3736 & 0.3970 & -5.0566 & 0.0033 &   -2.1070 &  2.0960 & -0.0067 &  0.0076 &  0.0116 &  0.0023 &  0.4325 &  0.0038 \\
7834.757019 & 20.3747 & 0.4070 & -5.0683 & 0.0034 &   -1.1920 &  2.1710 & -0.0048 &  0.0077 &  0.0109 &  0.0023 &  0.4326 &  0.0042 \\
7835.759584 & 20.3745 & 0.3010 & -5.0727 & 0.0023 &   -3.5170 &  1.4930 & -0.0068 &  0.0076 &  0.0118 &  0.0022 &  0.4328 &  0.0034 \\
7853.627296 & 20.3659 & 0.5220 & -5.0555 & 0.0051 &   -2.1720 &  1.6630 & -0.0065 &  0.0076 &  0.0105 &  0.0024 &  0.4315 &  0.0051 \\
7855.677883 & 20.3680 & 0.4130 & -5.0784 & 0.0036 &   -2.9100 &  1.5050 & -0.0057 &  0.0077 &  0.0103 &  0.0023 &  0.4323 &  0.0043 \\
7856.640626 & 20.3665 & 0.3130 & -5.0739 & 0.0024 &   -1.0680 &  1.6580 & -0.0041 &  0.0077 &  0.0086 &  0.0022 &  0.4321 &  0.0037 \\
7857.651311 & 20.3666 & 0.3460 & -5.0686 & 0.0027 &   -1.5960 &  1.7810 & -0.0051 &  0.0077 &  0.0103 &  0.0022 &  0.4323 &  0.0032 \\
7858.646404 & 20.3668 & 0.6360 & -5.0658 & 0.0065 &   -3.1240 &  1.9320 & -0.0069 &  0.0079 &  0.0130 &  0.0026 &  0.4322 &  0.0058 \\
7861.685921 & 20.3700 & 0.3420 & -5.0728 & 0.0027 &   -1.1100 &  1.4830 & -0.0051 &  0.0077 &  0.0099 &  0.0022 &  0.4321 &  0.0035 \\
7863.694299 & 20.3686 & 0.3780 & -5.0613 & 0.0030 &   -0.5310 &  1.6430 & -0.0044 &  0.0078 &  0.0083 &  0.0023 &  0.4319 &  0.0029 \\
7864.713291 & 20.3673 & 0.2600 & -5.0722 & 0.0019 &   -2.4400 &  1.6160 & -0.0060 &  0.0076 &  0.0109 &  0.0021 &  0.4326 &  0.0027 \\
7865.707339 & 20.3667 & 0.3900 & -5.0749 & 0.0033 &   -3.9520 &  1.1390 & -0.0072 &  0.0076 &  0.0111 &  0.0022 &  0.4325 &  0.0038 \\
7878.736396 & 20.3688 & 0.4670 & -5.0691 & 0.0042 &   -4.7510 &  1.3750 & -0.0088 &  0.0076 &  0.0141 &  0.0024 &  0.4325 &  0.0041 \\
7879.734625 & 20.3669 & 0.2980 & -5.0739 & 0.0022 &   -3.0720 &  1.6880 & -0.0072 &  0.0076 &  0.0131 &  0.0022 &  0.4322 &  0.0027 \\
7880.613739 & 20.3672 & 0.2790 & -5.0886 & 0.0021 &   -2.6570 &  1.9230 & -0.0062 &  0.0077 &  0.0113 &  0.0022 &  0.4325 &  0.0031 \\
7881.723476 & 20.3657 & 0.4010 & -5.0779 & 0.0035 &   -0.6990 &  2.0300 & -0.0049 &  0.0075 &  0.0101 &  0.0023 &  0.4327 &  0.0043 \\
7890.699985 & 20.3650 & 0.6740 & -5.0772 & 0.0073 &    0.9480 &  2.0700 & -0.0022 &  0.0078 &  0.0053 &  0.0026 &  0.4333 &  0.0069 \\
7893.614961 & 20.3652 & 0.3240 & -5.0774 & 0.0025 &   -2.6070 &  1.3780 & -0.0063 &  0.0075 &  0.0109 &  0.0022 &  0.4320 &  0.0038 \\
7894.674592 & 20.3660 & 0.2740 & -5.0762 & 0.0020 &   -1.1710 &  1.7830 & -0.0048 &  0.0075 &  0.0098 &  0.0022 &  0.4323 &  0.0030 \\
7895.676314 & 20.3663 & 0.2840 & -5.0797 & 0.0021 &   -1.6790 &  1.6100 & -0.0055 &  0.0075 &  0.0105 &  0.0022 &  0.4330 &  0.0033 \\
7896.717562 & 20.3664 & 0.3200 & -5.0768 & 0.0025 &   -2.8960 &  1.6630 & -0.0061 &  0.0077 &  0.0116 &  0.0022 &  0.4323 &  0.0034 \\
7897.682800 & 20.3667 & 0.2850 & -5.0752 & 0.0022 &   -2.1230 &  1.8220 & -0.0056 &  0.0076 &  0.0107 &  0.0022 &  0.4325 &  0.0027 \\
7898.607133 & 20.3669 & 0.2660 & -5.0787 & 0.0019 &   -1.2140 &  2.0460 & -0.0049 &  0.0077 &  0.0109 &  0.0022 &  0.4327 &  0.0035 \\
7915.448771 & 20.3722 & 0.3280 & -5.0812 & 0.0027 &   -2.9740 &  1.2890 & -0.0070 &  0.0074 &  0.0116 &  0.0022 &  0.4333 &  0.0039 \\
7916.675070 & 20.3716 & 0.3790 & -5.0804 & 0.0032 &   -2.2290 &  1.8810 & -0.0062 &  0.0076 &  0.0111 &  0.0023 &  0.4331 &  0.0042 \\
7928.530624 & 20.3756 & 0.6160 & -5.0513 & 0.0060 &   -2.4740 &  1.5390 & -0.0058 &  0.0075 &  0.0095 &  0.0026 &  0.4297 &  0.0045 \\
7929.626756 & 20.3765 & 0.6340 & -5.0629 & 0.0066 &    2.0280 &  2.2670 & -0.0011 &  0.0077 &  0.0065 &  0.0026 &  0.4292 &  0.0045 \\
7930.600989 & 20.3797 & 0.5740 & -5.0557 & 0.0048 &   -4.5470 &  1.4340 & -0.0086 &  0.0065 &  0.0125 &  0.0026 &  0.4297 &  0.0035 \\
7931.654422 & 20.3766 & 0.7500 & -5.0579 & 0.0071 &   -3.9220 &  1.8760 & -0.0112 &  0.0067 &  0.0138 &  0.0029 &  0.4294 &  0.0042 \\
7932.617009 & 20.3719 & 0.4150 & -5.0625 & 0.0035 &   -2.4920 &  1.6360 & -0.0068 &  0.0075 &  0.0109 &  0.0023 &  0.4309 &  0.0034 \\
7933.647373 & 20.3684 & 0.3450 & -5.0792 & 0.0028 &   -1.4420 &  2.1240 & -0.0066 &  0.0076 &  0.0118 &  0.0023 &  0.4320 &  0.0035 \\
7934.639703 & 20.3699 & 0.3090 & -5.0712 & 0.0024 &   -2.0400 &  1.7760 & -0.0062 &  0.0075 &  0.0113 &  0.0022 &  0.4322 &  0.0032 \\
7935.612600 & 20.3689 & 0.3280 & -5.0673 & 0.0025 &   -1.6660 &  1.7410 & -0.0056 &  0.0076 &  0.0108 &  0.0022 &  0.4322 &  0.0033 \\
7936.607996 & 20.3695 & 0.4670 & -5.0652 & 0.0043 &   -2.5980 &  1.2640 & -0.0061 &  0.0077 &  0.0099 &  0.0024 &  0.4330 &  0.0045 \\
7937.626468 & 20.3689 & 0.3330 & -5.0770 & 0.0027 &   -2.7690 &  1.6200 & -0.0052 &  0.0074 &  0.0103 &  0.0022 &  0.4323 &  0.0034 \\
7942.624235 & 20.3676 & 0.3020 & -5.0790 & 0.0024 &   -1.3390 &  1.6210 & -0.0052 &  0.0076 &  0.0095 &  0.0022 &  0.4325 &  0.0029 \\
7943.572969 & 20.3648 & 0.4140 & -5.0790 & 0.0036 &    0.2830 &  2.2350 & -0.0046 &  0.0075 &  0.0094 &  0.0023 &  0.4323 &  0.0038 \\
7944.543854 & 20.3668 & 0.2830 & -5.0719 & 0.0021 &   -3.3220 &  1.6510 & -0.0066 &  0.0075 &  0.0118 &  0.0022 &  0.4328 &  0.0031 \\
7949.554154 & 20.3690 & 0.4710 & -5.0705 & 0.0043 &   -2.7270 &  1.5050 & -0.0063 &  0.0077 &  0.0098 &  0.0023 &  0.4317 &  0.0044 \\
7950.537684 & 20.3691 & 0.3350 & -5.0786 & 0.0026 &   -1.5560 &  1.9290 & -0.0058 &  0.0076 &  0.0107 &  0.0022 &  0.4314 &  0.0035 \\
7952.619598 & 20.3658 & 0.4320 & -5.0736 & 0.0038 &   -1.6400 &  1.5400 & -0.0047 &  0.0074 &  0.0084 &  0.0023 &  0.4316 &  0.0040 \\
7953.586181 & 20.3655 & 0.3550 & -5.0652 & 0.0028 &   -1.9130 &  1.9050 & -0.0063 &  0.0076 &  0.0111 &  0.0022 &  0.4320 &  0.0028 \\
7954.564973 & 20.3677 & 0.2940 & -5.0758 & 0.0022 &   -0.4140 &  1.7310 & -0.0043 &  0.0075 &  0.0082 &  0.0022 &  0.4322 &  0.0030 \\
7956.459293 & 20.3676 & 0.2780 & -5.0855 & 0.0021 &   -2.7710 &  1.7350 & -0.0063 &  0.0074 &  0.0113 &  0.0022 &  0.4331 &  0.0037 \\
7969.461058 & 20.3653 & 0.3940 & -5.0775 & 0.0033 &   -1.3590 &  1.6300 & -0.0051 &  0.0074 &  0.0105 &  0.0023 &  0.4324 &  0.0041 \\
7971.474861 & 20.3664 & 0.3120 & -5.0673 & 0.0023 &   -1.8940 &  1.4020 & -0.0056 &  0.0075 &  0.0088 &  0.0022 &  0.4325 &  0.0031 \\
7972.406649 & 20.3650 & 0.4200 & -5.0759 & 0.0037 &   -1.6860 &  1.7200 & -0.0056 &  0.0076 &  0.0107 &  0.0023 &  0.4326 &  0.0045 \\
7973.367995 & 20.3665 & 0.4230 & -5.0723 & 0.0036 &   -1.2280 &  2.0120 & -0.0055 &  0.0074 &  0.0107 &  0.0023 &  0.4322 &  0.0049 \\
7974.564434 & 20.3640 & 0.5090 & -5.0691 & 0.0049 &   -1.9400 &  1.8380 & -0.0053 &  0.0075 &  0.0111 &  0.0024 &  0.4301 &  0.0047 \\
7975.389929 & 20.3642 & 0.3340 & -5.0815 & 0.0027 &   -2.8060 &  1.8120 & -0.0063 &  0.0074 &  0.0120 &  0.0022 &  0.4320 &  0.0037 \\
7976.517180 & 20.3651 & 0.3700 & -5.0623 & 0.0029 &   -2.3690 &  1.4480 & -0.0062 &  0.0074 &  0.0116 &  0.0022 &  0.4310 &  0.0033 \\
7984.509802 & 20.3659 & 0.3430 & -5.0760 & 0.0027 &   -3.2790 &  1.3930 & -0.0057 &  0.0075 &  0.0107 &  0.0022 &  0.4321 &  0.0033 \\
7989.419213 & 20.3685 & 0.3010 & -5.0761 & 0.0023 &    0.3160 &  2.0690 & -0.0035 &  0.0074 &  0.0081 &  0.0022 &  0.4310 &  0.0034 \\
7991.407297 & 20.3750 & 0.5790 & -5.0587 & 0.0048 &   -6.1480 &  1.4680 & -0.0104 &  0.0065 &  0.0155 &  0.0026 &  0.4300 &  0.0036 \\
7992.421070 & 20.3696 & 0.2920 & -5.0730 & 0.0022 &   -2.0140 &  1.5810 & -0.0054 &  0.0073 &  0.0104 &  0.0022 &  0.4312 &  0.0028 \\
7993.400517 & 20.3688 & 0.2930 & -5.0751 & 0.0022 &   -0.2570 &  2.2110 & -0.0045 &  0.0073 &  0.0108 &  0.0022 &  0.4316 &  0.0029 \\
7994.423305 & 20.3722 & 0.3210 & -5.0666 & 0.0024 &   -1.5750 &  1.7290 & -0.0046 &  0.0074 &  0.0096 &  0.0022 &  0.4319 &  0.0028 \\
7995.381212 & 20.3697 & 0.3450 & -5.0722 & 0.0027 &   -2.0920 &  1.8670 & -0.0055 &  0.0074 &  0.0112 &  0.0023 &  0.4326 &  0.0034 \\
7996.417124 & 20.3700 & 0.5730 & -5.0738 & 0.0059 &   -4.8350 &  1.4320 & -0.0081 &  0.0074 &  0.0122 &  0.0025 &  0.4328 &  0.0051 \\
7997.404485 & 20.3736 & 0.3940 & -5.0738 & 0.0033 &   -3.8020 &  1.2770 & -0.0071 &  0.0072 &  0.0119 &  0.0022 &  0.4325 &  0.0037 \\
7999.402305 & 20.3713 & 0.3700 & -5.0786 & 0.0031 &   -3.9480 &  1.0990 & -0.0071 &  0.0075 &  0.0116 &  0.0023 &  0.4328 &  0.0039 \\
8000.375013 & 20.3733 & 0.4370 & -5.0685 & 0.0039 &   -1.5250 &  1.6430 & -0.0050 &  0.0072 &  0.0099 &  0.0023 &  0.4326 &  0.0045 \\
8001.423596 & 20.3723 & 0.2910 & -5.0733 & 0.0022 &   -1.3580 &  1.9440 & -0.0045 &  0.0072 &  0.0104 &  0.0022 &  0.4306 &  0.0030 \\
8019.330908 & 20.3615 & 0.6530 & -5.0621 & 0.0071 &   -1.8030 &  1.6320 & -0.0043 &  0.0076 &  0.0087 &  0.0027 &  0.4328 &  0.0059 \\
8020.333101 & 20.3598 & 0.8020 & -5.0691 & 0.0099 &   -5.1400 &  2.2940 & -0.0092 &  0.0074 &  0.0167 &  0.0029 &  0.4324 &  0.0071 \\
8020.344177 & 20.3630 & 0.7230 & -5.0524 & 0.0081 &   -2.7730 &  2.0480 & -0.0065 &  0.0075 &  0.0140 &  0.0028 &  0.4328 &  0.0065 \\
8021.332643 & 20.3635 & 0.5900 & -5.0696 & 0.0061 &   -2.4310 &  1.4760 & -0.0044 &  0.0075 &  0.0068 &  0.0025 &  0.4325 &  0.0055 \\
8022.330786 & 20.3658 & 0.2610 & -5.0701 & 0.0018 &   -0.6500 &  1.9640 & -0.0048 &  0.0074 &  0.0093 &  0.0022 &  0.4319 &  0.0028 \\
8024.393432 & 20.3638 & 0.2520 & -5.0666 & 0.0017 &   -2.8970 &  1.8040 & -0.0062 &  0.0072 &  0.0118 &  0.0022 &  0.4325 &  0.0025 \\
8025.380660 & 20.3633 & 0.2610 & -5.0790 & 0.0019 &   -1.7710 &  1.6970 & -0.0051 &  0.0073 &  0.0097 &  0.0022 &  0.4334 &  0.0031 \\
8026.376037 & 20.3630 & 0.2790 & -5.0765 & 0.0020 &   -1.8400 &  1.6280 & -0.0052 &  0.0074 &  0.0098 &  0.0022 &  0.4328 &  0.0034 \\
8027.373334 & 20.3643 & 0.6860 & -5.0739 & 0.0079 &   -4.2790 &  1.7140 & -0.0086 &  0.0074 &  0.0140 &  0.0027 &  0.4333 &  0.0068 \\
8044.375945 & 20.3621 & 0.3330 & -5.0690 & 0.0026 &   -1.2910 &  1.5180 & -0.0044 &  0.0072 &  0.0084 &  0.0022 &  0.4320 &  0.0032 \\
8050.327410 & 20.3618 & 1.2730 & -5.0427 & 0.0194 &   -4.7300 &  3.1830 & -0.0089 &  0.0076 &  0.0125 &  0.0037 &  0.4326 &  0.0099 \\
8243.596266 & 20.3606 & 0.3510 & -5.0852 & 0.0028 &   -1.8040 &  1.7400 & -0.0054 &  0.0075 &  0.0100 &  0.0023 &  0.4331 &  0.0036 \\
8253.638641 & 20.3603 & 0.2750 & -5.0790 & 0.0020 &   -1.5820 &  1.7110 & -0.0054 &  0.0076 &  0.0100 &  0.0022 &  0.4326 &  0.0028 \\
8269.649711 & 20.3571 & 0.2600 & -5.0842 & 0.0019 &   -2.7090 &  1.5450 & -0.0060 &  0.0076 &  0.0111 &  0.0021 &  0.4332 &  0.0030 \\
8271.690393 & 20.3571 & 0.2550 & -5.0757 & 0.0018 &   -2.1490 &  1.7120 & -0.0055 &  0.0075 &  0.0100 &  0.0022 &  0.4320 &  0.0026 \\
8272.712209 & 20.3572 & 0.3450 & -5.0790 & 0.0027 &   -2.9670 &  1.3420 & -0.0060 &  0.0076 &  0.0106 &  0.0022 &  0.4323 &  0.0033 \\
8275.632635 & 20.3578 & 0.2340 & -5.0804 & 0.0016 &   -1.8780 &  1.6250 & -0.0060 &  0.0074 &  0.0105 &  0.0021 &  0.4326 &  0.0029 \\
8276.680466 & 20.3566 & 0.3420 & -5.0753 & 0.0027 &   -1.9320 &  1.7080 & -0.0057 &  0.0074 &  0.0107 &  0.0022 &  0.4318 &  0.0034 \\
8277.578443 & 20.3553 & 0.2920 & -5.0857 & 0.0022 &   -2.6630 &  1.6000 & -0.0067 &  0.0076 &  0.0113 &  0.0022 &  0.4323 &  0.0035 \\
8278.681214 & 20.3573 & 0.2590 & -5.0676 & 0.0018 &   -2.6830 &  1.7970 & -0.0066 &  0.0076 &  0.0118 &  0.0021 &  0.4312 &  0.0024 \\
8279.714701 & 20.3567 & 0.2110 & -5.0745 & 0.0014 &   -2.3650 &  1.9720 & -0.0064 &  0.0074 &  0.0120 &  0.0021 &  0.4321 &  0.0021 \\
\end{longtable}
\end{landscape}
}


\section{Stellar parameters}
\label{sec:star}
The main stellar parameters of HD\,164922 are collected in Table \ref{param}, either reported from the literature or specifically estimated in the present work. The astrometric parameters (positions, proper motions and parallax) are taken from the \textit{Gaia} Data Release 2 \citep{2018A&A...616A...1G,2018A&A...616A...2L}, that we used to evaluate the perspective acceleration in the previous section. 

To estimate the photospheric parameters, we produced a coadded spectrum from 212 HARPS-N spectra of our dataset with higher S/N. We also excluded spectra collected before JD 2456740 (March 2014), which are affected by a certain degree of defocusing (note that this issue had a negligible impact on the HARPS-N RVs, see \citealt{2017A&A...599A..90B}). 
The S/N of the coadded spectrum reached the high value of 2800. At this level we were mainly limited by systematic errors of the atmospheric parameters determination method rather than by the quality of this spectrum. 
We obtained a first determination of the atmospheric parameters using \texttt{CCFPams} \citep{2017MNRAS.469.3965M}.
By using the most recent release of the MOOG code \citep{1973ApJ...184..839S} and following the approach described in \cite{2011A&A...525A..35B} with the line list presented in \cite{2012MNRAS.427.2905B}, we measured the stellar effective temperature ($T_{\rm eff}$), surface gravity ($\log g$), iron abundance ($[\rm Fe/H]$) and microturbolence velocity $v_{\rm micro}$ from line equivalent widths (see Table \ref{param}). 
The projected rotational velocity was estimated with the same code and applying the spectral synthesis method after fixing the macroturbulence. Assuming $v_{\rm macro}=2.6$ km s$^{-1}$ from the relationship by \cite{2014MNRAS.444.3592D} based on asteroseismic analysis of main sequence stars observed by the NASA - {\it Kepler} satellite and depending on $T_{\rm eff}$ and $\log g$, we find a projected rotational velocity of $v \sin i=0.7\pm0.5$ km s$^{-1}$. 
This value is consistent with the determination by \cite{2016ApJS..225...32B}, obtained through the spectral synthesis modelling of HIRES data from a large sample of dwarfs and subgiants stars. However, since the velocity resolution of HARPS-N is about 2.0 km s$^{-1}$ we cannot assert that we resolve the actual stellar rotation velocity. 
We estimated the stellar luminosity by combining parallax
measurements with the bolometric corrections to obtain the bolometric magnitude.
By using a bolometric correction $BC=-0.164\pm0.11$  deduced here by following \cite{2010MNRAS.403.1592B}, we obtained $L/L_{\odot}=0.71\pm0.15$.
To obtain the mass, radius and age of our target we used
the web interface of the Bayesian tool PARAM\footnote{\url{http://stev.oapd.inaf.it/cgi-bin/param_1.3}} version1.3 \citep{2006A&A...458..609D,2014MNRAS.445.2758R}, based on the PARSEC isochrones \citep{2012MNRAS.427..127B}.
The input parameters are $T\rm_{eff} $, $[\rm Fe/H]$, and $\log g$
all used as priors to compute the probability density function
of the stellar parameters.
We found that the resulting fundamental parameters of this star are: age $=9.2\pm 2.5$ Gyr, stellar mass
$M= 0.93\pm0.05$ $M_{\odot}$ and radius $R=0.95\pm0.1$ $R_{\odot}$.

The systemic RV and the activity index of the Ca II H\&K lines, $\log R'_{\rm HK} $, are finally obtained from our HARPS-N dataset. The former is obtained from the orbital fit presented in Sect. \ref{subsec:3pl}. The evaluation of the $\log R'_{\rm HK}$ is not affected by the interstellar reddening being our target within 75 pc (see e.g. \citealt{2008ApJ...687.1264M}), and its small value, -5.075 $\pm$ 0.010, suggests that HD\,164922 is a quiet star.
According to the empirical relations by \cite{1984ApJ...279..763N} and \cite{2008ApJ...687.1264M} we should expect a rotation period ranging between 43 and 48 days. In Table \ref{param} we report the adopted $P_{\rm rot}$, $42.3^{+1.3}_{-0.7}$ days, as derived in Sect. \ref{sec:gpact}.

\begin{table}
\caption[]{Stellar parameters of HD\,164922.}
\label{param}
$$
\begin{array}{p{0.45\linewidth}cc}
\hline
\noalign{\smallskip}
Parameter & \mbox{Value} & \mbox{Ref.}\\
\noalign{\smallskip}
\hline
\hline
\noalign{\smallskip}
RA (J2015.5) [deg] & 270.630 \pm 0.016 & \mbox{a} \\
DEC (J2015.5) [deg] & 26.310 \pm 0.024 & \mbox{a} \\
pmRA [mas/yr] & 389.653 \pm 0.033 & \mbox{a} \\
pmDEC [mas/yr] & -602.313 \pm 0.433 & \mbox{a} \\
Parallax [mas] & 45.422 \pm 0.031 & \mbox{a} \\
\noalign{\smallskip}
Distance [pc] & 22.00^{+0.01}_{-0.02} & \mbox{b}  \\
\noalign{\smallskip}
Spectral Type &  \mbox{G9V} & \mbox{c} \\
m$_{\rm V}$ & 6.99 & \mbox{c} \\
B-V & 0.799 \pm 0.005 & \mbox{d} \\
$\dot{v_{\rm r}}$ [m s$^{-1}$ yr$^{-1}$] & 0.26 \pm 0.01  & \mbox{e} \\ 
Systemic RV [km s$^{-1}$] & 20.3634 \pm0.0007  & \mbox{e} \\
$T_{\rm eff}$ [K] & 5390 \pm 30 & \mbox{e} \\
$\log g$ [cm s$^{-2}$] & 4.46 \pm 0.13 & \mbox{e} \\
$ [$Fe/H$] $ [dex] & 0.18 \pm 0.08 & \mbox{e} \\
$v_{\rm micro}$ [km s$^{-1}$] & 0.80 \pm 0.03 & \mbox{e} \\
$v \sin i$ (km s$^{-1}$) & < 2.0 & \mbox{e} \\
$\log R'_{\rm HK} $ &  -5.075\pm0.010  & \mbox{e} \\
\noalign{\smallskip}
$P_{\rm rot}$ [d] & 42.3^{+1.3}_{-0.7}  & \mbox{e} \\
\noalign{\smallskip}
$L/L_{\odot}$ &  0.71 \pm 0.15  &  \mbox{e} \\
Mass [$M_{\odot}$] & 0.926^{+0.029}_{-0.035} & \mbox{e} \\
\noalign{\smallskip}
Radius [$R_{\odot}$] & 0.946^{+0.014}_{-0.021} & \mbox{e} \\
\noalign{\smallskip}
Age [Gyr] & 9.58^{+1.99}_{-1.55} & \mbox{e} \\
\noalign{\smallskip}
\hline
\end{array}
$$
\tablebib{(a) \cite{2018A&A...616A...1G}; (b) \cite{2018AJ....156...58B}; (c) \cite{2006ApJ...646..505B}; (d) \cite{2007A&A...474..653V}; (e) This work. 
}
\end{table}


\section{HARPS-N time series analysis} \label{sec:tsa}
In this Section we present a first evaluation of the HARPS-N time series of HD\,164922, obtained as described in Sect. \ref{sec:obs}.
This investigation was performed in preparation of the GP analysis presented in Sect. \ref{sec:gpanalisi}

\subsection{Periodogram analysis} \label{sec:gls}
\begin{figure}
\centering
\includegraphics[width=9cm,angle=0]{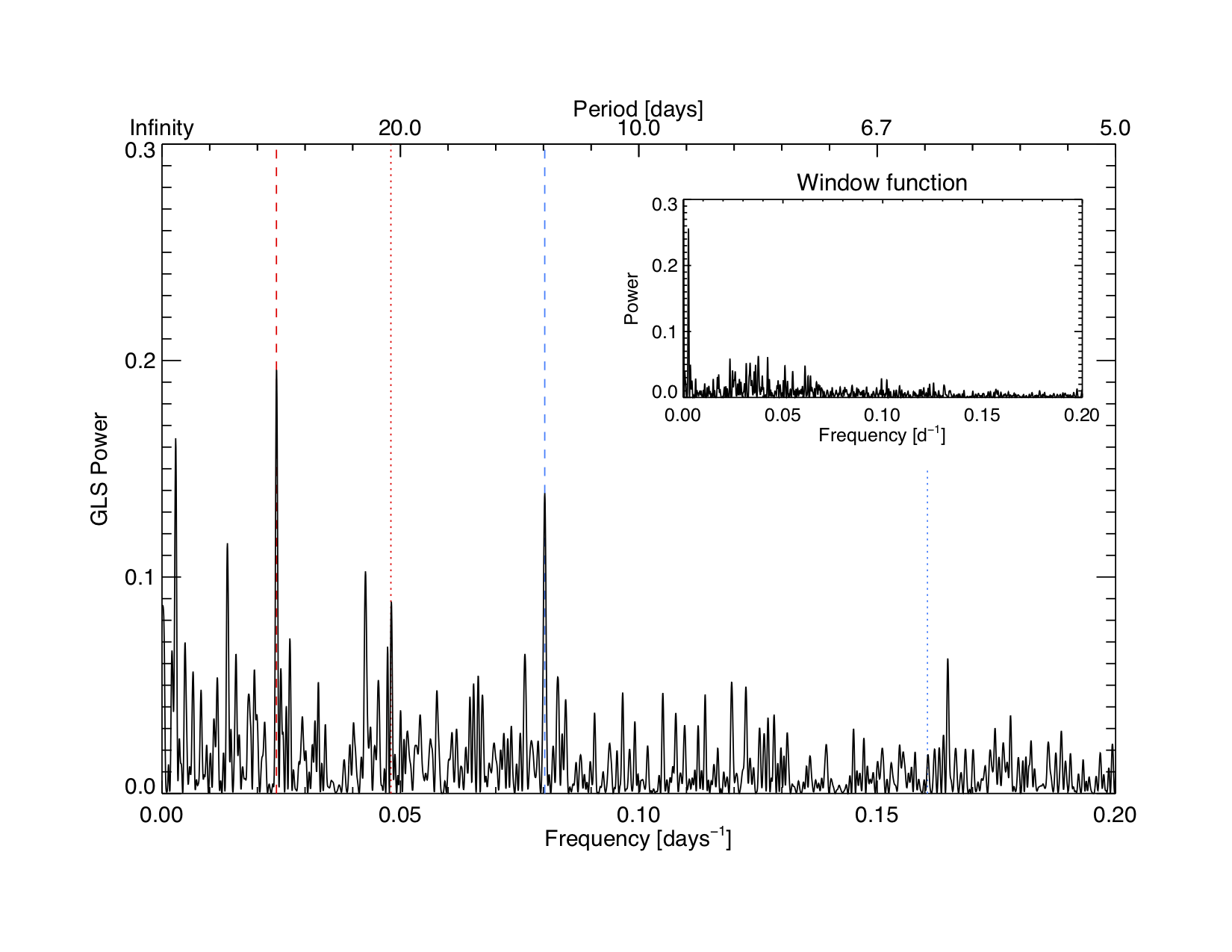}
\caption{\label{fig:gls_res_hn} Generalized Lomb-Scargle periodogram of the HARPS-N RV residuals of the two-planets fit. Dashed red and blue lines indicate the periodicities of 41.63 and 12.46 days with the corresponding harmonics (dotted lines). In the inset, the corresponding window function.}
\end{figure}

\begin{figure}
\centering
\includegraphics[width=9cm,angle=0]{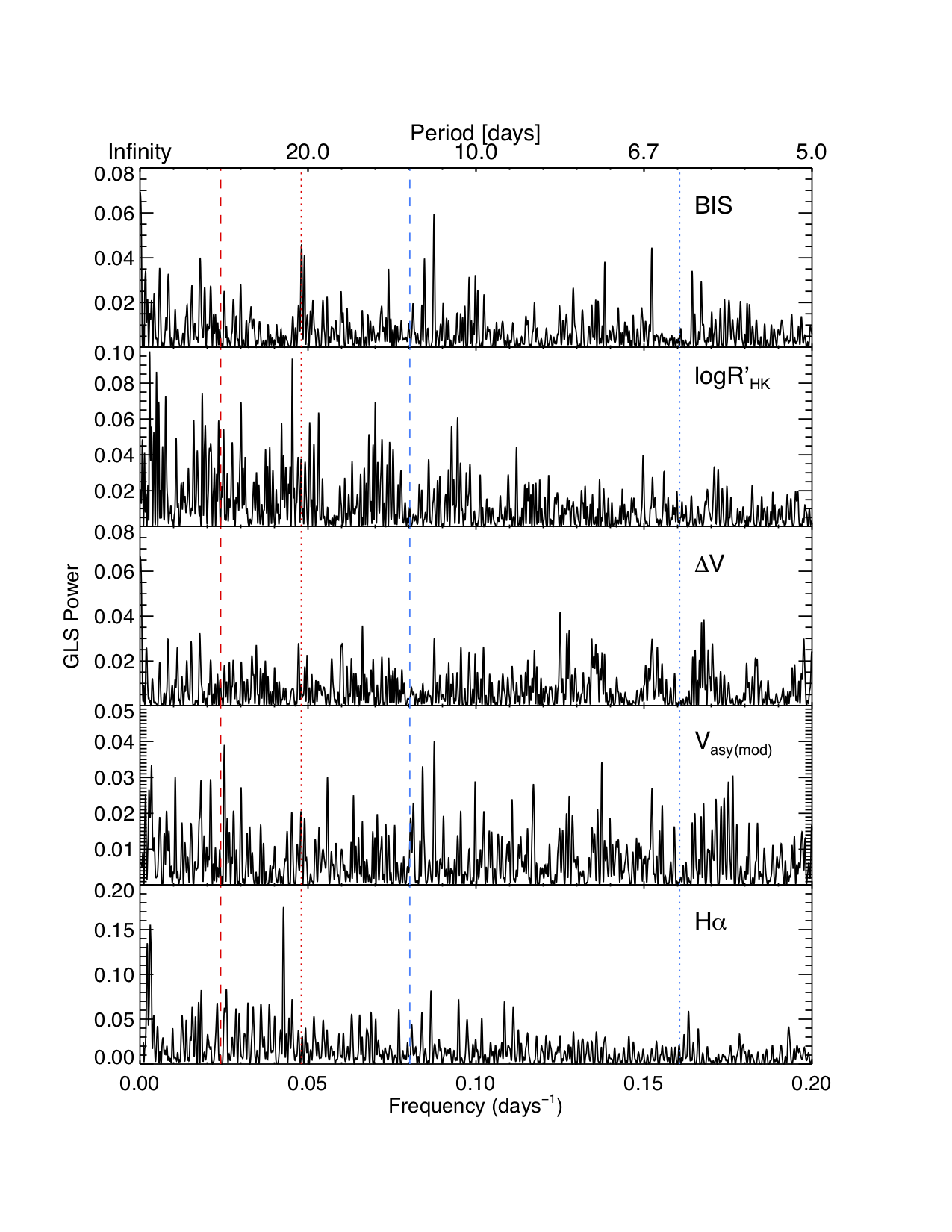}
\caption{\label{fig:glsact} Generalized Lomb-Scargle periodogram of the HARPS-N time series of activity indicators (bisector span, $\log$R$^{\rm \prime}_{\rm HK}$, $\Delta V$, V$_{\rm{asy(mod)}}$ and H$\alpha$). Reference lines have the same color code than in Fig. \ref{fig:gls_res_hn}}
\end{figure}

We carried out a preliminary two-planets fit of the HARPS-N RVs time series using the \texttt{MPFIT IDL} package\footnote{A non-linear least squares curve fitting presented by \cite{2009ASPC..411..251M}, available at \url{http://purl.com/net/mpfit}}, implemented as in \cite{2011A&A...533A..90D}, starting from the orbital solution reported by F2016 and depicted by a gray solid line in Fig. \ref{fig:rvharps}. Subsequently, we computed the Generalized Lomb-Scargle (\texttt{GLS}) periodogram \citep{2009A&A...496..577Z} of the RV residuals, shown in Fig. \ref{fig:gls_res_hn} together with the corresponding window function, in the inset plot. Except for the 1-year signal located at the lower frequencies, two main periodicities are visible: the dominant one is found at 0.0239 days$^{-1}$ (i.e., 41.63 d in the time domain, red dashed line), the second one at 0.0803 days$^{-1}$ (12.46 d, blue dashed line), both of them with a False Alarm Probability (FAP) less than $10^{-5}$, evaluated with 10\,000 bootstrap random permutations. 
The large peak at $\sim 41.6$ d could be the signature of the stellar rotation, as expected by the empirical activity-rotation relationships (Sect. \ref{sec:star}), while the second peak could suggest the presence of a short period planetary companion. For sake of completeness, the periodicity at about 0.165 days$^{-1}$ shows a high value of the FAP (15\%).
In order to understand whether the main periodicities are related to the stellar activity, we evaluated the \texttt{GLS} for the BIS, the $\log$R$^{\rm \prime}_{\rm HK}$ and the available asymmetry indices, as presented by \cite{2018A&A...616A.155L}. 
The \texttt{GLS} of all these parameters are reported in Fig. \ref{fig:glsact}, where the dashed lines follow the same color code as in Fig. \ref{fig:gls_res_hn}. No significant peak can be found close to the 12.46 d periodicity. Based on the FAP evaluation, none of the periodogram peaks appears significant, with the only exception of the 23.4 days in the periodogram of H$\alpha$ (FAP=0.01\%). 

We finally evaluated the Spearman correlation coefficient between the RV residuals of the two-planets fit and the considered activity indices but no correlation was found. This might be explained by the dominant contribution of the faculae to the RVs rather than starspots (e.g., \citealt{2016A&A...593A...5D} and \citealt{2019MNRAS.487.1082C} for the solar case).

\subsection{GP regression of the $\log$R$^{\rm \prime}_{\rm HK}$  index}
\label{sec:gpact}
To confirm the hypothesis that the rotation period of the star is about 42 days, we performed a GP regression analysis of the time series of the $\log$R$^{\rm \prime}_{\rm HK}$ index. We used a quasi-periodic model for the correlated signal of stellar origin. We first removed all the spectra with S/N < 20 in the region of the Ca II lines, to work with a more reliable determination of the $\log$R$^{\rm \prime}_{\rm HK}$ value. The Monte Carlo analysis has been done with the open source Bayesian inference tool \textsc{MultiNestv3.10} (e.g. \citealt{feroz13}), through the \textsc{pyMultiNest} \texttt{python} wrapper \citep{buchner14}, including the publicly available GP python module  \texttt{GEORGEv0.2.1} \citep{2015ITPAM..38..252A}. 
The formalism used for the GP regression is the same described in \cite{2016A&A...593A.117A}, with the quasi-periodic kernel described by the following covariance matrix:
\begin{eqnarray} \label{eq:eqgpkernel}
k(t, t^{\prime}) = h^2\cdot\exp\large[-\frac{(t-t^{\prime})^2}{2\lambda^2} - \frac{sin^{2}(\pi(t-t^{\prime})/\theta)}{2w^2}\large] + \nonumber \\
+\, (\sigma^{2}_{\log {\rm R}^{\rm \prime}_{\rm HK}}(t))\cdot\delta_{t, t^{\prime}}
\end{eqnarray}
where $t$ and $t^{\prime} (t)$ indicate two different epochs, $\sigma_{\log {\rm R}^{\rm \prime}_{\rm HK}}$ is the uncertainty of the $\log$R$^{\rm \prime}_{\rm HK}$ index at the time $t$ added quadratically to the diagonal of the covariance matrix, and $\delta_{t, t^{\prime}}$ is the Kronecker delta. Finally, $h$, $\lambda$, $\theta$ and $w$ are the hyper-parameters: $\theta$ represents the periodic time-scale of the modeled signal, and corresponds to the stellar rotation period; $h$ denotes the scale amplitude of the correlated signal; $w$ describes the level of high-frequency variation within a complete stellar rotation; $\lambda$ represents the decay timescale in days of the correlations and can be physically related to the active region lifetimes (see e.g. \citealt{2014MNRAS.443.2517H}).
We performed a run using a uniform prior on $\theta$ between 15 and 100 days, large enough to span the range of the possible rotation periods. The resulting posterior distribution for $\theta$ shows a maximum centered at $42.3^{+1.3}_{-0.7}$ days, which is close to the expected rotation period as derived from the empirical relation with the value of $\log$R$^{\rm \prime}_{\rm HK}$. Also the maximum a posteriori probability (MAP) value of $\theta$ is 42.2 days. 
The hyper-parameter $\lambda$ is $33.6^{+7.5}_{-14.7}$ days (MAP = 35.3 days). This value is comparable to $\theta$, thus it indicates that the correlated signal due to the stellar activity changes within a single rotation period and is well described by a quasi-periodic signal modulated over the stellar rotation period. 
To complete our analysis on the activity indicators, we run similar models both for the BIS and the V$_{\rm asy(mod)}$ with large uniform priors on $\theta$ (from 0 to 100 days) but no periodicity is found.
 
In conclusion, our GP analysis of the HARPS-N time series of the $\log$R$^{\rm \prime}_{\rm HK}$ activity index returned clear evidence for the presence of a 42-d periodicity, that we assume as the rotation period of our target.
This period is also very close to the signal of 41.7 days reported by F2016 in the periodogram of their RV data (with an amplitude of 1.9 ms$^{-1}$) and interpreted as a possible signature of the stellar rotation. However, as their analysis was focused on the detection of new planetary companions, they did not investigate in detail the nature of that signal.

\subsection{Comparison with data from the literature} \label{sec:literature}
As a final check before starting the GP modelling of our RV data, we analysed the rich RV dataset available in the literature for HD\,164922.
We considered the official HIRES/Keck data release \citep{2017AJ....153..208B} and the Levy/APF RVs published by F2016.
Both HIRES and Levy spectrographs adopted the Iodine cell method \citep{1996PASP..108..500B} to obtain high precision radial velocity measurements. With this technique a rich forest of I$_{\rm 2}$ lines is overimposed on the stellar spectrum as a stable reference in the wavelength range between 5000 and 6200 \si{\angstrom}.
Similarly to F2016, we considered the HIRES dataset as two separate time series because of the major upgrade of the spectrograph in 2004, labelling them as ``pre'' and ``post'' the modification.
The HIRES$_{\rm pre}$ dataset includes 51 spectra collected through 8 years with a mean RV error of 1.23 m s$^{-1}$.
The HIRES$_{\rm post}$ dataset includes 248 RVs spanning 10 years with a mean internal error of 1.11 m s$^{-1}$. Both of them clearly show the modulation of the long period giant planet. The perspective acceleration in the HIRES data is already taken into account by the standard RV extraction pipeline (e.g., \citealt{2010ApJ...721.1467H}). 
The 73 data from the Levy spectrograph show an RV error of 2.13 m s$^{-1}$ and do not allow to sample the full orbit of the planet b, since they were specifically collected to recover the planet c (the time span is 1.28 years).

\begin{figure}
\centering
\includegraphics[width=9cm,angle=0]{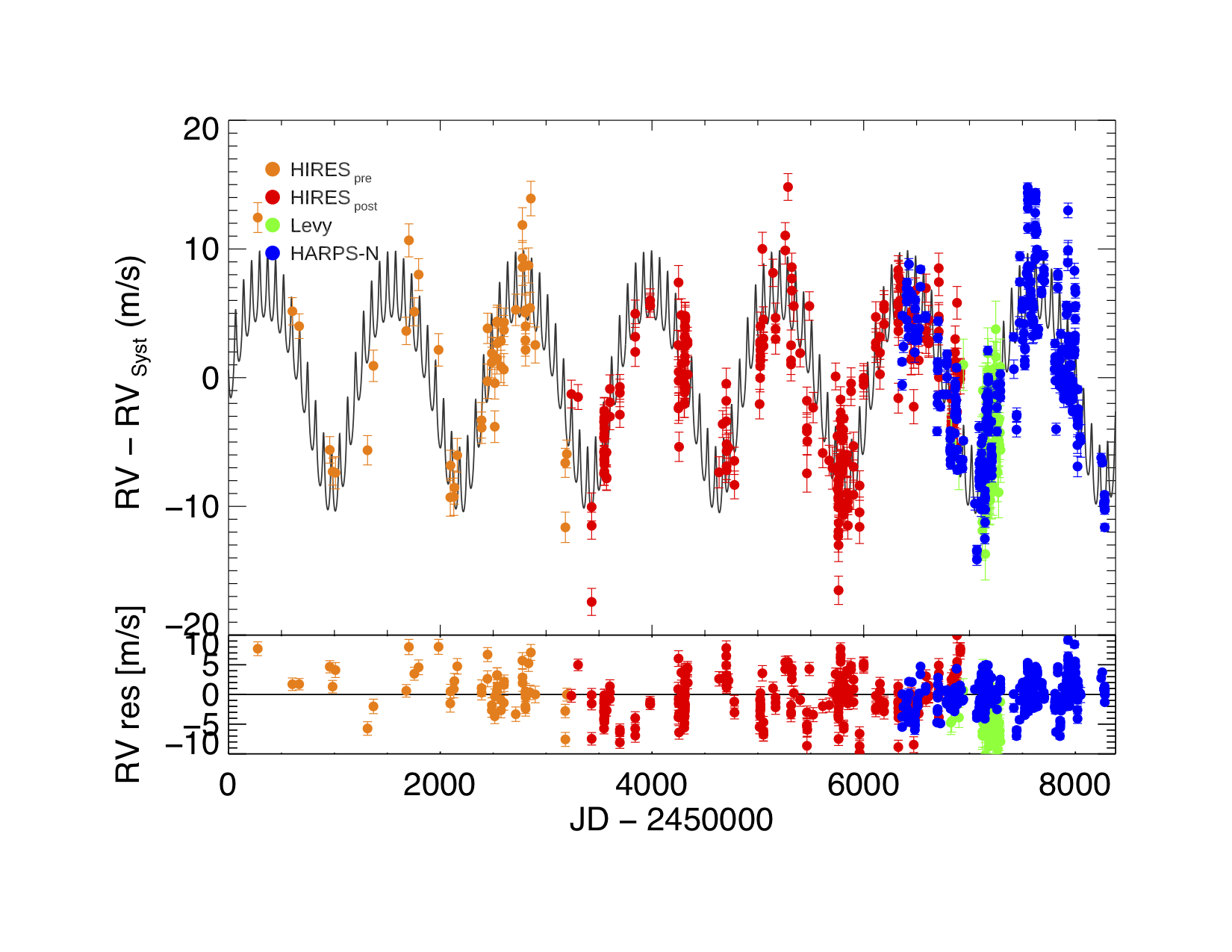}
\caption{\label{fig:rvall} The full RV dataset available in the literature for HD\,164922. The gray solid line represents the two-planet model obtained from the solution in F2016 with the corresponding residuals (lower panel). The color code is defined in the legend.}
\end{figure}

\begin{figure}
\centering
\includegraphics[width=9cm,angle=0]{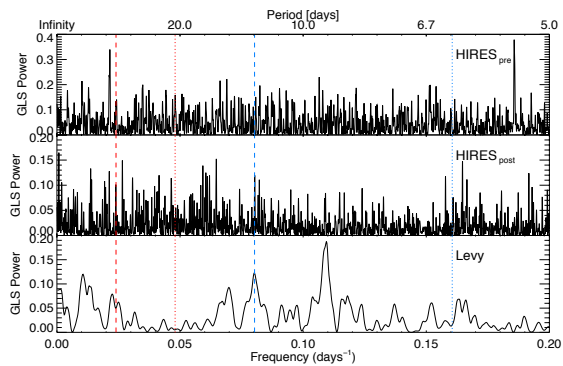}
\caption{\label{fig:glslit} \texttt{GLS} of the RV residuals for the three datasets available in the literature (HIRES ``pre'' and ``post'' the major upgrades and Levy). Vertical lines follow the same color code as in Fig. \ref{fig:gls_res_hn}}
\end{figure}

The full HIRES+Levy+HARPS-N dataset for HD\,164922 is probably one of the richest high-precision RV collections known in the literature, since it is composed by 684 data. It is shown in Fig. \ref{fig:rvall} together with the two-planet fit (gray solid line) obtained as in Sect. \ref{sec:gls}.
The \texttt{GLS} of the corresponding residuals is shown separately for the HIRES$_{\rm pre}$, HIRES$_{\rm post}$ and Levy datasets in Fig. \ref{fig:glslit}.
As a reference, we overplotted the location of the periodicities found in the HARPS-N residuals with the corresponding harmonics (same color code as in Fig. \ref{fig:gls_res_hn}). We observe a periodicity close to 41.63 d in the HIRES$_{\rm pre}$ dataset, and some power both in the HIRES$_{\rm post}$ and Levy \texttt{GLS}, even if the peaks are not significantly higher than the background noise. This is expected because of the typical lack of coherence of the stellar rotation modulation. The signal at 12.46 d is not recovered in the HIRES$_{\rm pre}$ data because of the sparse sampling, while it is recognized among the peaks in the HIRES$_{\rm post}$ \texttt{GLS}. In the Levy \texttt{GLS} the periodicity at 12.46 d is clearly found, while the dominant peak (at about 9.13 days) seems to be spurious after an inspection with 10\,000 bootstrap random permutations, similarly for the peak at 5.3 days in the HIRES$_{\rm pre}$ \texttt{GLS} (FAP > 5\%).

Due to the large time span of the dataset from the literature, useful to constrain the orbit of planet b, we included those data in the GP modelling presented in the next section. 
As a final note, during the preparation of the present work, we learned of an alternative RV extraction of HIRES data from \citealt{2019MNRAS.484L...8T}. After the comparison between the two datasets, that showed a typical difference of the RVs of about 0.5 $\ms$, we decided to proceed with the data from the official pipeline by \cite{2017AJ....153..208B}.

\section{Analysis of the radial velocity datasets}
\label{sec:gpanalisi}
For the RV analysis we used the same tools described in Section \ref{sec:gpact}, with 800 random walkers and a sampling efficiency of 0.8 The covariance matrix is similar to Eq. \ref{eq:eqgpkernel}, but in this case we added quadratically an uncorrelated noise term defined as: 
\begin{equation} \label{eqn:rv}
(\sigma^{2}_{RV}(t)\,+\,\sigma_{j}^{2})\cdot\delta_{t, t^{\prime}}.
\end{equation}
In this case, $\sigma_{RV}(t)$ is the RV uncertainty at the time $t$ and $\sigma_{j}$ is the additional noise we used to take into account instrumental effects and further noise sources.
We performed a GP analysis of the data from by HIRES (``pre'' and ``post''), Levy, and HARPS-N data. The final, multi-instrument dataset is composed by 684 RVs covering a time span of nearly 22 years, for which a GP regression including two or more Keplerian signals represents a huge computational effort (from a few weeks to months).

\subsection{Two-planet model} \label{subsec:2pl}
First of all, we explored the model composed by two Keplerian signals (i.e., planet b and c) including the correlated term due to the stellar activity, modelled by a GP quasi-periodic kernel. The priors for the GP hyper-parameters and for the planetary orbital parameters are reported in Table \ref{Table:postgprv}. For $\lambda$ and $w$ we chose the same conservative priors used to model the $\log$R$^{\rm \prime}_{\rm HK}$ time series in Sect. \ref{sec:gpact}, while we extended the prior on $\theta$ in the range 30-50 days. We consider this a choice justified by the large temporal extent of the dataset, since we can expect a change in the measured value of the stellar rotation period over 22 years, according to the evolution of the activity cycle.

Concerning the two Keplerian signals we chose not to impose Gaussian priors centered on the orbital solution obtained by F2016: this would be too restrictive, given the inclusion of so many new data extending the time span by more than 5 years. Moreover, since the HIRES data are included in our analysis, it would be improper to adopt an informative prior based on the same data. Finally, we included a linear term as a free parameter of the model to investigate the presence of a long term additional signal.
The results are reported in Table \ref{Table:postgprv}.
We first notice that the introduction of the HIRES datasets have necessarily impacted on the activity description of the star, with respect to the analysis performed in Sect. \ref{sec:gpact} for HARPS-N data only. On a 22 years timespan, we could expect rotation period variations due to differential rotation and/or variation of the magnetic activity cycle, so the hyper-parameter $\theta$ changed slightly from the expected 42.3 to 41 days. This can also explain the higher uncertainty on $\theta$, with respect to the GP regression in the case of the HARPS-N time series only, obtained in five years of intensive monitoring. 

\begin{figure}[htbp]
\centering
\includegraphics[width=9cm,angle=0]{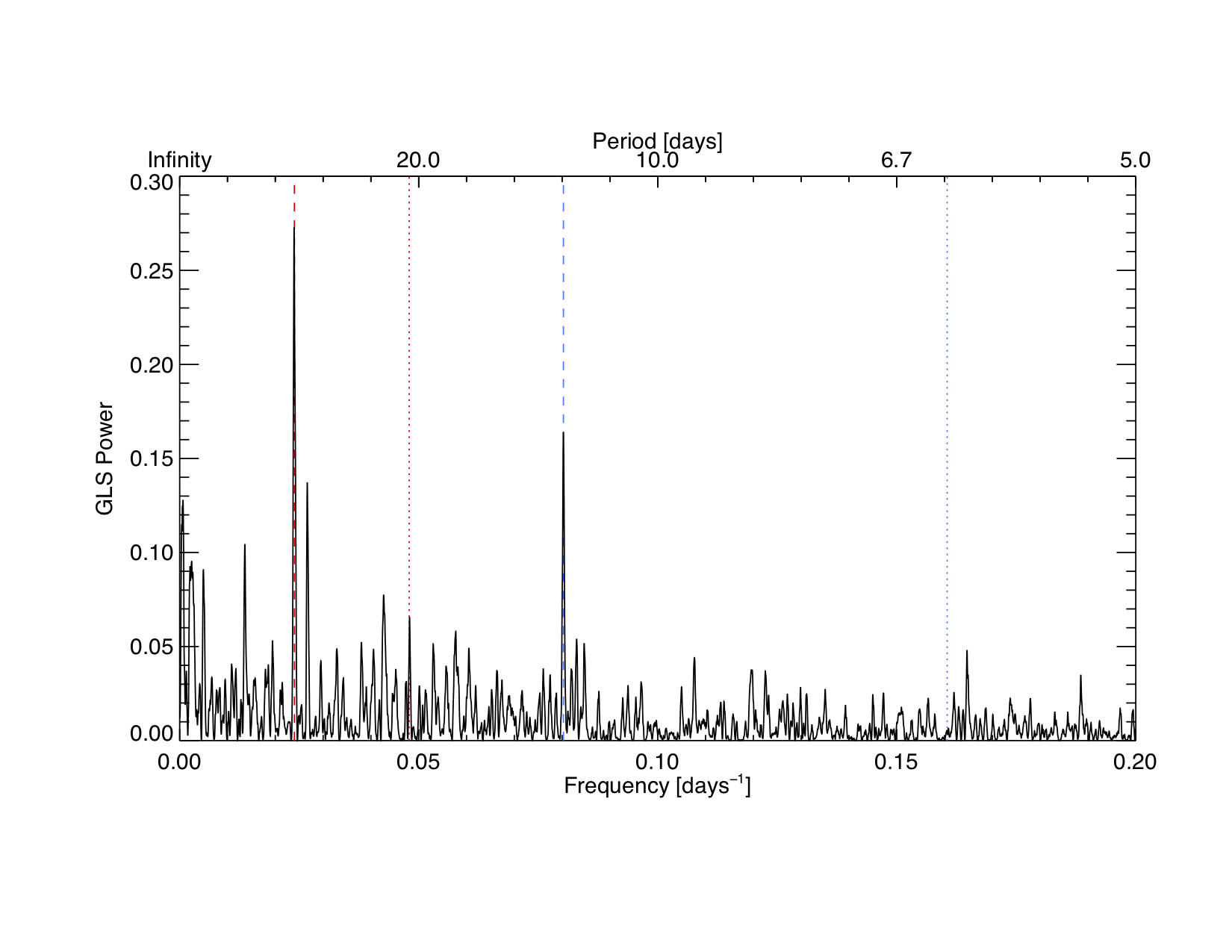}
\caption{\label{fig:res_2p} \texttt{GLS} of the RV residuals of the two-planets fit for the full dataset of HD\,164922. Same color code as Fig. \ref{fig:gls_res_hn}}
\end{figure}
The \texttt{GLS} of the residuals (Fig. \ref{fig:res_2p}) shows the same periodicities identified in Fig. \ref{fig:gls_res_hn}, i.e., the signature of the stellar activity (0.02394$\pm 2 \cdot 10^{-5}$ days$^{-1}$ corresponding to 41.77$\pm$0.03 days) and the signal at 0.08027$\pm 2 \cdot 10^{-5}$ days$^{-1}$, corresponding to a period of 12.457$\pm$0.030 days. 
If we subtract from these residuals the model of the stellar activity as obtained by the GP regression, the final time series does not show any periodicity, not even the one at $\sim$ 12.46 days.
In this case, the flexibility of the GP regression,
likely implied that the activity-induced term has absorbed the candidate third planetary signal. For this reason we modelled the data including one additional Keplerian signal to investigate the nature of the periodicity at $\sim$ 12.46 days, as described in the next sub-section.

\subsection{Three-planet model} \label{subsec:3pl}
The presence in our data of the robust additional signal at nearly 12.46 days led us to include a third Keplerian signal in the global stellar activity+planet model.
We kept fixed the priors for both the activity and the two known planets with respect to the model presented in Sect. \ref{subsec:2pl}, while the priors for the parameters of the candidate planet d are reported in Table \ref{Table:postgprv}. Despite the very precise determination of the signal in the \texttt{GLS} periodogram in Fig. \ref{fig:res_2p}, we use a uniform prior for the orbital period in a relatively wide range with the specific aim to avoid biasing the solution toward that value. 

Table \ref{Table:postgprv} shows the result of the fit. The description of the correlated activity term is consistent with the one obtained for the two-planet model. The orbital period of the planetary candidate is $12.458\pm 0.003$ days, the RV semi-amplitude is $1.3 \pm 0.2$ m s$^{-1}$ (6$\sigma$ significance), and the minimum mass $m_{\rm d}\sin i_d =4\pm1$ M$_{\oplus}$ is 4$\sigma$ significant. 

The best-fit values of the Keplerian parameters are consistent with those by F2016 within 1-$\sigma$, except for the eccentricity of planet b that we find to be lower and significantly consistent with zero ($e_b = 0.08^{+0.06}_{-0.05}$ with respect to $0.126 \pm 0.049$ in F2016). A more constrained eccentricity consistent with zero is also found for planet c (i.e., $0.12^{+0.11}_{-0.09}$ versus $0.22 \pm 0.13$ from F2016). 
The model returned a small but significant acceleration term
which is found to be correlated to the RV offsets associated to the four instruments. 
We evaluated the difference of the proper motion components from Hipparcos and \textit{Gaia} DR2 with respect to the reference proper motion from the catalog of accelerations by \cite{2018ApJS..239...31B}. 
In both cases, the $\Delta \mu_{\alpha,\delta}$ are lower than 1 mas/yr and compatible with zero within $1\sigma$, hinting at a non-existent or negligible anomaly from a purely linear motion on the plane of the sky.
\begin{figure}
\centering
\includegraphics[width=9cm,angle=0]{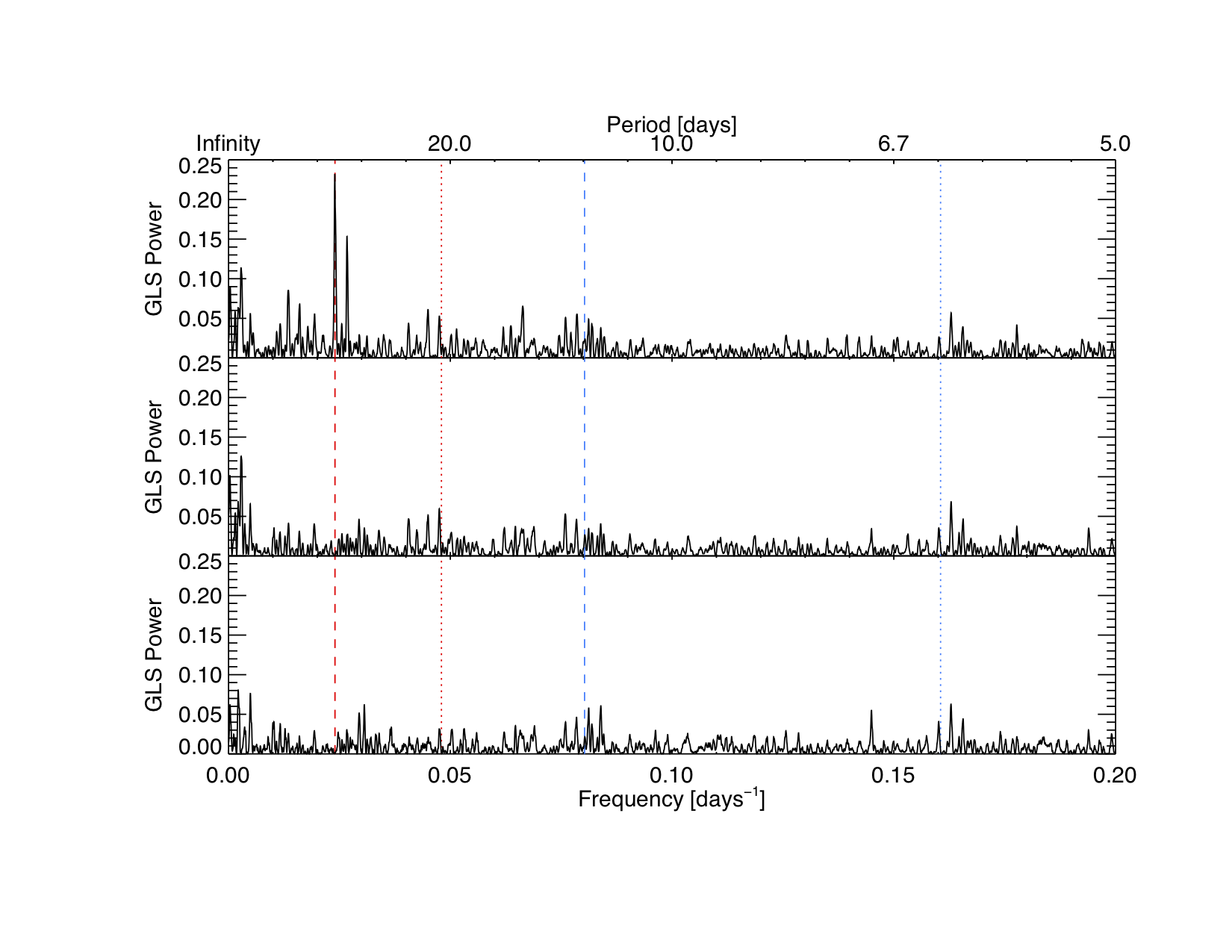}
\caption{\label{fig:res_3p} Upper panel: \texttt{GLS} of the RV residuals for the GP regression including three-planets, after removing the instrumental offsets and the planetary signals. Middle and lower panels show the \texttt{GLS} of the RV residuals after two iterations of the prewhitening procedure.}
\end{figure}

\begin{figure}
\centering
\includegraphics[width=9cm,angle=0]{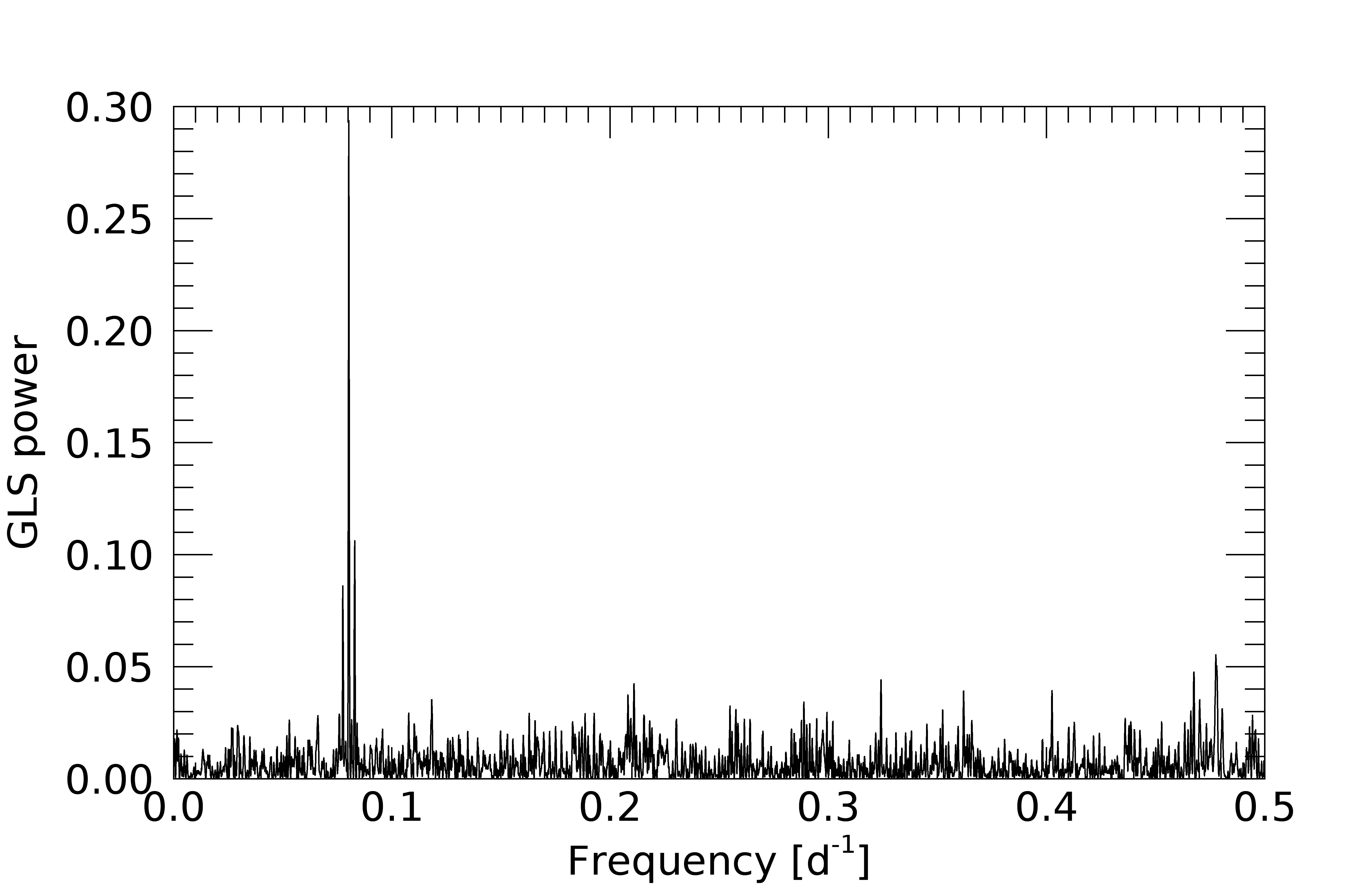}
\caption{\label{fig:res_pld} \texttt{GLS} of the RV residuals after removing planets b and c and the stellar contribution as fitted by the GP model. The signal of planet d at $\sim 12.46$ days is clearly recovered.}
\end{figure}

Upper panel of Fig. \ref{fig:res_3p} shows the \texttt{GLS} of the resulting RV residuals, after removing the three planetary signals and the instrumental offsets from the original dataset. Except for the periodicity at $\sim 42$ days and its first harmonics, we notice a peak close to 1-yr signal (0.0028 d$^{-1}$, which was also present in the two-planet fit residuals in Fig. \ref{fig:res_2p}). We performed a pre-whitening operation (where the signal of the most prominent periodicity in the \texttt{GLS} is subtracted from the time series and the main signal in the \texttt{GLS} is evaluated for the residuals and subtracted again until no peaks are above 3$\sigma$ from the noise) and after the iterative subtraction of the 42 days (\texttt{GLS} in the middle panel of Fig. \ref{fig:res_3p}) and the 1yr signals, the resulting periodogram shows no significant periodicities (lower panel).
The origin of the 1-yr periodicity will be discussed in Sect. \ref{sec:stability}.  As pointed out in the previous sub-section, the signal of the third planet at about 12.46 days is absorbed in the GP regression, while we demonstrated that it is actually present in our data and must be included in the model. Indeed, the \texttt{GLS} of the RV residuals after the removal of the two known planets and the stellar contribution clearly show the signal of planet d (Fig. \ref{fig:res_pld}).
\begin{figure}
\centering
\includegraphics[width=9.5cm,angle=0]{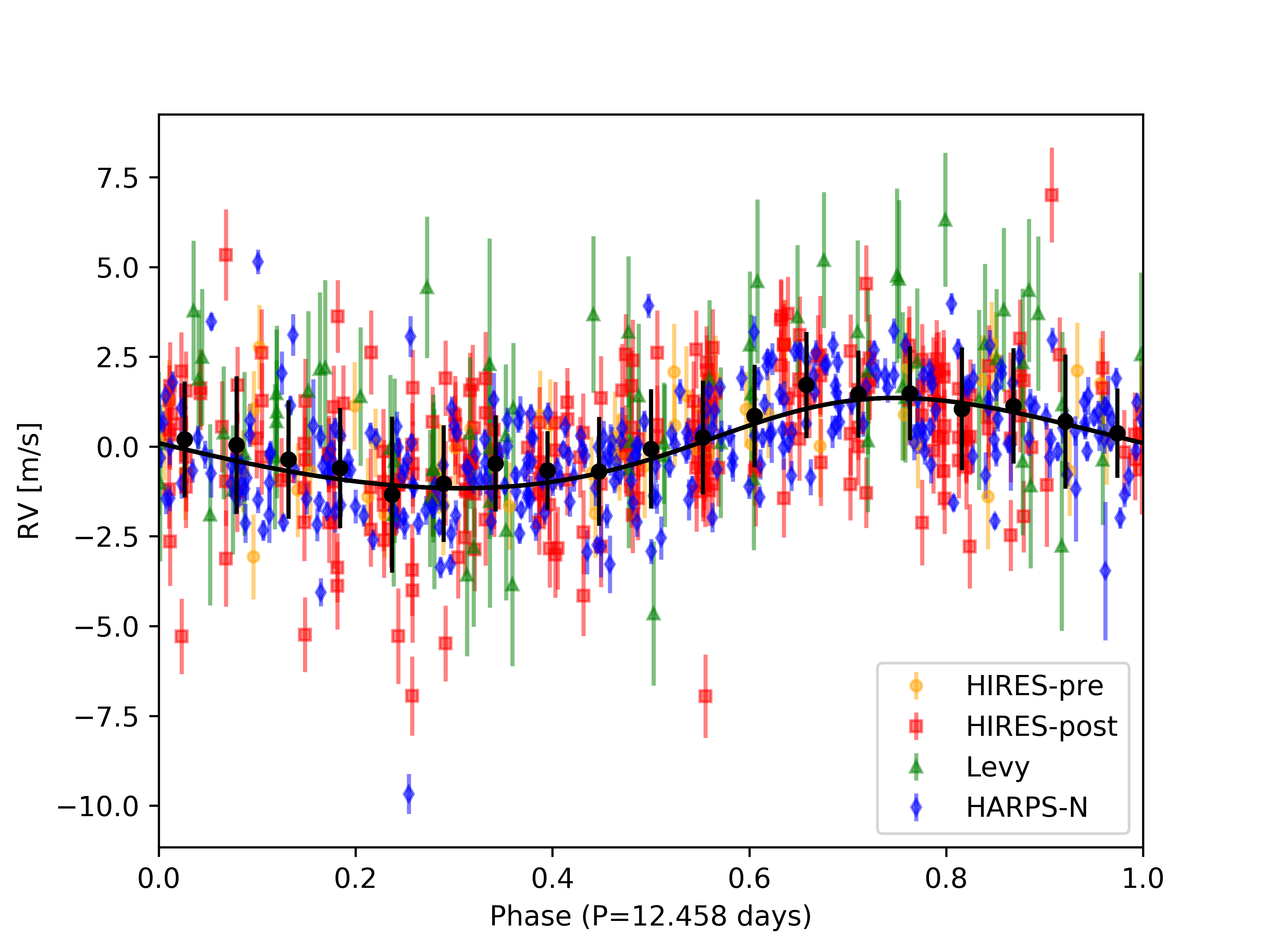}
\caption{\label{fig:phased} Spectroscopic orbit of planet d. The RV residuals from HIRES (``pre'' and ``post'' upgrade), Levy/APF, and HARPS-N are phase-folded at the orbital period 12.458 days (color code in the legend, same as in Fig. \ref{fig:rvall}). The residuals are calculated by subtracting the activity-induced term and the signals of planet b and c from the original data, using the results in Tab. \ref{Table:postgprv}. The black solid line represents the best-fit spectroscopic orbit of planet d, while black dots indicate the binned RVs and the error bars represent the RMS of the data in each bin.}
\end{figure}
\begin{figure}
\centering
\includegraphics[width=9.5cm,angle=0]{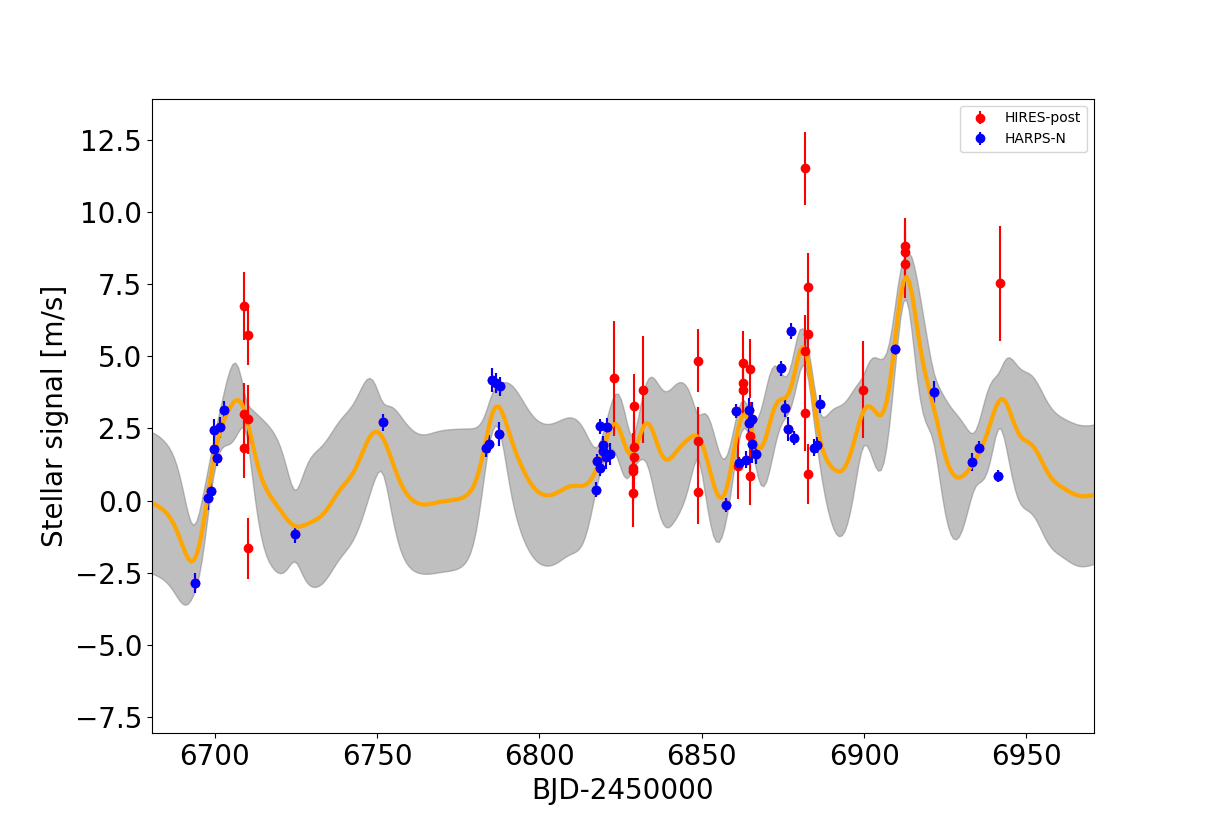}
\caption{\label{fig:GP_activity} Zoom on the GP quasi-periodic modelling of the stellar signal contribution to the RV times series. The orange line represents the best-fit curve, while the gray shaded area is the confidence interval of $1\sigma$. In this part of the sampling, HARPS-N (blue dots) and HIRES$_{\rm post}$ (red dots) data overlap.}
\end{figure}
The RV phased diagram of the newly detected planet companion is shown in Fig.\ref{fig:phased}. This plot shows the crucial contribution of HARPS-N (blue diamonds) to recover the small amplitude $K_{\rm d}$, which is of the same order of the RV uncertainties of the HIRES datasets. Finally, we show in Fig. \ref{fig:GP_activity} a portion of the time series where both HARPS-N (blue dots) and HIRES$_{\rm post}$ (red dots) data are present. Here the stellar contribution to the RVs is fitted by our GP quasi-periodic model (orange line, while the shaded region indicates the 1$\sigma$ confidence interval). 

To verify that the GP regression does not alter the planetary parameters of the newly detected planet, we set up simple simulations. We assume that our determined planetary parameters of the two known planets are accurate and that the planet we discovered is well retrieved. We  considered only the HARPS-N residuals after removing the three Keplerians, with the activity signal left in the data. We then injected the signal of the third planet in the data using our best-fit solution, then creating 3 mock datasets by randomly shifting each data point within the error bars (including the jitter term we found for HARPS-N, added in quadrature). Each dataset is fitted with a GP regression and one Keplerian to see if the GP interferes with the planet signals (with the priors in common same as those used for the analysis of the real data). In the three cases, both the orbital period and the RV semi-amplitude are recovered within the errors and in full agreement with the values reported in Table \ref{Table:postgprv}. This demonstrates that the signal of planet d is retrieved accurately and precisely, with no interference with the GP. 

\begin{table*}  
  \caption[]{Priors and percentiles (16$^{\rm th}$, 50$^{\rm th}$, and 84$^{\rm th}$) of the posterior distributions concerning the analysis of the full RV dataset using a GP quasi-periodic kernel.}
         \label{Table:postgprv}
          \small
         \centering
   \begin{tabular} {c c c c }
            \hline
            \noalign{\smallskip}
            \textbf{Jump parameter} & \textbf{Prior} & \multicolumn{2}{c}{\textbf{Best-fit values}} \\
            \hline
            \noalign{\smallskip}  
            & & N=2 planets & N=3 planets \textbf{(Adopted solution)}  \\
            \hline
            \noalign{\smallskip}  
            $h$ [\ms] & $\mathcal{U}$ (0, 10) & 2.7$\pm$0.2  & 2.5$\pm$0.2    \\
            \noalign{\smallskip}
            $\lambda$ [days] &  $\mathcal{U}$ (0, 2000) & 29$\pm$9 & 23$^{+8}_{-7}$   \\
            \noalign{\smallskip}
            $w$ & $\mathcal{U}$ (0, 1) & 0.20$^{+0.05}_{-0.04}$ & 0.3$\pm$0.1  \\
            \noalign{\smallskip}
            $\theta$ [days] & $\mathcal{U}$ (30, 50) & 42$^{+8}_{-4}$  & 41$^{+7}_{-9}$   \\
            \noalign{\smallskip}
            \hline
            \noalign{\smallskip}
            $K_{\rm b}$ [\ms] & $\mathcal{U}$ (0, 10) & 6.7$\pm$0.3 & 6.7$\pm$0.3   \\
            \noalign{\smallskip}
            $P_{\rm b}$  [days] & $\mathcal{U} (1150, 1250)$ & 1206$\pm$5  & 1207$^{+4}_{-5}$  \\
            \noalign{\smallskip}
            $T_{\rm 0,b}$  [BJD-2\,450\,000] & $\mathcal{U}$ (6700, 8050) & 7965$^{+21}_{-20}$ & 7978$^{+20}_{-26}$  \\
            \noalign{\smallskip}
            $\sqrt e_{\rm b}\cos\omega_{\rm b}$ & $\mathcal{U}$ (-1, 1) & -0.196$^{+0.182}_{-0.112}$ & -0.24$^{+0.15}_{-0.10}$ \\
            \noalign{\smallskip}
            $\sqrt e_{\rm b}\sin\omega_{\rm b}$ & $\mathcal{U}$ (-1, 1) & 0.004$^{+0.137}_{-0.143}$ & 0.10$^{+0.11}_{-0.15}$   \\
            \noalign{\smallskip}
            $e_{\rm b}$ & derived & 0.06$^{+0.05}_{-0.04}$  & 0.08$^{+0.06}_{-0.05}$   \\
            \noalign{\smallskip}
            $\omega_{\rm b}$ [rad] & derived & 0.27$^{+2.56}_{-3.10}$ & 2.03$^{+0.83}_{-4.79}$  \\
            \noalign{\smallskip}
            $m_{\rm b} \sin i_{\rm b}$ [M$_{\oplus}$] & derived &  & 116$^{+10}_{-12}$ \\
            \noalign{\smallskip}
            $a_{\rm b}$ [au] & derived &  & 2.16$\pm$0.03 \\
            \noalign{\smallskip}
            \hline
            \noalign{\smallskip}
            $K_{\rm c}$ [\ms] & $\mathcal{U}$ (0, 5) & 2.2$\pm$0.3 & 2.2$\pm$0.2  \\
            \noalign{\smallskip}
            $P_{\rm c}$  [days] & $\mathcal{U} (71, 81)$ & 75.73$\pm$0.06 & 75.74$^{+0.06}_{-0.05}$  \\
            \noalign{\smallskip}
            $T_{\rm 0,c}$  [BJD-2\,450\,000] & $\mathcal{U}$ (7500,7590) & 7565.2$^{+2.7}_{-2.9}$ & 7564.7$^{+2.7}_{-3.2}$ \\
            \noalign{\smallskip}
            $\sqrt e_{\rm c}\cos\omega_{\rm c}$ & $\mathcal{U}$ (-1, 1) & 0.09$^{+0.23}_{-0.25}$ &  0.13$\pm$0.22     \\
            \noalign{\smallskip}
            $\sqrt e_{\rm c}\sin\omega_{\rm c}$ & $\mathcal{U}$ (-1, 1) & -0.013$^{+0.245}_{-0.247}$ & -0.05$^{+0.18}_{-0.30}$  \\
            \noalign{\smallskip}
            $e_{\rm c}$ & derived & 0.09$^{+0.10}_{-0.07}$ & 0.12$^{+0.11}_{-0.09}$  \\
            \noalign{\smallskip}
            $\omega_{\rm c}$ [rad] & derived & -0.10$^{+1.77}_{-1.66}$ & -0.50$^{+1.89}_{-1.34}$  \\
            \noalign{\smallskip}
            $m_{\rm c} \sin i_{\rm c}$ [M$_{\oplus}$] & derived &  & 13$^{+3}_{-2}$ \\
            \noalign{\smallskip}
            $a_{\rm c}$ [au] & derived &  & 0.341$\pm$0.004 \\
            \noalign{\smallskip}
            \hline
            \noalign{\smallskip}
            $K_{\rm d}$ [\ms] & $\mathcal{U}$ (0,5) & & 1.3$\pm$0.2   \\
            \noalign{\smallskip}
            $P_{\rm d}$  [days] & $\mathcal{U}$ (10,15) & &  12.458$\pm0.003$  \\
            \noalign{\smallskip}
            $T_{\rm 0,d}$  [BJD-2\,450\,000] & $\mathcal{U}$ (7500,7518) & & 7503.9$^{+0.4}_{-0.8}$ \\
            \noalign{\smallskip}
            $\sqrt e_{\rm d}\cos\omega_{\rm d}$ & $\mathcal{U}$ (-1, 1) &  &  0.03$^{+0.32}_{-0.26}$    \\
            \noalign{\smallskip}
            $\sqrt e_{\rm d}\sin\omega_{\rm d}$ & $\mathcal{U}$ (-1, 1) &  &  -0.20$^{+0.27}_{-0.18}$  \\
            \noalign{\smallskip}
            $e_{\rm d}$ & derived &  &  0.12$^{+0.13}_{-0.08}$    \\
            \noalign{\smallskip}
            $\omega_{\rm d}$ [rad] & derived &  & -0.83$^{+2.29}_{-1.34}$   \\
            \noalign{\smallskip}
            $m_{\rm d} \sin i_{\rm d}$ [M$_{\oplus}$] & derived &  & 4$\pm$1 \\
            \noalign{\smallskip}
            $a_{\rm d}$ [au] & derived &  & 0.103$\pm0.001$ \\    
            \noalign{\smallskip}
            \hline
            \noalign{\smallskip}
            $\dot{\gamma}$ [\ms d$^{-1}$] & $\mathcal{U}$ (-1, 1) & 0.0006$\pm0.0002$ & 0.0006$^{+0.0003}_{-0.0002}$ \\ 
            \noalign{\smallskip}
            \hline
            \noalign{\smallskip}
            Uncorrelated jitter & &\\
            \noalign{\smallskip}
            $\sigma_{\rm jit, Keck-pre}$ [\ms] & $\mathcal{U}$ (0, 5) & 1.1$^{+0.5}_{-0.6}$ & 1.3$\pm$0.6 \\ 
            \noalign{\smallskip}
            $\sigma_{\rm jit, Keck-post}$ [\ms] & $\mathcal{U}$ (0, 5) & 1.6$\pm$0.1 & 1.7$\pm$0.1  \\ 
            \noalign{\smallskip}
            $\sigma_{\rm jit, HARPS-N}$ [\ms] & $\mathcal{U}$ (0, 5) & 1.3$^{+0.5}_{-0.6}$ & 1.3$^{+0.4}_{-0.5}$ \\ 
            \noalign{\smallskip}
            $\sigma_{\rm jit, APF}$ [\ms] & $\mathcal{U}$ (0, 5) & 1.23$\pm$0.09  & 1.3$\pm$0.1 \\ 
            \noalign{\smallskip}
            \hline
            \noalign{\smallskip}
            RV offset & &\\
            \noalign{\smallskip}
            $\gamma_{\rm Keck-pre}$ [\ms] & $\mathcal{U}$ (-5, 5) & 0.4$^{+0.8}_{-0.7}$ & 0.6$^{+0.6}_{-0.8}$  \\ 
            \noalign{\smallskip}
            $\gamma_{\rm Keck-post}$ [\ms] & $\mathcal{U}$ (-5, 5) & 0.1$\pm$0.5 &  0.2$\pm$0.4 \\ 
            \noalign{\smallskip}
            $\gamma_{\rm HARPS-N}$ [\ms] & $\mathcal{U}$ (20'000, 20'700) & 20'363.2$\pm$0.7  & 20'363.4$^{+0.5}_{-0.7}$ \\
            \noalign{\smallskip}
            $\gamma_{\rm APF}$ [\ms] & $\mathcal{U}$ (-10, 10) & 2.1$^{+0.8}_{-0.9}$ & 2.1$\pm$0.6  \\
            \hline
            \noalign{\smallskip}
            Bayesian evidence ln$\mathcal{Z}$ & & -1634.2$\pm0.1$  & -1626.8$\pm$0.1 \\ 
            \noalign{\smallskip}
            $\Delta$ ln$\mathcal{Z}$ with respect to N=2 pl. model & &  & +7.4 \\
            \hline
     \end{tabular} 
\end{table*}

\subsection{Additional tests}
Because the 42-days periodicity persisted through the HIRES/HARPS-N observation time span, we performed a test to confirm that this signature is actually related to stellar activity. Specifically, we analysed the RV residuals of the three-planets fit with the kernel regression (KR) technique, recently applied by \cite{2018A&A...616A.155L} on a sample of GAPS targets characterised by slow rotation and weak stellar activity. \cite{2018A&A...616A.155L} show that the KR can model the dependence of the RVs as a function of time and the activity indices (e.g., CCF asymmetry indicators and the $\log$R$^{\rm \prime}_{\rm HK}$), allowing to evaluate the impact of stellar activity on the observed RV time series.

\begin{table} 
  \caption[]{Summary of the results of the kernel regression (KR) technique on the HARPS-N RV residuals of the three-planets fit as function of time and of activity indicators. The index showing a better correlation is reported for each observing season together with the RV dispersion before ($\sigma$) and after ($\sigma_{\rm KR res}$) the correction.}
         \label{tab:KR}
          \centering
   \begin{tabular}{c c c c c}
            \hline
            \noalign{\smallskip}
            Season     &  BJD - 2'450'000 &  Best index & $\sigma$ & $\sigma_{\rm KR res}$\\
             & & & [m s$^{-1}$] & [m s$^{-1}$] \\
            \noalign{\smallskip}
            \hline
            \noalign{\smallskip}
            1 & < 6660 & BIS & 2.49 & 1.08 \\
                        \noalign{\smallskip}
            2 & 6660 - 7000 & V$\rm _{asy(mod)}$ & 1.46 & 0.75 \\
            \noalign{\smallskip}
            3 & 7000 - 7350 & BIS & 2.09 & 0.89 \\
            \noalign{\smallskip}
            4 & 7350 - 7750 & $\log$R$^{\rm \prime}_{\rm HK}$ & 2.39 & 1.0 \\
            \noalign{\smallskip}
            5 & 7750 - 8100 & BIS & 3.27 & 1.29 \\
            \noalign{\smallskip}
            6 & > 8100 & $\log$R$^{\rm \prime}_{\rm HK}$ & 1.41 & 0.45 \\
            \noalign{\smallskip}
            \hline
     \end{tabular}  
\end{table}

\begin{figure}[htbp]
\centering
\includegraphics[width=6cm,angle=270]{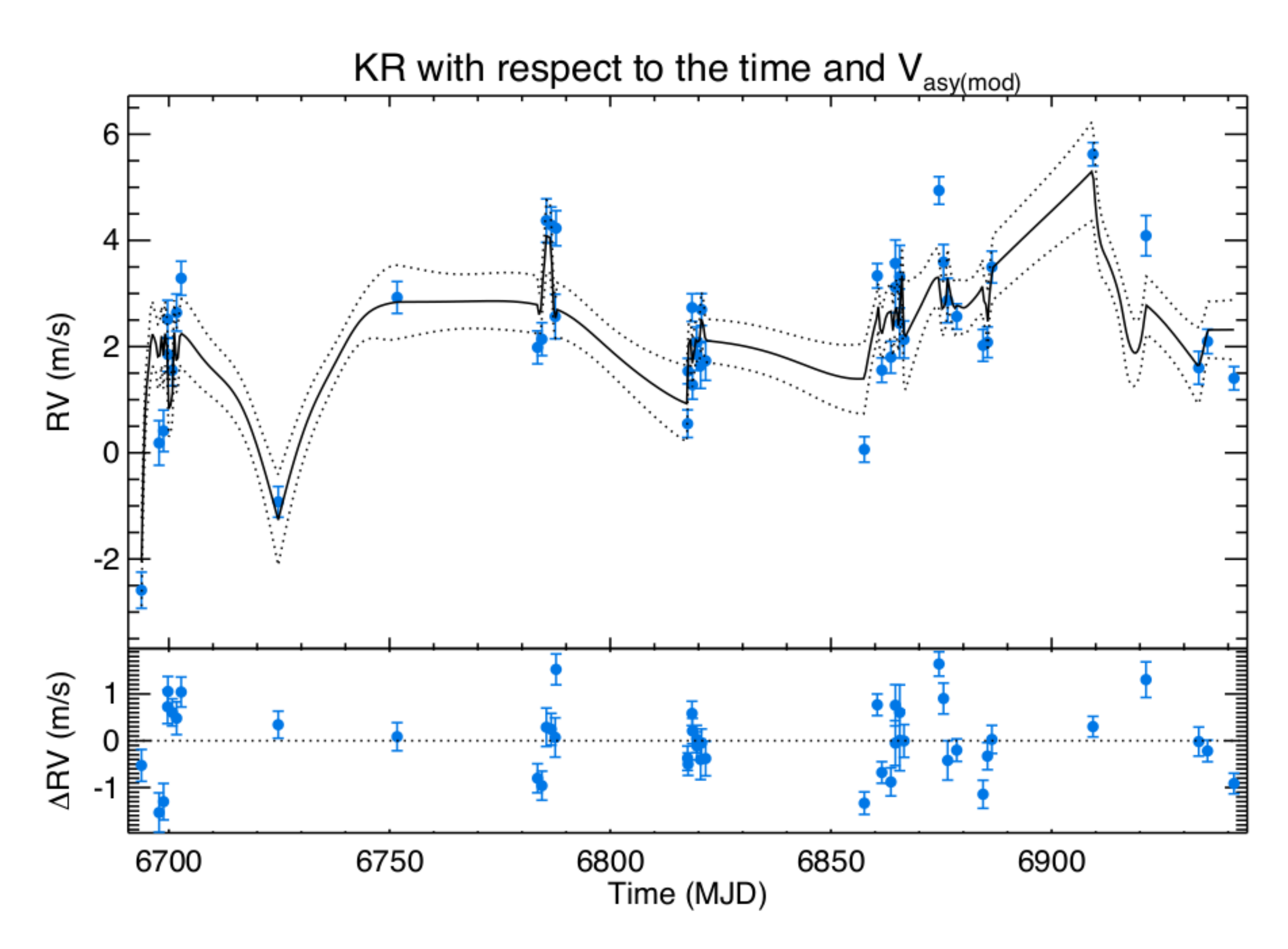}
\caption{\label{fig:KR} \textit{Top panel}: HARPS-N RV residuals of the three-planet fit for HD\,164922 (second season, same timespan as in Fig. \ref{fig:GP_activity}) as modelled through the KR with respect to the time and the V$\rm _{asy(mod)}$; \textit{Bottom panel}: Corresponding residuals.}
\end{figure}

We apply this method to the HARPS-N dataset with the following activity indicators: BIS, $\Delta V$, V$\rm _{asy(mod)}$, CCF Full Width at Half Maximum (FWHM), CCF contrast and $\log$R$^{\rm \prime}_{\rm HK}$.
We subdivided the dataset into the six observing seasons, to take into account possible variation of the activity behaviour over the years, which can be better described by different indices. 
With the KR method a locally linear model is used to fit the RVs for a well defined time $t_k$. In this way, KR takes into account the temporal variation of the correlation between the RVs and the activity indices, which is expected to be non-linear \citep{2007A&A...473..983D}. The significance level of the KR modelling for each indicator is estimated through the Fisher-Snedecor statistics, allowing to identify which indicator better represents the properties of the stellar activity for each observing season. 
According to Table \ref{tab:KR}, the best regression is obtained with CCF indicators, while the chromospheric index is preferable only in the fourth and (in the less sampled) sixth season. If we remove the contribution of the activity from the RVs, by using the best model obtained by the KR (see, e.g., the modelling relative to the second season in Fig. \ref{fig:KR}), we observe a remarkable reduction of the RV dispersion (from 2.73 to 0.72 ms$^{-1}$ for the whole time series, while the decreasing of the dispersion for each season is reported in Table \ref{tab:KR}) as well as of the power of all the periodicities identified in Fig. \ref{fig:res_3p}. 
The result of the KR modelling provides an independent confirmation that the RV signal observed at $\sim$ 42 days is ascribable to the stellar activity and validates the solution of the GP analysis of the $\log$R$^{\rm \prime}_{\rm HK}$ index presented in Sect. \ref{sec:gpact}.

We also employed the Stacked Bayesian General Lomb-Scargle periodogram (\texttt{SBGLS}), proposed by \cite{2017A&A...601A.110M} to discriminate stellar activity from genuine planetary signals of the \texttt{GLS}.
Since the activity signals are supposed to be unstable, showing phase incoherence, this method allows to evaluate possible phase/period/amplitude variations of the signal of interest in the classical \texttt{GLS} periodogram calculated as a function of time, i.e., with the increasing number of data.
\begin{figure}
\centering
\includegraphics[width=9.5cm,angle=0]{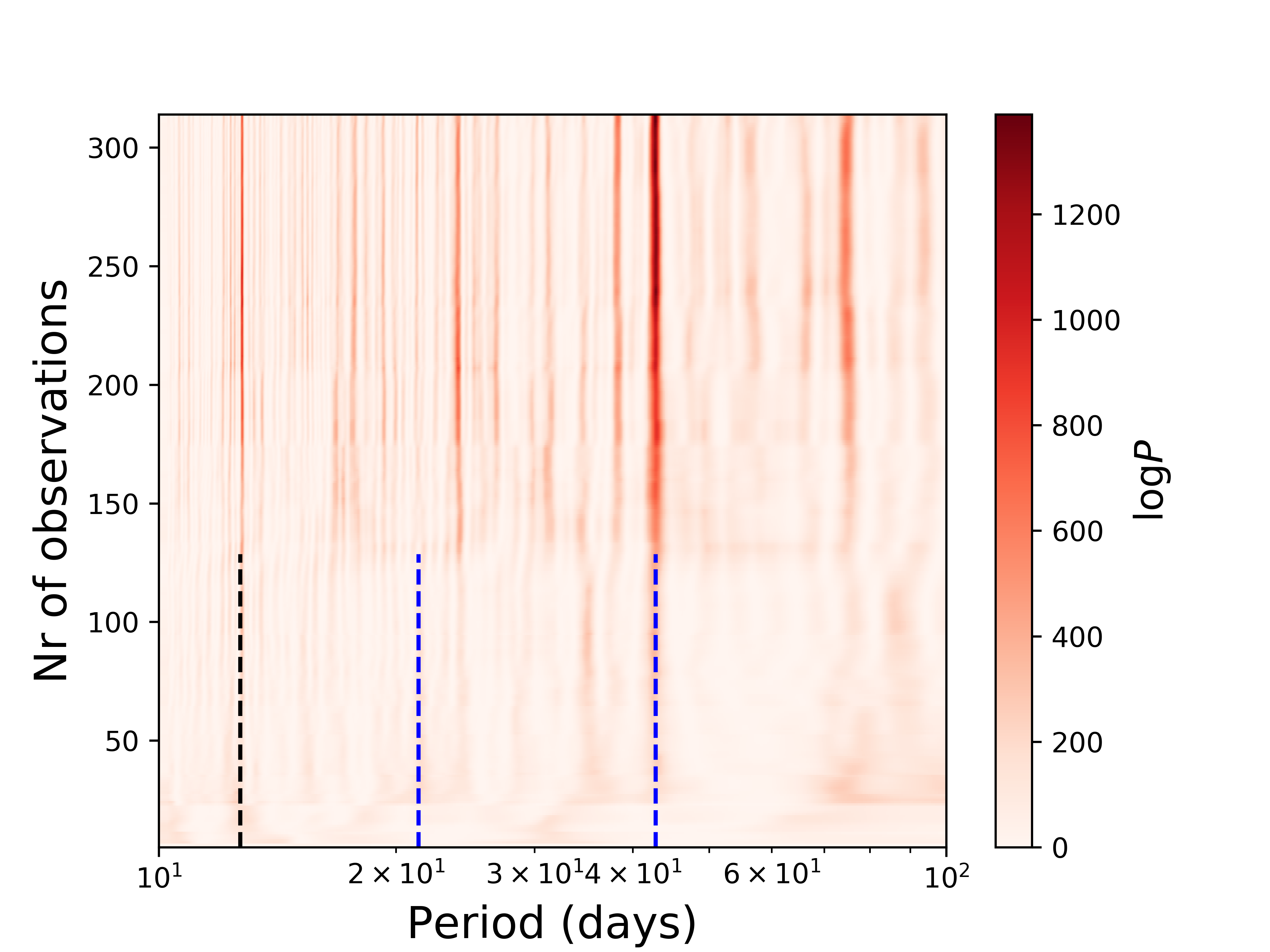}
\caption{Stacked Bayesian General Lomb-Scargle periodogram applied to the HARPS-N residuals of the two-planets fit. The colour bar indicates the Bayesian proability of the signals. The rotation period with its first harmonics (blue dashed lines) and the orbital period of planet d (black dashed line) are also indicated.} 
\label{fig:gls_time}%
\end{figure}
Fig. \ref{fig:gls_time} shows the \texttt{SBGLS} for the HARPS-N time series of HD\,164922 after the removal of the two-planets model, where periodicities are plotted with respect to the increasing number of observations, while the third coordinate represents the Bayesian proability of a specific signal. After gathering $\sim 100$ data points, both the signal of planet d and the rotation period become clearly distinguishable and are seen to strengthen with the increasing number of observations. After 150 data, also the first harmonics of the rotation period can be seen. While the 42 days signal is highly significant but not clearly peaked (i.e., the distribution of the power is spread around 42 days, indicated with a blue dashed line together with its first harmonics), the signature of planet d is strictly pointed at 12.46 days marked with a black dashed line, showing a high Bayesian probability.

\section{System architecture} \label{sec:architecture}
In this section we evaluate the stability of the HD\,164922 system and put detection thresholds on additional planets. With this information and the evaluation of the water snow line, we try to infer which formation and migration scenarios could have shaped the current orbital architecture.

\subsection{Orbital stability} \label{sec:stability}
\begin{figure*}
\centering
\includegraphics[width=6cm,angle=0]{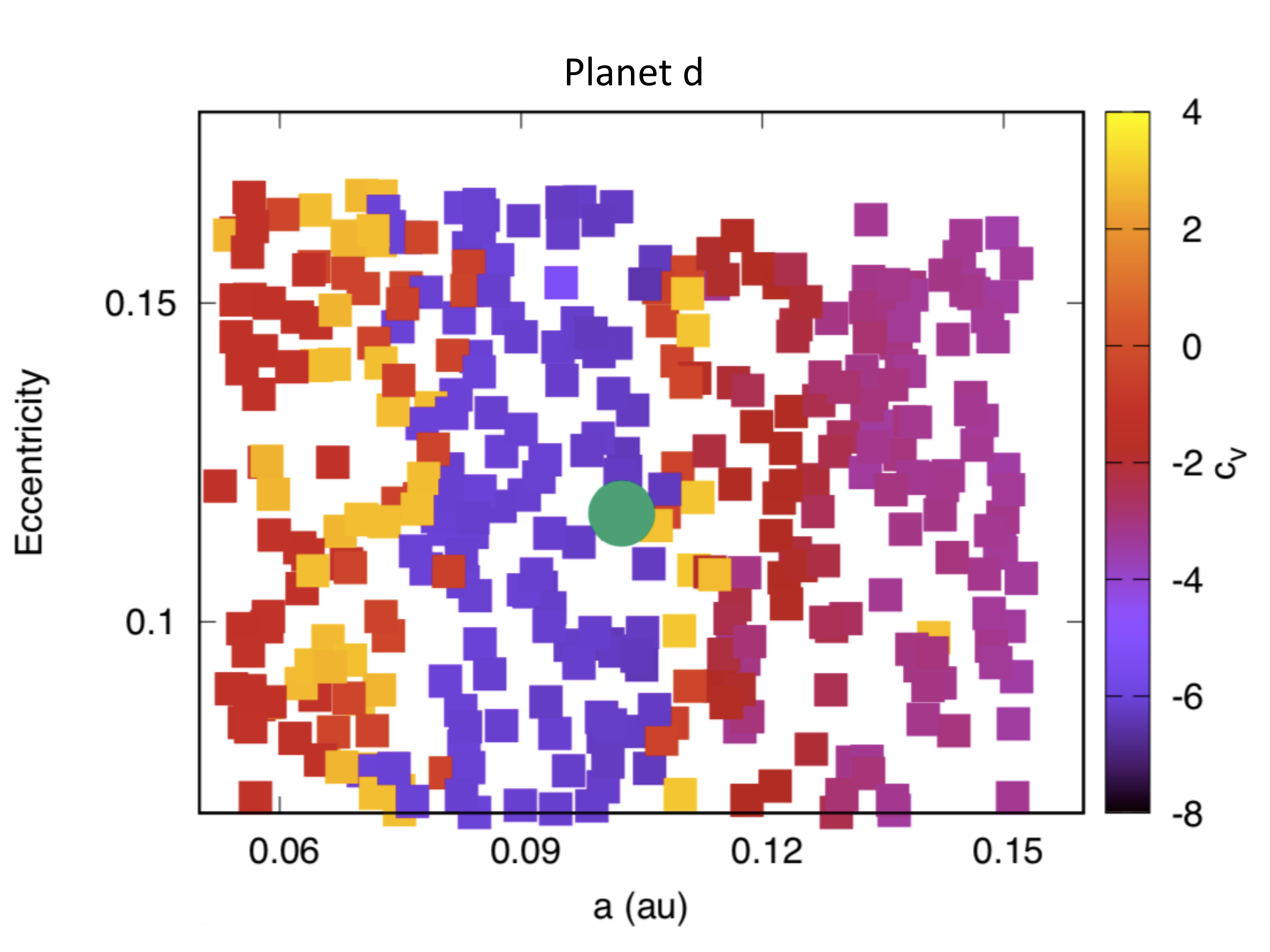}
\includegraphics[width=6cm,angle=0]{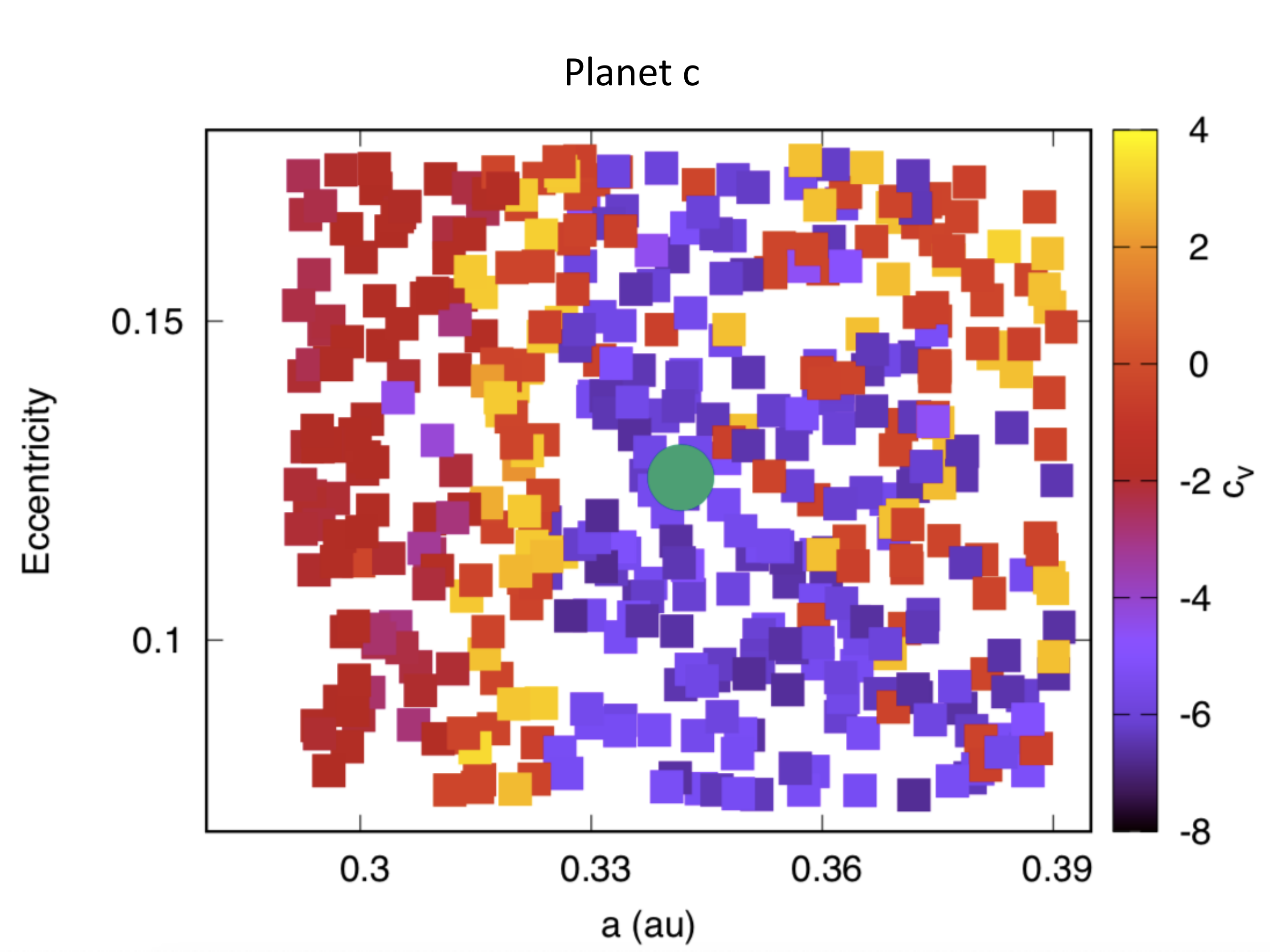}
\includegraphics[width=6cm,angle=0]{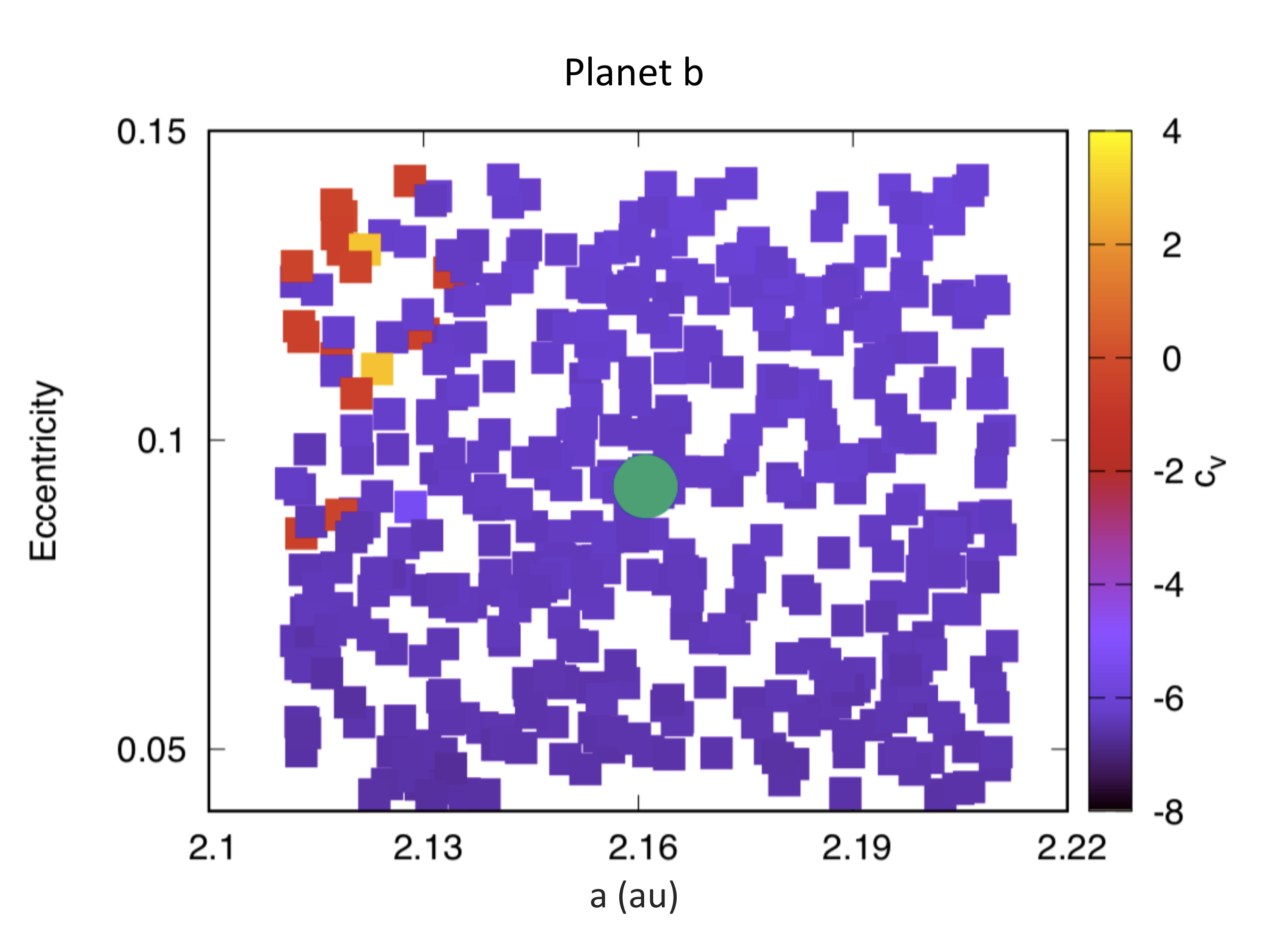}
\caption{Stability of the three planetary companions of the HD\,164922 system as a function of the initial eccentricity and semi-major axis. Colour scale indicates the value of the orbital diffusion speed ($c_V$) ranging from black, representing stable regions, to yellow, which indicates a higher degree of instability for planets with given separation and eccentricity. Nominal positions of the planets are identified with a green circle. From left to right, the planets are represented as a function of the distance from the host star.}
\label{fig:stability}%
\end{figure*}

The long term stability of the HD\,164922 system has been tested as in \cite{2017A&A...599A..90B}, starting from the stellar and planetary parameters in Tables \ref{param} and \ref{Table:postgprv} through the exploration of the phase space
around the nominal positions (semi-major axis vs eccentricity) of the three planet companions. We exploited the Frequency Map Analysis (FMA; \citealt{lask93, sine96, marz03}) tool to evaluate a possible chaotic behaviour of the orbits with short term numerical integrations by analysing the variation of the secular frequencies of the system. We measured the diffusion speed in the phase space, identified by the parameter $c_V$, as the dispersion of the main secular frequency of the system over running windows covering the integration timespan. The results are reported in Figure \ref{fig:stability} for each planet, starting, left to right, from the super-Earth close to the star, then the Neptune-mass planet and finally the Saturn-mass planet in the outer region. 
In each of the three models, two of the planets are kept on fixed orbits, the nominal ones, while the orbital parameters of the third planet are randomly changed within a wide interval of values to explore its range of stability in semi-major axis and eccentricity.
In each plot, we show the nominal position of the planet and the
stability properties of the region around it. The two inner low-mass planets lie within narrow stable zones while the stability region for the outer planet is significantly wider. It means that its orbit can be varied
on a broader range without affecting the overall dynamical stability of the system. 
With the same method we investigated the possible stable zones for an additional fourth unseen planet. We set the three known planets on their nominal orbits and randomly select the initial orbital elements of the fourth potential planet adopting for it a mass equal to 1 M$_{\oplus}$.
We focused on the regions between planet d and c (0.1 - 0.3 au) and between planet c and b (0.4 - 1.5 au).
We did not consider orbits for the fourth planet with semi-major axis smaller than 0.1 au due to the excessive computing load related to the short orbital period.

We found a narrow stable region ranging between 0.18 and 0.21 au, corresponding to an interval of orbital periods from $\sim$30 and $\sim$36 days, therefore, a putative planet with a period of $\sim$42 days should not be permitted, further strengthening our interpretation of the RV signal at this period as the rotation period of the star. A second allowable region is located between 0.6 and 1.4 au (300-500 days), including the Habitable Zone (HZ) that we evaluated to be 0.95-1.68 au for the conservative HZ, and 0.75-1.76 au in the optimistic case. The calculation was performed by using the public tool\footnote{\url{http://depts.washington.edu/naivpl/sites/default/files/hz.shtml}} based on \cite{2013ApJ...765..131K,2014ApJ...787L..29K}, assuming a planet of 1 M$_{\oplus}$ and adopting $T_{\rm eff}$ and stellar luminosity from Table \ref{param}. 

As mentioned in Sect. \ref{subsec:3pl}, the \texttt{GLS} of the HARPS-N RVs reveals the presence of a periodicity of $\sim 357$ days, that we analysed on the light of this latter result. Rather than being due to a planet, this signal has more probably the same origin of the spurious 1-yr signal first reported and characterised by \cite{2015ApJ...808..171D} by using HARPS data. For HARPS-N data this feature is also suspected in the \texttt{GLS} periodograms presented in Barbato et al. 2020 (submitted). As in Fig. 3 from \cite{2015ApJ...808..171D}, we verified that the RV signal is in opposition of phase with the Earth orbital motion around the Sun, showing a correlation with the Barycentric Earth Radial Velocity (BERV). In the case of HARPS, CCD imperfections lead to a deformation of specific spectral lines passing by the crossing block stitching of the detector. The long observation baseline of HD\,164922 with HARPS-N allows to detect this spurious signal since we sample a large variation of the BERV ($\pm 20$ km s$^{-1}$). However, the treatment and/or the removal of this signal is beyond the scope of the present paper.

\subsection{Detectability limits}
\begin{figure}
\begin{center}
\centering
\includegraphics[width=9.0cm]{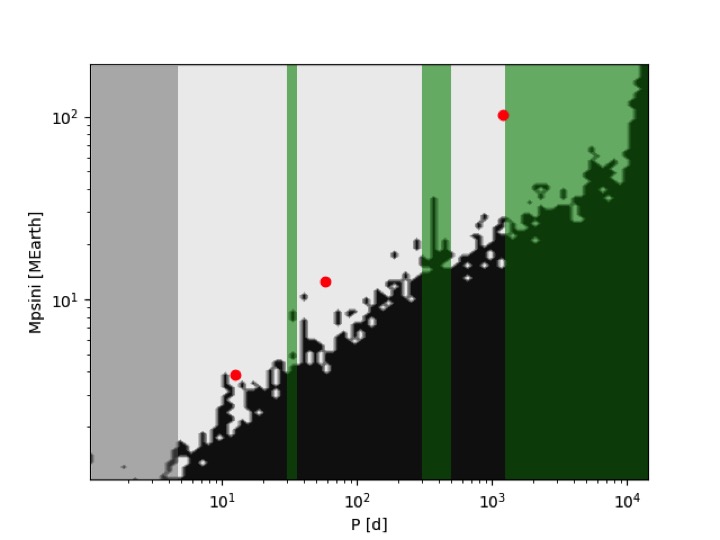}
\caption{Detection map for the HARPS-N RV residuals of the three-planet fit for the HD\,164922 system. The white part corresponds to the area in the period - minimum mass space where additional signals could be detected if present in the data, while the black region corresponds to the area where the detection probability is negligible. The red circles mark the position in the parameter space of the three planets orbiting HD\,164922. The green areas denote the stability regions for additional companions, while the gray area indicates the part of the system not explored by our orbital simulations. }
\label{fig:det_lim}
\end{center} 
\end{figure}

We computed the detection threshold for the HARPS-N dataset following the Bayesian approach from \cite{2014MNRAS.441.1545T}. We applied this technique on the RV residuals after subtracting the signals of the three planets and including in the model the GP to take into account the stellar activity contribution. The resulting detectability map is shown in Fig. \ref{fig:det_lim}. Planets b, c and d (red circles) are located above the detection threshold, as expected. HD\,164922 d, is just above the minimum mass detection threshold, which in the interval 12 < $P_{\rm orb}$ < 13 days corresponds to m$_p \sin i$ = 2.3 M$_{\oplus}$. We can also compute the minimum mass threshold in the orbitally-stable regions identified in Sect. \ref{sec:stability}: we obtain a minimum mass limit of 5.0 M$_{\oplus}$ for the 30-36 days region, 18 M$_{\oplus}$ for the 300-500 days region, and a few tens of M$_{\oplus}$ for the P > 1200 d region.
In the figure we indicated with green areas the allowed regions for additional companions in the system, as defined in Sect. \ref{sec:stability}, while the inner part of the system that we did not explored is represented in gray.

\subsection{Snow line position of the HD\,164922 system}
\label{sec:snowline}   
In order to investigate the planetary migration scenario of the HD\,164922 system, we estimated the position of the water snow line, namely, the distance from the central star where the midplane temperature of the proto-planetary disk drops down to the sublimation temperature of the water ice \citep{hayashi1981}, at the classical value of 170\ K.
Out of the snow line the surface density of the disk increases by a factor of about three due to the presence of water ice, leading to an increase of the isolation mass (i.e. the mass a body can accrete solely through planetesimal accretion; \citealt{lissauer1987, lissauerandstevenson2007}) and of the size of protoplanetary cores. For this reason, the region around the snow line provides a favourable location for forming planets \citep{stevensonetal1988, morbidellietal2015}.
In order to evaluate the snow line position for the HD\,164922 system, we considered a simple Minimum Mass Solar Nebula (MMSN) disk model as in \cite{kennedyandkenyon2008}, where the surface density of the disk, $\Sigma$, changes with time and the distance from the central star ($a$) as:
\begin{equation} 
\Sigma (a,t)= \Sigma_0 \eta f_{ice} \frac{M_\star}{M_\odot} a_{au}^{-3/2},
\end{equation}
where $\Sigma_0=100$\ kg\ m$^{-2}$, $f_{ice}$ represents the increasing of $\Sigma$ due to ice condensation outside the snow line, and $M_\star$ is the stellar mass. The factor $\eta$ provides a range of disk masses for a fixed stellar mass (``relative disk mass'', \citealt{kennedyandkenyon2008}) and is varied to account for the observed range of disk masses at fixed stellar mass. Observations (e.g., \citealt{nattaetal2000}) stated that this parameter ranges between from 0.5 up to 5 for disk masses between 0.01 and 0.1\ M$_\star$.
In the case of HD\,164922 we considered $\eta =4$. The midplane temperature of accreting protoplanetary disk irradiated by the central star is given by:
\begin{equation}
    T^4_{\rm mid} =T^4_{\rm mid,acc}+T^4_{\rm irr}, 
\end{equation}
where $T_{\rm mid,acc}$ is the temperature profile of the midplane due to the mass accretion and the corresponding viscous heating of the disk, and $T_{\rm irr}$ is the temperature profile of the disk due to the irradiation heating \citep{sasselovandlecar2000, lecaretal2006, kennedyandkenyon2008}. For $T_{\rm irr}$ we consider the scale law obtained by \cite{sasselovandlecar2000}\footnote{$T_{\rm irr}(a)=T_0 a^{-3/7}$, with $T_0\propto M_\star^{3/(10-k)}$ (e.g., for M$_\star=0.5$ M$_\odot$, $T_0$ is equal to 140 K), where the opacity $k$ is $0\leq k< 2$.}, while for the contribution due to the accretion we consider the standard relation reported in Eq. 1 in \cite{belletal1997} and the dust opacity according to \cite{hubeny1990}\footnote{$T^4_{\rm mid,acc} \sim \frac{3}{8} \tau \ T^4_{\rm eff,acc}$, where $T_{\rm eff,acc}$ is the effective temperature of a given ring of a disk, defined as $T^4_{\rm eff,acc}= \frac{3}{8 \pi}\frac{GM_\star \dot{M}}{\sigma a^3} \left(1-\sqrt{\frac{R_\star}{a}}\right)$, where $\dot{M}$ is the accretion rate, $R_\star$ is the stellar radius, and $\sigma$ is the Stefan's constant.}.
For simplicity, we assume that the dust opacities are constant throughout the disk.
To evaluate $T_{\rm mid,acc}$, we obtained the stellar parameters of HD\,164922 in the pre main sequence (PMS) phase at about 1 Myr of age, exploiting the PARSEC evolutionary tracks. We adopted the set of models with a metallicity of [Fe/H]$=0.18$ (from Table \ref{param}), corresponding to X= 0.73, Y=0.25 and Z=0.02 \citep{asplundetal2004}. From this set of models we selected the parameters for masses of 0.90, 0.95 and 1.00 M$_\odot$ at the age of 1 Myr. Stellar parameters were obtained interpolating the values for masses, temperatures and radii of the three chosen models and they resulted to be R$_{\rm 1Myr}=2.31$\ R$_\odot$, T$_{\rm eff, 1Myr}=4443.3$\ K. For the accretion rate, we considered the typical value for the MMSN of $10^{-8}$\ M$_\odot$ yr$^{-1}$.

\begin{figure}
\begin{center}
\centering
\includegraphics[width=9.0cm]{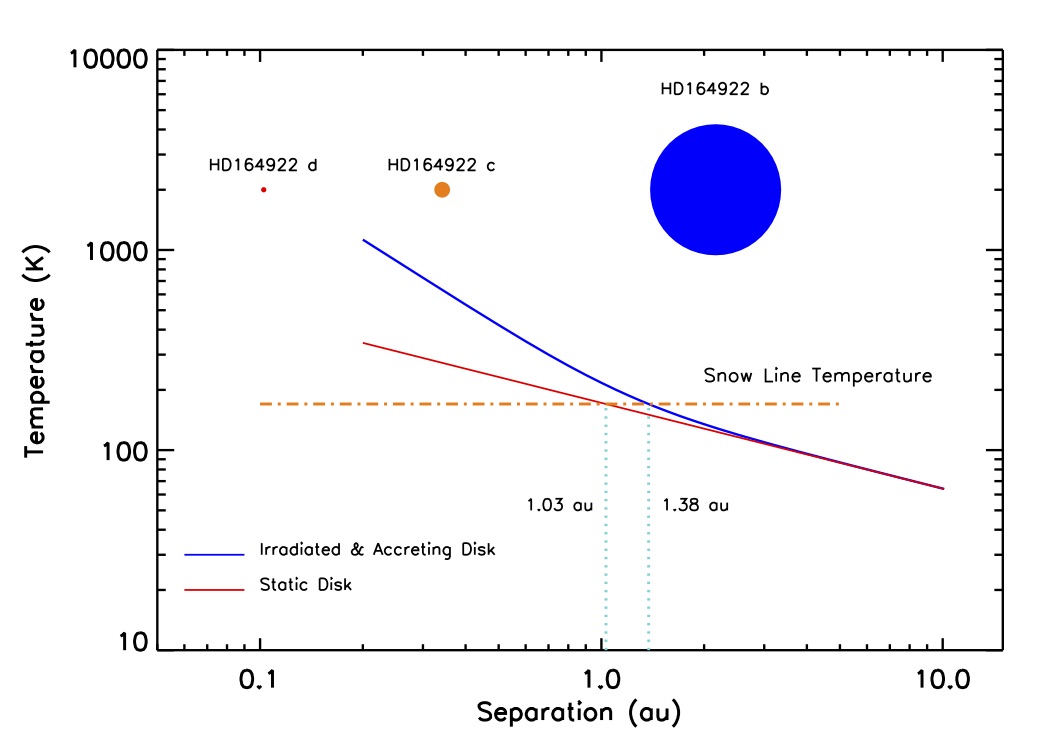}
\caption{Temperature profile for the midplane of a static disk (solid red line) and an accreting and irradiated protoplanetary disk (solid blue line). The horizontal dashed line shows the position of the snow line considering a value of 170\ K for the sublimation temperature of water ice. The edges of the corresponding snow lines are indicated with dotted vertical lines and their distances. The separation of the planets in the system is also shown as a reference. The size of the symbols is proportional to the planetary masses.}
\label{f:tprofdisk}
\end{center} 
\end{figure}

Both the passive and the accreting disk temperature profiles are shown in Figure \ref{f:tprofdisk}, together with the location of the snow line and the position of the three planets of the HD\,164922 system. The snow line position is found to be at 1.03 au for the passive disk and at 1.38 au for the accreting and irradiated disk, thus located about halfway between planet c and b. 

\subsection{Discussion}

According to our GP regression, the eccentricities of the three planetary companions of HD\,164922 are all consistent with zero (Table \ref{Table:postgprv}) and the RV analysis over a time span of 22 years excludes the presence of outer companions with a mass larger than $\sim 30$ M$\oplus$ up to 5 au (with a period of about 4000 days, cf Fig. \ref{fig:det_lim}), a result corroborated by the negligible value of the $\Delta \mu_{\alpha,\delta}$.
The lack of a massive perturbing body and the low planet eccentricities indicate that a migration through the protoplanetary disk could have acted in shaping the present system architecture.
Planet b formed beyond the water snow line. Due to its relatively low mass it is supposed to have quickly migrated closer to the star until the disk dissipation. 
The formation of the two low-mass planets occurred within the snow-line, we can consider two possible scenarios: the in-situ formation that requires a large amount of solids in the protoplanetary disk within 1 au \citep{2012ApJ...751..158H} and the inside-out model \citep{2014ApJ...780...53C}, that invokes pebbles formation and their consequent drifting toward the inner region of the disk, stimulating then the planet formation process. 

According to Eq. 3 in \cite{2012ApJ...751..158H}, the minimum mass of HD\,164922 d allows in principle to accrete gas from the disk, but its orbital distance also enables the photoevaporation of the same gas (cf Fig. 15 in \citealt{2012ApJ...751..158H}). No high-volatile atmosphere is then expected for this planet.
By applying the tidal model of \cite{Leconteetal10} with a tidal time lag $\Delta t_{\rm p} = (3/2) (k_{2} Q^{\prime}_{\rm d} n)^{-1}$, where $Q^{\prime}_{\rm d}$ is the modified tidal quality factor, a parameterization of the tidal response of the planet's interior,
we find a heat flux across the surface of planet d of 0.8 W~m$^{-2}$ for a radius of 1.3~R$_{\odot}$ and of 5 W~m$^{-2}$ for a radius of 2.4~R$_{\odot}$. However, the value of $Q_{\rm d}^{\prime}$ appropriate for a planet with a radius of 2.4~R$_{\odot}$ could be much larger leading to a proportional reduction of the heat flux. For example, adopting $Q^{\prime}_{\rm d} = 1.4 \times 10^{5}$, that is estimated to be appropriate for Uranus \citep{2014ARA&A..52..171O}, we obtain a flux of only $0.05$~W~m$^{-2}$. In the case of a rocky planet with  $Q^{\prime}_{\rm d}$ similar to that of the Earth, planet d may host a significant volcanic activity powered by  tidal dissipation in its interior. For comparison, in the case of the volcanic Jupiter moon Io, the heat flux is $2.0-2.6$~W~m$^{-2}$ \citep[see][]{Rathbunetal04}. The timescale for the tidal decay of the eccentricity of the orbit of planet d is longer than $\approx 6$~Gyr even for $Q^{\prime}_{\rm d} =1400$ and $R_{\rm d} = 2.4$~R$_{\odot}$, indicating that its possible eccentricity could be primordial, that is, not necessarily excited by the other planets in the system. The original eccentricity of planet d could have been higher, although not high enough to cross the other orbits or destabilize the system. However, the unknown radius and internal structure of planet d prevent us from drawing sound conclusions on its evolution.

Our interpretation and comprehension of the planetary
system architecture, formation and evolution rely on the accuracy of the inferred stellar parameters. Unfortunately,
the fundamental parameters can be better constrained only by an appropriate asteroseismic campaign. At present we can only predict that the oscillation spectrum of this star
should be characterized by approximately a large frequency separation of
$\Delta \nu=140.6,/ \mu$Hz and a  frequency of maximum power of
 $\nu_{max}\simeq 3306,/ \mu$Hz obtained
by the scaling relation calibrated on solar values given by
\cite{1995A&A...293...87K}.
 The values are interesting when compared to the values of the Sun
$\Delta \nu=135,/ \mu$Hz and a  frequency of maximum power of
 $\nu_{max}= 3050,/ \mu$Hz.
The hope is in the next future, to have the possibility to compare our findings with asteroseismic data collected for this star.

\section{Conclusions}
\label{sec:conc}
One of the largest high-precision RV datasets available in the literature allowed us to claim the presence of an additional hot super-Earth with minimum mass $m_d \sin i_d$=4$\pm$1 M$_{\oplus}$ in the HD\,164922 planetary system, known to host a sub-Neptune-mass planet and a distant Saturn-mass planet. We monitored this target in the framework of the GAPS programme focused on finding close-in low-mass companions in systems with outer giant planets. 
Thanks to the dedicated monitoring and the high precision RV measurements ensured by HARPS-N, we took full advantage of the GP regression analysis. With such a technique we could robustly model the stellar activity as a quasi-periodic signal both from the $\log$R$^{\rm \prime}_{\rm HK}$ index and the RV time series. The subtraction of this contribution to the RVs allowed us to recover the low-amplitude Keplerian signal of the additional companion at a 6$\sigma$ level ($K_d=1.3\pm0.2 $ m s$^{-1}$).
The application of the Kernel Regression technique to the residuals of the three-planets fit confirmed that the signal content can be properly described as a function of time and the activity indices.

The analysis of the system architecture allowed us to infer that the most probable planet migration process is through the gas of the protoplanetary disk. 
The challenging detection of HD\,164922 d is an example of the observational effort invoked by \cite{2018AJ....156...92Z} to find low mass planets in systems with cold gaseous planets with state-of-the-art instrumentation, because of the very high probability that those systems could host low-mass planets in their interiors. Unfortunately, this target will not be observed with the NASA Transiting Exoplanets Survey Satellite (TESS, \citealt{2015JATIS...1a4003R}) satellite, at least in Cycle 2, to verify if it transits. Dedicated observations with the ESA CHarachterizing ExOPlanet Satellite (CHEOPS, \citealt{2013EPJWC..4703005B}) could be proposed, but they can be severely affected by the uncertainty on the transit time. For instance, at the beginning of 2021 it could be of the order of 0.9 days by using our ephemeris in Table \ref{Table:postgprv}. A former attempt to search for transits of planets b and c are reported in F2016 with APT photometry (about 1000 data spread over 10 years), with no evidence of such events or photometric variability between 1 and 100 days. However, their data cadence seems to be not suitable to detect a potential transit of planet d.

\begin{acknowledgements}
The authors acknowledge the anonymous Referee for her/his useful comments and suggestions. The authors acknowledge the support from INAF through the "WOW Premiale" funding scheme of the Italian Ministry of Education, University, and Research. MD acknowledges financial support from Progetto Premiale 2015 FRONTIERA (OB.FU. 1.05.06.11) funding scheme of the Italian Ministry of Education, University, and Research. We acknowledge financial support from the ASI-INAF agreement n.2018-16-HH.0. We acknowledge the computing centres of INAF - Osservatorio Astronomico di Trieste / Osservatorio Astrofisico di Catania, under the coordination of the CHIPP project, for the availability of computing resources and support. MB acknowledges support from the STFC research grant ST/000631/1.
This work has made use of data from the European Space Agency (ESA) mission
{\it Gaia} (\url{https://www.cosmos.esa.int/gaia}), processed by the {\it Gaia}
Data Processing and Analysis Consortium (DPAC,
\url{https://www.cosmos.esa.int/web/gaia/dpac/consortium}). Funding for the DPAC
has been provided by national institutions, in particular the institutions
participating in the {\it Gaia} Multilateral Agreement. 
\end{acknowledgements}


\begin{thebibliography}{93}
\expandafter\ifx\csname natexlab\endcsname\relax\def\natexlab#1{#1}\fi

\bibitem[{{Affer} {et~al.}(2016){Affer}, {Micela}, {Damasso}, {Perger},
  {Ribas}, {Su{\'a}rez Mascare{\~n}o}, {Gonz{\'a}lez Hern{\'a}ndez}, {Rebolo},
  {Poretti}, {Maldonado}, {Leto}, {Pagano}, {Scandariato}, {Zanmar Sanchez},
  {Sozzetti}, {Bonomo}, {Malavolta}, {Morales}, {Rosich}, {Bignamini},
  {Gratton}, {Velasco}, {Cenadelli}, {Claudi}, {Cosentino}, {Desidera},
  {Giacobbe}, {Herrero}, {Lafarga}, {Lanza}, {Molinari}, \&
  {Piotto}}]{2016A&A...593A.117A}
{Affer}, L., {Micela}, G., {Damasso}, M., {et~al.} 2016, \aap, 593, A117

\bibitem[{{Agnew} {et~al.}(2018){Agnew}, {Maddison}, \&
  {Horner}}]{2018MNRAS.481.4680A}
{Agnew}, M.~T., {Maddison}, S.~T., \& {Horner}, J. 2018, \mnras, 481, 4680

\bibitem[{{Agnew} {et~al.}(2017){Agnew}, {Maddison}, {Thilliez}, \&
  {Horner}}]{2017MNRAS.471.4494A}
{Agnew}, M.~T., {Maddison}, S.~T., {Thilliez}, E., \& {Horner}, J. 2017,
  \mnras, 471, 4494

\bibitem[{{Ambikasaran} {et~al.}(2015){Ambikasaran}, {Foreman-Mackey},
  {Greengard}, {Hogg}, \& {O'Neil}}]{2015ITPAM..38..252A}
{Ambikasaran}, S., {Foreman-Mackey}, D., {Greengard}, L., {Hogg}, D.~W., \&
  {O'Neil}, M. 2015, IEEE Transactions on Pattern Analysis and Machine
  Intelligence, 38 [\eprint[arXiv]{1403.6015}]

\bibitem[{{Armitage}(2003)}]{2003ApJ...582L..47A}
{Armitage}, P.~J. 2003, \apjl, 582, L47

\bibitem[{{Asplund} {et~al.}(2004){Asplund}, {Grevesse}, {Sauval}, {Allende
  Prieto}, \& {Kiselman}}]{asplundetal2004}
{Asplund}, M., {Grevesse}, N., {Sauval}, A.~J., {Allende Prieto}, C., \&
  {Kiselman}, D. 2004, \aap, 417, 751

\bibitem[{{Bailer-Jones} {et~al.}(2018){Bailer-Jones}, {Rybizki}, {Fouesneau},
  {Mantelet}, \& {Andrae}}]{2018AJ....156...58B}
{Bailer-Jones}, C.~A.~L., {Rybizki}, J., {Fouesneau}, M., {Mantelet}, G., \&
  {Andrae}, R. 2018, \aj, 156, 58

\bibitem[{{Barbato} {et~al.}(2018){Barbato}, {Sozzetti}, {Desidera}, {Damasso},
  {Bonomo}, {Giacobbe}, {Colombo}, {Lazzoni}, {Claudi}, {Gratton}, {LoCurto},
  {Marzari}, \& {Mordasini}}]{2018A&A...615A.175B}
{Barbato}, D., {Sozzetti}, A., {Desidera}, S., {et~al.} 2018, \aap, 615, A175

\bibitem[{{Batalha} {et~al.}(2013){Batalha}, {Rowe}, {Bryson}, {Barclay},
  {Burke}, {Caldwell}, {Christiansen}, {Mullally}, {Thompson}, {Brown},
  {Dupree}, {Fabrycky}, {Ford}, {Fortney}, {Gilliland}, {Isaacson}, {Latham},
  {Marcy}, {Quinn}, {Ragozzine}, {Shporer}, {Borucki}, {Ciardi}, {Gautier},
  {Haas}, {Jenkins}, {Koch}, {Lissauer}, {Rapin}, {Basri}, {Boss}, {Buchhave},
  {Carter}, {Charbonneau}, {Christensen-Dalsgaard}, {Clarke}, {Cochran},
  {Demory}, {Desert}, {Devore}, {Doyle}, {Esquerdo}, {Everett}, {Fressin},
  {Geary}, {Girouard}, {Gould}, {Hall}, {Holman}, {Howard}, {Howell},
  {Ibrahim}, {Kinemuchi}, {Kjeldsen}, {Klaus}, {Li}, {Lucas}, {Meibom},
  {Morris}, {Pr{\v{s}}a}, {Quintana}, {Sanderfer}, {Sasselov}, {Seader},
  {Smith}, {Steffen}, {Still}, {Stumpe}, {Tarter}, {Tenenbaum}, {Torres},
  {Twicken}, {Uddin}, {Van Cleve}, {Walkowicz}, \&
  {Welsh}}]{2013ApJS..204...24B}
{Batalha}, N.~M., {Rowe}, J.~F., {Bryson}, S.~T., {et~al.} 2013, \apjs, 204, 24

\bibitem[{{Bell} {et~al.}(1997){Bell}, {Cassen}, {Klahr}, \&
  {Henning}}]{belletal1997}
{Bell}, K.~R., {Cassen}, P.~M., {Klahr}, H.~H., \& {Henning}, T. 1997, \apj,
  486, 372

\bibitem[{{Benatti} {et~al.}(2016){Benatti}, {Claudi}, {Desidera}, {Gratton},
  {Lanza}, {Micela}, {Pagano}, {Piotto}, {Sozzetti}, {Boccato}, {Cosentino},
  {Covino}, {Maggio}, {Molinari}, {Poretti}, {Smareglia}, \& {GAPS
  Team}}]{2016frap.confE..69B}
{Benatti}, S., {Claudi}, R., {Desidera}, S., {et~al.} 2016, in Frontier
  Research in Astrophysics II (FRAPWS2016), 69

\bibitem[{{Benatti} {et~al.}(2017){Benatti}, {Desidera}, {Damasso},
  {Malavolta}, {Lanza}, {Biazzo}, {Bonomo}, {Claudi}, {Marzari}, {Poretti},
  {Gratton}, {Micela}, {Pagano}, {Piotto}, {Sozzetti}, {Boccato}, {Cosentino},
  {Covino}, {Maggio}, {Molinari}, {Smareglia}, {Affer}, {Andreuzzi},
  {Bignamini}, {Borsa}, {di Fabrizio}, {Esposito}, {Martinez Fiorenzano},
  {Messina}, {Giacobbe}, {Harutyunyan}, {Knapic}, {Maldonado}, {Masiero},
  {Nascimbeni}, {Pedani}, {Rainer}, {Scandariato}, \&
  {Silvotti}}]{2017A&A...599A..90B}
{Benatti}, S., {Desidera}, S., {Damasso}, M., {et~al.} 2017, \aap, 599, A90

\bibitem[{{Biazzo} {et~al.}(2012){Biazzo}, {D'Orazi}, {Desidera}, {Covino},
  {Alcal{\'a}}, \& {Zusi}}]{2012MNRAS.427.2905B}
{Biazzo}, K., {D'Orazi}, V., {Desidera}, S., {et~al.} 2012, \mnras, 427, 2905

\bibitem[{{Biazzo} {et~al.}(2011){Biazzo}, {Randich}, \&
  {Palla}}]{2011A&A...525A..35B}
{Biazzo}, K., {Randich}, S., \& {Palla}, F. 2011, \aap, 525, A35

\bibitem[{{Brandt}(2018)}]{2018ApJS..239...31B}
{Brandt}, T.~D. 2018, \apjs, 239, 31

\bibitem[{{Bressan} {et~al.}(2012){Bressan}, {Marigo}, {Girardi}, {Salasnich},
  {Dal Cero}, {Rubele}, \& {Nanni}}]{2012MNRAS.427..127B}
{Bressan}, A., {Marigo}, P., {Girardi}, L., {et~al.} 2012, \mnras, 427, 127

\bibitem[{{Brewer} {et~al.}(2016){Brewer}, {Fischer}, {Valenti}, \&
  {Piskunov}}]{2016ApJS..225...32B}
{Brewer}, J.~M., {Fischer}, D.~A., {Valenti}, J.~A., \& {Piskunov}, N. 2016,
  \apjs, 225, 32

\bibitem[{{Broeg} {et~al.}(2013){Broeg}, {Fortier}, {Ehrenreich}, {Alibert},
  {Baumjohann}, {Benz}, {Deleuil}, {Gillon}, {Ivanov}, {Liseau}, {Meyer},
  {Oloffson}, {Pagano}, {Piotto}, {Pollacco}, {Queloz}, {Ragazzoni}, {Renotte},
  {Steller}, \& {Thomas}}]{2013EPJWC..4703005B}
{Broeg}, C., {Fortier}, A., {Ehrenreich}, D., {et~al.} 2013, in European
  Physical Journal Web of Conferences, Vol.~47, European Physical Journal Web
  of Conferences, 03005

\bibitem[{{Bryan} {et~al.}(2019){Bryan}, {Knutson}, {Lee}, {Fulton}, {Batygin},
  {Ngo}, \& {Meshkat}}]{2019AJ....157...52B}
{Bryan}, M.~L., {Knutson}, H.~A., {Lee}, E.~J., {et~al.} 2019, \aj, 157, 52

\bibitem[{{Buchner} {et~al.}(2014){Buchner}, {Georgakakis}, {Nandra}, {Hsu},
  {Rangel}, {Brightman}, {Merloni}, {Salvato}, {Donley}, \&
  {Kocevski}}]{buchner14}
{Buchner}, J., {Georgakakis}, A., {Nandra}, K., {et~al.} 2014, \aap, 564, A125

\bibitem[{{Butler} {et~al.}(1996){Butler}, {Marcy}, {Williams}, {McCarthy},
  {Dosanjh}, \& {Vogt}}]{1996PASP..108..500B}
{Butler}, R.~P., {Marcy}, G.~W., {Williams}, E., {et~al.} 1996, \pasp, 108, 500

\bibitem[{{Butler} {et~al.}(2017){Butler}, {Vogt}, {Laughlin}, {Burt},
  {Rivera}, {Tuomi}, {Teske}, {Arriagada}, {Diaz}, {Holden}, \&
  {Keiser}}]{2017AJ....153..208B}
{Butler}, R.~P., {Vogt}, S.~S., {Laughlin}, G., {et~al.} 2017, \aj, 153, 208

\bibitem[{{Butler} {et~al.}(2006){Butler}, {Wright}, {Marcy}, {Fischer},
  {Vogt}, {Tinney}, {Jones}, {Carter}, {Johnson}, {McCarthy}, \&
  {Penny}}]{2006ApJ...646..505B}
{Butler}, R.~P., {Wright}, J.~T., {Marcy}, G.~W., {et~al.} 2006, \apj, 646, 505

\bibitem[{{Buzzoni} {et~al.}(2010){Buzzoni}, {Patelli}, {Bellazzini}, {Pecci},
  \& {Oliva}}]{2010MNRAS.403.1592B}
{Buzzoni}, A., {Patelli}, L., {Bellazzini}, M., {Pecci}, F.~F., \& {Oliva}, E.
  2010, \mnras, 403, 1592

\bibitem[{{Chatterjee} \& {Tan}(2014)}]{2014ApJ...780...53C}
{Chatterjee}, S. \& {Tan}, J.~C. 2014, \apj, 780, 53

\bibitem[{{Collier Cameron} {et~al.}(2019){Collier Cameron}, {Mortier},
  {Phillips}, {Dumusque}, {Haywood}, {Langellier}, {Watson}, {Cegla}, {Costes},
  {Charbonneau}, {Coffinet}, {Latham}, {Lopez-Morales}, {Malavolta},
  {Maldonado}, {Micela}, {Milbourne}, {Molinari}, {Saar}, {Thompson},
  {Buchschacher}, {Cecconi}, {Cosentino}, {Ghedina}, {Glenday}, {Gonzalez},
  {Li}, {Lodi}, {Lovis}, {Pepe}, {Poretti}, {Rice}, {Sasselov}, {Sozzetti},
  {Szentgyorgyi}, {Udry}, \& {Walsworth}}]{2019MNRAS.487.1082C}
{Collier Cameron}, A., {Mortier}, A., {Phillips}, D., {et~al.} 2019, \mnras,
  487, 1082

\bibitem[{{Cosentino} {et~al.}(2014){Cosentino}, {Lovis}, {Pepe}, {Cameron},
  {Latham}, {Molinari}, {Udry}, {Bezawada}, {Buchschacher}, {Figueira},
  {Fleury}, {Ghedina}, {Glenday}, {Gonzalez}, {Guerra}, {Henry}, {Hughes},
  {Maire}, {Motalebi}, \& {Phillips}}]{2014SPIE.9147E..8CC}
{Cosentino}, R., {Lovis}, C., {Pepe}, F., {et~al.} 2014, in \procspie, Vol.
  9147, Ground-based and Airborne Instrumentation for Astronomy V, 91478C

\bibitem[{{Covino} {et~al.}(2013){Covino}, {Esposito}, {Barbieri}, {Mancini},
  {Nascimbeni}, {Claudi}, {Desidera}, {Gratton}, {Lanza}, {Sozzetti}, {Biazzo},
  {Affer}, {Gandolfi}, {Munari}, {Pagano}, {Bonomo}, {Collier Cameron},
  {H{\'e}brard}, {Maggio}, {Messina}, {Micela}, {Molinari}, {Pepe}, {Piotto},
  {Ribas}, {Santos}, {Southworth}, {Shkolnik}, {Triaud}, {Bedin}, {Benatti},
  {Boccato}, {Bonavita}, {Borsa}, {Borsato}, {Brown}, {Carolo}, {Ciceri},
  {Cosentino}, {Damasso}, {Faedi}, {Mart{\'{\i}}nez Fiorenzano}, {Latham},
  {Lovis}, {Mordasini}, {Nikolov}, {Poretti}, {Rainer}, {Rebolo L{\'o}pez},
  {Scandariato}, {Silvotti}, {Smareglia}, {Alcal{\'a}}, {Cunial}, {Di
  Fabrizio}, {Di Mauro}, {Giacobbe}, {Granata}, {Harutyunyan}, {Knapic},
  {Lattanzi}, {Leto}, {Lodato}, {Malavolta}, {Marzari}, {Molinaro},
  {Nardiello}, {Pedani}, {Prisinzano}, \& {Turrini}}]{2013A&A...554A..28C}
{Covino}, E., {Esposito}, M., {Barbieri}, M., {et~al.} 2013, \aap, 554, A28

\bibitem[{{da Silva} {et~al.}(2006){da Silva}, {Girardi}, {Pasquini},
  {Setiawan}, {von der L{\"u}he}, {de Medeiros}, {Hatzes}, {D{\"o}llinger}, \&
  {Weiss}}]{2006A&A...458..609D}
{da Silva}, L., {Girardi}, L., {Pasquini}, L., {et~al.} 2006, \aap, 458, 609

\bibitem[{{Desidera} {et~al.}(2011){Desidera}, {Carolo}, {Gratton}, {Martinez
  Fiorenzano}, {Endl}, {Mesa}, {Barbieri}, {Bonavita}, {Cecconi}, {Claudi},
  {Cosentino}, {Marzari}, \& {Scuderi}}]{2011A&A...533A..90D}
{Desidera}, S., {Carolo}, E., {Gratton}, R., {et~al.} 2011, \aap, 533, A90

\bibitem[{{Desort} {et~al.}(2007){Desort}, {Lagrange}, {Galland}, {Udry}, \&
  {Mayor}}]{2007A&A...473..983D}
{Desort}, M., {Lagrange}, A.-M., {Galland}, F., {Udry}, S., \& {Mayor}, M.
  2007, \aap, 473, 983

\bibitem[{{Doyle} {et~al.}(2014){Doyle}, {Davies}, {Smalley}, {Chaplin}, \&
  {Elsworth}}]{2014MNRAS.444.3592D}
{Doyle}, A.~P., {Davies}, G.~R., {Smalley}, B., {Chaplin}, W.~J., \&
  {Elsworth}, Y. 2014, \mnras, 444, 3592

\bibitem[{{Dravins} {et~al.}(1999){Dravins}, {Lindegren}, \&
  {Madsen}}]{1999A&A...348.1040D}
{Dravins}, D., {Lindegren}, L., \& {Madsen}, S. 1999, \aap, 348, 1040

\bibitem[{{Dumusque}(2016)}]{2016A&A...593A...5D}
{Dumusque}, X. 2016, \aap, 593, A5

\bibitem[{{Dumusque} {et~al.}(2015){Dumusque}, {Pepe}, {Lovis}, \&
  {Latham}}]{2015ApJ...808..171D}
{Dumusque}, X., {Pepe}, F., {Lovis}, C., \& {Latham}, D.~W. 2015, \apj, 808,
  171

\bibitem[{{Dumusque} {et~al.}(2011{\natexlab{a}}){Dumusque}, {Santos}, {Udry},
  {Lovis}, \& {Bonfils}}]{2011A&A...527A..82D}
{Dumusque}, X., {Santos}, N.~C., {Udry}, S., {Lovis}, C., \& {Bonfils}, X.
  2011{\natexlab{a}}, \aap, 527, A82

\bibitem[{{Dumusque} {et~al.}(2011{\natexlab{b}}){Dumusque}, {Udry}, {Lovis},
  {Santos}, \& {Monteiro}}]{2011A&A...525A.140D}
{Dumusque}, X., {Udry}, S., {Lovis}, C., {Santos}, N.~C., \& {Monteiro},
  M.~J.~P.~F.~G. 2011{\natexlab{b}}, \aap, 525, A140

\bibitem[{{Egeland} {et~al.}(2017){Egeland}, {Soon}, {Baliunas}, {Hall},
  {Pevtsov}, \& {Bertello}}]{2017ApJ...835...25E}
{Egeland}, R., {Soon}, W., {Baliunas}, S., {et~al.} 2017, \apj, 835, 25

\bibitem[{{Feroz} {et~al.}(2013){Feroz}, {Hobson}, {Cameron}, \&
  {Pettitt}}]{feroz13}
{Feroz}, F., {Hobson}, M.~P., {Cameron}, E., \& {Pettitt}, A.~N. 2013, ArXiv
  e-prints [\eprint[arXiv]{1306.2144}]

\bibitem[{{Figueira} {et~al.}(2013){Figueira}, {Santos}, {Pepe}, {Lovis}, \&
  {Nardetto}}]{2013A&A...557A..93F}
{Figueira}, P., {Santos}, N.~C., {Pepe}, F., {Lovis}, C., \& {Nardetto}, N.
  2013, \aap, 557, A93

\bibitem[{{Fulton} {et~al.}(2016){Fulton}, {Howard}, {Weiss}, {Sinukoff},
  {Petigura}, {Isaacson}, {Hirsch}, {Marcy}, {Henry}, {Grunblatt}, {Huber},
  {von Braun}, {Boyajian}, {Kane}, {Wittrock}, {Horch}, {Ciardi}, {Howell},
  {Wright}, \& {Ford}}]{2016ApJ...830...46F}
{Fulton}, B.~J., {Howard}, A.~W., {Weiss}, L.~M., {et~al.} 2016, \apj, 830, 46

\bibitem[{{Gaia Collaboration} {et~al.}(2018){Gaia Collaboration}, {Brown},
  {Vallenari}, {Prusti}, {de Bruijne}, {Babusiaux}, {Bailer-Jones}, {Biermann},
  {Evans}, {Eyer}, {Jansen}, {Jordi}, {Klioner}, {Lammers}, {Lindegren},
  {Luri}, {Mignard}, {Panem}, {Pourbaix}, {Randich}, {Sartoretti}, {Siddiqui},
  {Soubiran}, {van Leeuwen}, {Walton}, {Arenou}, {Bastian}, {Cropper},
  {Drimmel}, {Katz}, {Lattanzi}, {Bakker}, {Cacciari}, {Casta{\~n}eda},
  {Chaoul}, {Cheek}, {De Angeli}, {Fabricius}, {Guerra}, {Holl}, {Masana},
  {Messineo}, {Mowlavi}, {Nienartowicz}, {Panuzzo}, {Portell}, {Riello},
  {Seabroke}, {Tanga}, {Th{\'e}venin}, {Gracia-Abril}, {Comoretto},
  {Garcia-Reinaldos}, {Teyssier}, {Altmann}, {Andrae}, {Audard},
  {Bellas-Velidis}, {Benson}, {Berthier}, {Blomme}, {Burgess}, {Busso},
  {Carry}, {Cellino}, {Clementini}, {Clotet}, {Creevey}, {Davidson}, {De
  Ridder}, {Delchambre}, {Dell'Oro}, {Ducourant},
  {Fern{\'a}ndez-Hern{\'a}ndez}, {Fouesneau}, {Fr{\'e}mat}, {Galluccio},
  {Garc{\'\i}a-Torres}, {Gonz{\'a}lez-N{\'u}{\~n}ez}, {Gonz{\'a}lez-Vidal},
  {Gosset}, {Guy}, {Halbwachs}, {Hambly}, {Harrison}, {Hern{\'a}ndez},
  {Hestroffer}, {Hodgkin}, {Hutton}, {Jasniewicz}, {Jean-Antoine-Piccolo},
  {Jordan}, {Korn}, {Krone-Martins}, {Lanzafame}, {Lebzelter}, {L{\"o}ffler},
  {Manteiga}, {Marrese}, {Mart{\'\i}n-Fleitas}, {Moitinho}, {Mora}, {Muinonen},
  {Osinde}, {Pancino}, {Pauwels}, {Petit}, {Recio-Blanco}, {Richards},
  {Rimoldini}, {Robin}, {Sarro}, {Siopis}, {Smith}, {Sozzetti}, {S{\"u}veges},
  {Torra}, {van Reeven}, {Abbas}, {Abreu Aramburu}, {Accart}, {Aerts},
  {Altavilla}, {{\'A}lvarez}, {Alvarez}, {Alves}, {Anderson}, {Andrei},
  {Anglada Varela}, {Antiche}, {Antoja}, {Arcay}, {Astraatmadja}, {Bach},
  {Baker}, {Balaguer-N{\'u}{\~n}ez}, {Balm}, {Barache}, {Barata}, {Barbato},
  {Barblan}, {Barklem}, {Barrado}, {Barros}, {Barstow}, {Bartholom{\'e}
  Mu{\~n}oz}, {Bassilana}, {Becciani}, {Bellazzini}, {Berihuete}, {Bertone},
  {Bianchi}, {Bienaym{\'e}}, {Blanco-Cuaresma}, {Boch}, {Boeche}, {Bombrun},
  {Borrachero}, {Bossini}, {Bouquillon}, {Bourda}, {Bragaglia}, {Bramante},
  {Breddels}, {Bressan}, {Brouillet}, {Br{\"u}semeister}, {Brugaletta},
  {Bucciarelli}, {Burlacu}, {Busonero}, {Butkevich}, {Buzzi}, {Caffau},
  {Cancelliere}, {Cannizzaro}, {Cantat-Gaudin}, {Carballo}, {Carlucci},
  {Carrasco}, {Casamiquela}, {Castellani}, {Castro-Ginard}, {Charlot},
  {Chemin}, {Chiavassa}, {Cocozza}, {Costigan}, {Cowell}, {Crifo}, {Crosta},
  {Crowley}, {Cuypers}, {Dafonte}, {Damerdji}, {Dapergolas}, {David}, {David},
  {de Laverny}, {De Luise}, {De March}, {de Martino}, {de Souza}, {de Torres},
  {Debosscher}, {del Pozo}, {Delbo}, {Delgado}, {Delgado}, {Di Matteo},
  {Diakite}, {Diener}, {Distefano}, {Dolding}, {Drazinos}, {Dur{\'a}n},
  {Edvardsson}, {Enke}, {Eriksson}, {Esquej}, {Eynard Bontemps}, {Fabre},
  {Fabrizio}, {Faigler}, {Falc{\~a}o}, {Farr{\`a}s Casas}, {Federici},
  {Fedorets}, {Fernique}, {Figueras}, {Filippi}, {Findeisen}, {Fonti},
  {Fraile}, {Fraser}, {Fr{\'e}zouls}, {Gai}, {Galleti}, {Garabato},
  {Garc{\'\i}a-Sedano}, {Garofalo}, {Garralda}, {Gavel}, {Gavras}, {Gerssen},
  {Geyer}, {Giacobbe}, {Gilmore}, {Girona}, {Giuffrida}, {Glass}, {Gomes},
  {Granvik}, {Gueguen}, {Guerrier}, {Guiraud}, {Guti{\'e}rrez-S{\'a}nchez},
  {Haigron}, {Hatzidimitriou}, {Hauser}, {Haywood}, {Heiter}, {Helmi}, {Heu},
  {Hilger}, {Hobbs}, {Hofmann}, {Holland}, {Huckle}, {Hypki}, {Icardi},
  {Jan{\ss}en}, {Jevardat de Fombelle}, {Jonker}, {Juh{\'a}sz}, {Julbe},
  {Karampelas}, {Kewley}, {Klar}, {Kochoska}, {Kohley}, {Kolenberg},
  {Kontizas}, {Kontizas}, {Koposov}, {Kordopatis}, {Kostrzewa-Rutkowska},
  {Koubsky}, {Lambert}, {Lanza}, {Lasne}, {Lavigne}, {Le Fustec}, {Le
  Poncin-Lafitte}, {Lebreton}, {Leccia}, {Leclerc}, {Lecoeur-Taibi},
  {Lenhardt}, {Leroux}, {Liao}, {Licata}, {Lindstr{\o}m}, {Lister}, {Livanou},
  {Lobel}, {L{\'o}pez}, {Managau}, {Mann}, {Mantelet}, {Marchal}, {Marchant},
  {Marconi}, {Marinoni}, {Marschalk{\'o}}, {Marshall}, {Martino}, {Marton},
  {Mary}, {Massari}, {Matijevi{\v{c}}}, {Mazeh}, {McMillan}, {Messina},
  {Michalik}, {Millar}, {Molina}, {Molinaro}, {Moln{\'a}r}, {Montegriffo},
  {Mor}, {Morbidelli}, {Morel}, {Morris}, {Mulone}, {Muraveva}, {Musella},
  {Nelemans}, {Nicastro}, {Noval}, {O'Mullane}, {Ord{\'e}novic},
  {Ord{\'o}{\~n}ez-Blanco}, {Osborne}, {Pagani}, {Pagano}, {Pailler},
  {Palacin}, {Palaversa}, {Panahi}, {Pawlak}, {Piersimoni}, {Pineau}, {Plachy},
  {Plum}, {Poggio}, {Poujoulet}, {Pr{\v{s}}a}, {Pulone}, {Racero}, {Ragaini},
  {Rambaux}, {Ramos-Lerate}, {Regibo}, {Reyl{\'e}}, {Riclet}, {Ripepi}, {Riva},
  {Rivard}, {Rixon}, {Roegiers}, {Roelens}, {Romero-G{\'o}mez}, {Rowell},
  {Royer}, {Ruiz-Dern}, {Sadowski}, {Sagrist{\`a} Sell{\'e}s}, {Sahlmann},
  {Salgado}, {Salguero}, {Sanna}, {Santana-Ros}, {Sarasso}, {Savietto},
  {Schultheis}, {Sciacca}, {Segol}, {Segovia}, {S{\'e}gransan}, {Shih},
  {Siltala}, {Silva}, {Smart}, {Smith}, {Solano}, {Solitro}, {Sordo}, {Soria
  Nieto}, {Souchay}, {Spagna}, {Spoto}, {Stampa}, {Steele},
  {Steidelm{\"u}ller}, {Stephenson}, {Stoev}, {Suess}, {Surdej}, {Szabados},
  {Szegedi-Elek}, {Tapiador}, {Taris}, {Tauran}, {Taylor}, {Teixeira},
  {Terrett}, {Teyssand ier}, {Thuillot}, {Titarenko}, {Torra Clotet}, {Turon},
  {Ulla}, {Utrilla}, {Uzzi}, {Vaillant}, {Valentini}, {Valette}, {van Elteren},
  {Van Hemelryck}, {van Leeuwen}, {Vaschetto}, {Vecchiato}, {Veljanoski},
  {Viala}, {Vicente}, {Vogt}, {von Essen}, {Voss}, {Votruba}, {Voutsinas},
  {Walmsley}, {Weiler}, {Wertz}, {Wevers}, {Wyrzykowski}, {Yoldas},
  {{\v{Z}}erjal}, {Ziaeepour}, {Zorec}, {Zschocke}, {Zucker}, {Zurbach}, \&
  {Zwitter}}]{2018A&A...616A...1G}
{Gaia Collaboration}, {Brown}, A.~G.~A., {Vallenari}, A., {et~al.} 2018, \aap,
  616, A1

\bibitem[{{Gomes da Silva} {et~al.}(2011){Gomes da Silva}, {Santos}, {Bonfils},
  {Delfosse}, {Forveille}, \& {Udry}}]{2011A&A...534A..30G}
{Gomes da Silva}, J., {Santos}, N.~C., {Bonfils}, X., {et~al.} 2011, \aap, 534,
  A30

\bibitem[{{Hansen} \& {Murray}(2012)}]{2012ApJ...751..158H}
{Hansen}, B. M.~S. \& {Murray}, N. 2012, \apj, 751, 158

\bibitem[{{Hatzes} {et~al.}(2000){Hatzes}, {Cochran}, {McArthur}, {Baliunas},
  {Walker}, {Campbell}, {Irwin}, {Yang}, {K{\"u}rster}, {Endl}, {Els},
  {Butler}, \& {Marcy}}]{2000ApJ...544L.145H}
{Hatzes}, A.~P., {Cochran}, W.~D., {McArthur}, B., {et~al.} 2000, \apjl, 544,
  L145

\bibitem[{{Hayashi}(1981)}]{hayashi1981}
{Hayashi}, C. 1981, Progress of Theoretical Physics Supplement, 70, 35

\bibitem[{{Haywood} {et~al.}(2014){Haywood}, {Collier Cameron}, {Queloz},
  {Barros}, {Deleuil}, {Fares}, {Gillon}, {Lanza}, {Lovis}, {Moutou}, {Pepe},
  {Pollacco}, {Santerne}, {S{\'e}gransan}, \& {Unruh}}]{2014MNRAS.443.2517H}
{Haywood}, R.~D., {Collier Cameron}, A., {Queloz}, D., {et~al.} 2014, \mnras,
  443, 2517

\bibitem[{{Howard} {et~al.}(2010){Howard}, {Johnson}, {Marcy}, {Fischer},
  {Wright}, {Bernat}, {Henry}, {Peek}, {Isaacson}, {Apps}, {Endl}, {Cochran},
  {Valenti}, {Anderson}, \& {Piskunov}}]{2010ApJ...721.1467H}
{Howard}, A.~W., {Johnson}, J.~A., {Marcy}, G.~W., {et~al.} 2010, \apj, 721,
  1467

\bibitem[{{Hubeny}(1990)}]{hubeny1990}
{Hubeny}, I. 1990, \apj, 351, 632

\bibitem[{Hunter {et~al.}(2012)Hunter, Macgregor, Szabo, Wellington, \&
  Bellgard}]{YABI}
Hunter, A., Macgregor, A., Szabo, T., Wellington, C., \& Bellgard, M. 2012,
  Source Code Biol. Med., 7, 1

\bibitem[{{Izidoro} {et~al.}(2015){Izidoro}, {Raymond}, {Morbidelli},
  {Hersant}, \& {Pierens}}]{2015ApJ...800L..22I}
{Izidoro}, A., {Raymond}, S.~N., {Morbidelli}, A.~r., {Hersant}, F., \&
  {Pierens}, A. 2015, \apjl, 800, L22

\bibitem[{{Kennedy} \& {Kenyon}(2008)}]{kennedyandkenyon2008}
{Kennedy}, G.~M. \& {Kenyon}, S.~J. 2008, \apj, 673, 502

\bibitem[{{Kjeldsen} \& {Bedding}(1995)}]{1995A&A...293...87K}
{Kjeldsen}, H. \& {Bedding}, T.~R. 1995, \aap, 293, 87

\bibitem[{{Kopparapu} {et~al.}(2013){Kopparapu}, {Ramirez}, {Kasting}, {Eymet},
  {Robinson}, {Mahadevan}, {Terrien}, {Domagal-Goldman}, {Meadows}, \&
  {Deshpande}}]{2013ApJ...765..131K}
{Kopparapu}, R.~K., {Ramirez}, R., {Kasting}, J.~F., {et~al.} 2013, \apj, 765,
  131

\bibitem[{{Kopparapu} {et~al.}(2014){Kopparapu}, {Ramirez}, {SchottelKotte},
  {Kasting}, {Domagal-Goldman}, \& {Eymet}}]{2014ApJ...787L..29K}
{Kopparapu}, R.~K., {Ramirez}, R.~M., {SchottelKotte}, J., {et~al.} 2014,
  \apjl, 787, L29

\bibitem[{{Lanza} {et~al.}(2018){Lanza}, {Malavolta}, {Benatti}, {Desidera},
  {Bignamini}, {Bonomo}, {Esposito}, {Figueira}, {Gratton}, {Scandariato},
  {Damasso}, {Sozzetti}, {Biazzo}, {Claudi}, {Cosentino}, {Covino}, {Maggio},
  {Masiero}, {Micela}, {Molinari}, {Pagano}, {Piotto}, {Poretti}, {Smareglia},
  {Affer}, {Boccato}, {Borsa}, {Boschin}, {Giacobbe}, {Knapic}, {Leto},
  {Maldonado}, {Mancini}, {Martinez Fiorenzano}, {Messina}, {Nascimbeni},
  {Pedani}, \& {Rainer}}]{2018A&A...616A.155L}
{Lanza}, A.~F., {Malavolta}, L., {Benatti}, S., {et~al.} 2018, \aap, 616, A155

\bibitem[{{Laskar}(1993)}]{lask93}
{Laskar}, J. 1993, Physica D Nonlinear Phenomena, 67, 257

\bibitem[{{Lecar} {et~al.}(2006){Lecar}, {Podolak}, {Sasselov}, \&
  {Chiang}}]{lecaretal2006}
{Lecar}, M., {Podolak}, M., {Sasselov}, D., \& {Chiang}, E. 2006, \apj, 640,
  1115

\bibitem[{{Leconte} {et~al.}(2010){Leconte}, {Chabrier}, {Baraffe}, \&
  {Levrard}}]{Leconteetal10}
{Leconte}, J., {Chabrier}, G., {Baraffe}, I., \& {Levrard}, B. 2010, \aap, 516,
  A64

\bibitem[{{Lindegren} {et~al.}(2018){Lindegren}, {Hern{\'a}ndez}, {Bombrun},
  {Klioner}, {Bastian}, {Ramos-Lerate}, {de Torres}, {Steidelm{\"u}ller},
  {Stephenson}, {Hobbs}, {Lammers}, {Biermann}, {Geyer}, {Hilger}, {Michalik},
  {Stampa}, {McMillan}, {Casta{\~n}eda}, {Clotet}, {Comoretto}, {Davidson},
  {Fabricius}, {Gracia}, {Hambly}, {Hutton}, {Mora}, {Portell}, {van Leeuwen},
  {Abbas}, {Abreu}, {Altmann}, {Andrei}, {Anglada}, {Balaguer-N{\'u}{\~n}ez},
  {Barache}, {Becciani}, {Bertone}, {Bianchi}, {Bouquillon}, {Bourda},
  {Br{\"u}semeister}, {Bucciarelli}, {Busonero}, {Buzzi}, {Cancelliere},
  {Carlucci}, {Charlot}, {Cheek}, {Crosta}, {Crowley}, {de Bruijne}, {de
  Felice}, {Drimmel}, {Esquej}, {Fienga}, {Fraile}, {Gai}, {Garralda},
  {Gonz{\'a}lez-Vidal}, {Guerra}, {Hauser}, {Hofmann}, {Holl}, {Jordan},
  {Lattanzi}, {Lenhardt}, {Liao}, {Licata}, {Lister}, {L{\"o}ffler},
  {Marchant}, {Martin-Fleitas}, {Messineo}, {Mignard}, {Morbidelli}, {Poggio},
  {Riva}, {Rowell}, {Salguero}, {Sarasso}, {Sciacca}, {Siddiqui}, {Smart},
  {Spagna}, {Steele}, {Taris}, {Torra}, {van Elteren}, {van Reeven}, \&
  {Vecchiato}}]{2018A&A...616A...2L}
{Lindegren}, L., {Hern{\'a}ndez}, J., {Bombrun}, A., {et~al.} 2018, \aap, 616,
  A2

\bibitem[{{Lissauer}(1987)}]{lissauer1987}
{Lissauer}, J.~J. 1987, \icarus, 69, 249

\bibitem[{{Lissauer} \& {Stevenson}(2007)}]{lissauerandstevenson2007}
{Lissauer}, J.~J. \& {Stevenson}, D.~J. 2007, Protostars and Planets V, 591

\bibitem[{{Lovis} {et~al.}(2011){Lovis}, {Dumusque}, {Santos}, {Bouchy},
  {Mayor}, {Pepe}, {Queloz}, {S{\'e}gransan}, \& {Udry}}]{2011arXiv1107.5325L}
{Lovis}, C., {Dumusque}, X., {Santos}, N.~C., {et~al.} 2011
  [\eprint[arXiv]{1107.5325}]

\bibitem[{{Malavolta} {et~al.}(2017){Malavolta}, {Lovis}, {Pepe}, {Sneden}, \&
  {Udry}}]{2017MNRAS.469.3965M}
{Malavolta}, L., {Lovis}, C., {Pepe}, F., {Sneden}, C., \& {Udry}, S. 2017,
  \mnras, 469, 3965

\bibitem[{{Mamajek} \& {Hillenbrand}(2008)}]{2008ApJ...687.1264M}
{Mamajek}, E.~E. \& {Hillenbrand}, L.~A. 2008, \apj, 687, 1264

\bibitem[{{Mandell} {et~al.}(2007){Mandell}, {Raymond}, \&
  {Sigurdsson}}]{2007ApJ...660..823M}
{Mandell}, A.~M., {Raymond}, S.~N., \& {Sigurdsson}, S. 2007, \apj, 660, 823

\bibitem[{{Markwardt}(2009)}]{2009ASPC..411..251M}
{Markwardt}, C.~B. 2009, in Astronomical Society of the Pacific Conference
  Series, Vol. 411, Astronomical Data Analysis Software and Systems XVIII, ed.
  D.~A. {Bohlender}, D.~{Durand}, \& P.~{Dowler}, 251

\bibitem[{{Marzari} {et~al.}(2003){Marzari}, {Tricarico}, \& {Scholl}}]{marz03}
{Marzari}, F., {Tricarico}, P., \& {Scholl}, H. 2003, \mnras, 345, 1091

\bibitem[{{Matsumura} {et~al.}(2013){Matsumura}, {Ida}, \&
  {Nagasawa}}]{2013ApJ...767..129M}
{Matsumura}, S., {Ida}, S., \& {Nagasawa}, M. 2013, \apj, 767, 129

\bibitem[{{Mayor} {et~al.}(2004){Mayor}, {Udry}, {Naef}, {Pepe}, {Queloz},
  {Santos}, \& {Burnet}}]{2004A&A...415..391M}
{Mayor}, M., {Udry}, S., {Naef}, D., {et~al.} 2004, \aap, 415, 391

\bibitem[{{Morbidelli} {et~al.}(2015){Morbidelli}, {Lambrechts}, {Jacobson}, \&
  {Bitsch}}]{morbidellietal2015}
{Morbidelli}, A., {Lambrechts}, M., {Jacobson}, S., \& {Bitsch}, B. 2015,
  \icarus, 258, 418

\bibitem[{{Mortier} \& {Collier Cameron}(2017)}]{2017A&A...601A.110M}
{Mortier}, A. \& {Collier Cameron}, A. 2017, \aap, 601, A110

\bibitem[{{Nardetto} {et~al.}(2006){Nardetto}, {Mourard}, {Kervella},
  {Mathias}, {M{\'e}rand}, \& {Bersier}}]{2006A&A...453..309N}
{Nardetto}, N., {Mourard}, D., {Kervella}, P., {et~al.} 2006, \aap, 453, 309

\bibitem[{{Natta} {et~al.}(2000){Natta}, {Grinin}, \&
  {Mannings}}]{nattaetal2000}
{Natta}, A., {Grinin}, V., \& {Mannings}, V. 2000, Protostars and Planets IV,
  559

\bibitem[{{Noyes} {et~al.}(1984){Noyes}, {Hartmann}, {Baliunas}, {Duncan}, \&
  {Vaughan}}]{1984ApJ...279..763N}
{Noyes}, R.~W., {Hartmann}, L.~W., {Baliunas}, S.~L., {Duncan}, D.~K., \&
  {Vaughan}, A.~H. 1984, \apj, 279, 763

\bibitem[{{Ogilvie}(2014)}]{2014ARA&A..52..171O}
{Ogilvie}, G.~I. 2014, \araa, 52, 171

\bibitem[{{Pepe} {et~al.}(2002){Pepe}, {Mayor}, {Galland}, {Naef}, {Queloz},
  {Santos}, {Udry}, \& {Burnet}}]{2002A&A...388..632P}
{Pepe}, F., {Mayor}, M., {Galland}, F., {et~al.} 2002, \aap, 388, 632

\bibitem[{{Queloz} {et~al.}(2001){Queloz}, {Henry}, {Sivan}, {Baliunas},
  {Beuzit}, {Donahue}, {Mayor}, {Naef}, {Perrier}, \&
  {Udry}}]{2001A&A...379..279Q}
{Queloz}, D., {Henry}, G.~W., {Sivan}, J.~P., {et~al.} 2001, \aap, 379, 279

\bibitem[{{Rathbun} {et~al.}(2004){Rathbun}, {Spencer}, {Tamppari}, {Martin},
  {Barnard}, \& {Travis}}]{Rathbunetal04}
{Rathbun}, J.~A., {Spencer}, J.~R., {Tamppari}, L.~K., {et~al.} 2004, \icarus,
  169, 127

\bibitem[{{Ricker} {et~al.}(2015){Ricker}, {Winn}, {Vanderspek}, {Latham},
  {Bakos}, {Bean}, {Berta-Thompson}, {Brown}, {Buchhave}, {Butler}, {Butler},
  {Chaplin}, {Charbonneau}, {Christensen-Dalsgaard}, {Clampin}, {Deming},
  {Doty}, {De Lee}, {Dressing}, {Dunham}, {Endl}, {Fressin}, {Ge}, {Henning},
  {Holman}, {Howard}, {Ida}, {Jenkins}, {Jernigan}, {Johnson}, {Kaltenegger},
  {Kawai}, {Kjeldsen}, {Laughlin}, {Levine}, {Lin}, {Lissauer}, {MacQueen},
  {Marcy}, {McCullough}, {Morton}, {Narita}, {Paegert}, {Palle}, {Pepe},
  {Pepper}, {Quirrenbach}, {Rinehart}, {Sasselov}, {Sato}, {Seager},
  {Sozzetti}, {Stassun}, {Sullivan}, {Szentgyorgyi}, {Torres}, {Udry}, \&
  {Villasenor}}]{2015JATIS...1a4003R}
{Ricker}, G.~R., {Winn}, J.~N., {Vanderspek}, R., {et~al.} 2015, Journal of
  Astronomical Telescopes, Instruments, and Systems, 1, 014003

\bibitem[{{Rodrigues} {et~al.}(2014){Rodrigues}, {Girardi}, {Miglio},
  {Bossini}, {Bovy}, {Epstein}, {Pinsonneault}, {Stello}, {Zasowski}, {Allende
  Prieto}, {Chaplin}, {Hekker}, {Johnson}, {M{\'e}sz{\'a}ros}, {Mosser},
  {Anders}, {Basu}, {Beers}, {Chiappini}, {da Costa}, {Elsworth},
  {Garc{\'{\i}}a}, {Garc{\'{\i}}a P{\'e}rez}, {Hearty}, {Maia}, {Majewski},
  {Mathur}, {Montalb{\'a}n}, {Nidever}, {Santiago}, {Schultheis}, {Serenelli},
  \& {Shetrone}}]{2014MNRAS.445.2758R}
{Rodrigues}, T.~S., {Girardi}, L., {Miglio}, A., {et~al.} 2014, \mnras, 445,
  2758

\bibitem[{{S{\'a}nchez} {et~al.}(2018){S{\'a}nchez}, {de El{\'\i}a}, \&
  {Darriba}}]{2018MNRAS.481.1281S}
{S{\'a}nchez}, M.~B., {de El{\'\i}a}, G.~C., \& {Darriba}, L.~A. 2018, \mnras,
  481, 1281

\bibitem[{{Sasselov} \& {Lecar}(2000)}]{sasselovandlecar2000}
{Sasselov}, D.~D. \& {Lecar}, M. 2000, \apj, 528, 995

\bibitem[{{Sneden}(1973)}]{1973ApJ...184..839S}
{Sneden}, C. 1973, \apj, 184, 839

\bibitem[{{Stevenson} \& {Lunine}(1988)}]{stevensonetal1988}
{Stevenson}, D.~J. \& {Lunine}, J.~I. 1988, \icarus, 75, 146

\bibitem[{{Tal-Or} {et~al.}(2019){Tal-Or}, {Trifonov}, {Zucker}, {Mazeh}, \&
  {Zechmeister}}]{2019MNRAS.484L...8T}
{Tal-Or}, L., {Trifonov}, T., {Zucker}, S., {Mazeh}, T., \& {Zechmeister}, M.
  2019, \mnras, 484, L8

\bibitem[{{Th{\'e}bault} {et~al.}(2002){Th{\'e}bault}, {Marzari}, \&
  {Scholl}}]{2002A&A...384..594T}
{Th{\'e}bault}, P., {Marzari}, F., \& {Scholl}, H. 2002, \aap, 384, 594

\bibitem[{{Tuomi} {et~al.}(2014){Tuomi}, {Jones}, {Barnes},
  {Anglada-Escud{\'e}}, \& {Jenkins}}]{2014MNRAS.441.1545T}
{Tuomi}, M., {Jones}, H. R.~A., {Barnes}, J.~R., {Anglada-Escud{\'e}}, G., \&
  {Jenkins}, J.~S. 2014, \mnras, 441, 1545

\bibitem[{{van Leeuwen}(2007)}]{2007A&A...474..653V}
{van Leeuwen}, F. 2007, \aap, 474, 653

\bibitem[{{{\v{S}}idlichovsk{\'y}} \& {Nesvorn{\'y}}(1996)}]{sine96}
{{\v{S}}idlichovsk{\'y}}, M. \& {Nesvorn{\'y}}, D. 1996, Celestial Mechanics
  and Dynamical Astronomy, 65, 137

\bibitem[{{Weiss} {et~al.}(2018){Weiss}, {Marcy}, {Petigura}, {Fulton},
  {Howard}, {Winn}, {Isaacson}, {Morton}, {Hirsch}, {Sinukoff}, {Cumming},
  {Hebb}, \& {Cargile}}]{2018AJ....155...48W}
{Weiss}, L.~M., {Marcy}, G.~W., {Petigura}, E.~A., {et~al.} 2018, \aj, 155, 48

\bibitem[{{Zechmeister} \& {K{\"u}rster}(2009)}]{2009A&A...496..577Z}
{Zechmeister}, M. \& {K{\"u}rster}, M. 2009, \aap, 496, 577

\bibitem[{{Zhu} \& {Wu}(2018)}]{2018AJ....156...92Z}
{Zhu}, W. \& {Wu}, Y. 2018, \aj, 156, 92

\end{thebibliography}
\end{document}